\documentclass[%
 aip,
 jmp,%
 amsmath,amssymb,
 preprint,%
]{revtex4-1}

\usepackage{graphicx}
\usepackage{dcolumn}
\usepackage{bm}
\usepackage{epstopdf, epsfig}
\usepackage{color}
\usepackage{caption}
\usepackage{subcaption}
\usepackage{amsmath}
\usepackage{gensymb}
\usepackage{multirow}
\usepackage{hyperref}
\hypersetup{%
 bookmarksnumbered,
 pdfstartview=FitH,
 unicode,
 pdfborder = {0 0 0},
 colorlinks,
 citecolor=blue,
 filecolor=Darkgreen,
 linkcolor=blue,
 urlcolor=cyan!50!black!90
}

\AppendGraphicsExtensions{.tif}

\begin{document}

\preprint{AIP/123-QED}

\title[Attached flow structure and streamwise energy spectra]{Attached flow structure and streamwise energy spectra in a
  turbulent boundary layer:\\}

\author{S. Srinath$^{1,4}$, J. C. Vassilicos$^{1,2}$, C. Cuvier$^{1,4}$, J. -P. Laval$^{3,4}$, M. Stanislas$^{1}$ and J. -M. Foucaut$^{1,4}$}
\affiliation{%
$^1$Centrale Lille, F-59000 Lille, France \\ 
$^2$Department of Aeronautics, Imperial College London, London SW7 2AZ, United Kingdom \\ 
$^3$CNRS, FRE 3723, F-59650 Villeneuve d'Ascq, France \\
$^4$Univ. Lille, FRE 3723 Laboratoire de M\'ecanique de Lille, F-59000 Lille, France
}

\date{\today}

\begin{abstract}
On the basis of (i) Particle Image Velocimetry data of a Turbulent
Boundary Layer with large field of view and good spatial resolution
and (ii) a mathematical relation between the energy spectrum and
specifically modeled flow structures, we show that the scalings of the
streamwise energy spectrum $E_{11}(k_{x})$ in a wavenumber range
directly affected by the wall are determined by wall-attached eddies
but are not given by the Townsend-Perry attached eddy model's
prediction of these spectra, at least at the Reynolds numbers
$Re_{\tau}$ considered here which are between $10^{3}$ and
$10^{4}$. Instead, we find $E_{11}(k_{x}) \sim k_{x}^{-1-p}$ where $p$
varies smoothly with distance to the wall from negative values in the
buffer layer to positive values in the inertial layer. The exponent
$p$ characterises the turbulence levels inside wall-attached streaky
structures conditional on the length of these structures.
\end{abstract}

\keywords{Turbulent Boundary Layers, wall-attached flow structures, Particle Image Velocimetry}
\maketitle

\section{\label{sec:Introduction}Introduction:\protect\\}

In the past forty years, the turbulence spectrum of velocity
fluctuations in wall turbulence has received considerable attention as
it gives valuable insight into the behaviour of wall-bounded flows by
indicating the distribution of energy across scales. Spectral scaling
laws built on ideas initiated by Townsend [\onlinecite{townsend1976structure}] , in
particular the attached eddy hypothesis, have seen consistent
development over the years (see Refs. Perry \& Chong [\onlinecite{perry1982mechanism}], Perry \textit{et al.} [\onlinecite{perry1986theoretical}], Perry \& Li [\onlinecite{perry1990experimental}], Marusic \textit{et al.} [\onlinecite{marusic1997similarity}] and Marusic \& Kunkel [\onlinecite{marusic2003streamwise}]). Perry \& Abell [\onlinecite{perry1977asymptotic}] and Perry \textit{et al.} [\onlinecite{perry1986theoretical}] showed how Townsend's
attached eddy hypothesis implies that the energy spectrum
$E_{11}(k_{x})$ of the turbulent streamwise fluctuating velocity at a
distance $y$ from the wall scales as $E_{11}(k_{x}) \sim U_{\tau}^{2}
k_{x}^{-1}$ in the range $1/\delta \ll k_{x} \ll 1/y$ where $U_{\tau}$
is the friction velocity and $\delta$ is the boundary layer thickness. Nickels \textit{et al.} [\onlinecite{nickels2005evidence}] stressed the use of overlap arguments to
deduce the -1 power law behaviour. That is, a $k_x^{-1}$ region in the
spectra would exist where the inner scaling (based on $y$ and
$U_\tau$) and outer scaling (based on $\delta$ and $U_\tau$) are
simultaneously valid over the same wavenumber range. Nickels \textit{et al.} [\onlinecite{nickels2007some}] stated that it is necessary to take
measurements surprisingly close to the wall to observe a $k_x^{-1}$
behaviour and thought this was the reason why Morrison \textit{et al.}
[\onlinecite{morrison2004scaling}] and McKeon \& Morrison [\onlinecite{mckeon2007asymptotic}] did not
observe any $-1$ region in their spectra as their measurements were
not close enough to the wall. However, recent experiments by Vallikivi \textit{et al.} [\onlinecite{vallikivi2015spectral}] do not show an overlap region and these authors infer that the $k_x^{-1}$ region cannot be expected even at very high Reynolds numbers.

The present work looks at the basis for the $k_x^{-1}$ range in flat
plate turbulent boundary layers from a new perspective. Using Particle
Image Velocimetry (PIV) and a simple model which can in principle be
applied to various wall-bounded turbulent flows, we show how, in the
turbulent boundary layer, a power-law spectral range exists but is not
a Townsend-Perry $k_x^{-1}$ range and how it can be accounted for by
taking only streamwise lengths and intensities of wall-attached
structures into account.

This paper is organized as follows.  In sections \ref{sec: simplest
  possible model} and \ref{sec:Modified TP range} we provide a model
for the streamwise energy spectrum. The experimental set-up of the
flat plate boundary layer is presented in section \ref{sec:Exp
  setup}. Our data set is validated in \ref{sec:Validation of
  experimental data} and the method for educing the wall-attached flow
structures relevant to our model is described in section
\ref{sec:Structure detection}.
The main results of the paper are in \ref{sec:Lengths} and
\ref{sec:spectra} followed by a discussion in \ref{sec:discussion}. We
conclude in section \ref{Conclusion}.

\section{A simple model for the spectral signature of the Townsend-Perry attached eddy range of wavenumbers} \label{sec: simplest possible model}

As already mentioned in the introduction, Perry \& Abell [\onlinecite{perry1977asymptotic}], Perry \& Chong
[\onlinecite{perry1982mechanism}] and Perry \textit{et al.}
[\onlinecite{perry1986theoretical}] showed how Townsend's attached eddy
hypothesis implies $E_{11}(k_{x}) \sim U_{\tau}^{2} k_{x}^{-1}$ in the
range $1/\delta \ll k_{x} \ll 1/y$.  Perry \textit{et al.}
[\onlinecite{perry1986theoretical}] also
developed a flow structure model for this spectral range in terms of
specific attached eddies of varying sizes randomly distributed in
space and with a number density that is inversely proportional to
size. In this paper we attempt to distill such a type of model to its
bare essentials. \textcolor{black}{These bare essentials are that flow
  structures are primarily objects with clear spatial boundaries. In
  section \ref{sec:Resuls and discussion} we model these boundaries with on-off functions in the
  expectation that the spectral signature in the attached eddy
  wavenumber range is dominated by these sharp gradient, effectively
  on-off, behaviours. The concomitant expectation is that the
  additional superimposed velocity fluctuations fill the content of a
  predominantly higher frequency spectral range. In this section we
  show that the streamwise energy spectrum's $k_{x}^{-1}$ spectral
  range can be captured by simple on-off representations of elongated
  streaky structures of varying sizes as long as their number density
  has a space-filling power law dependence on size.}

We therefore assume that the attached eddies responsible for the
$k_{x}^{-1}$ spectral range have a long streaky structure footprint on
the 1D streamwise fluctuating velocity signals at a distance $y$ from
the wall.  We also assume that these streaky structures can be modeled
as simple on-off functions and that it is sufficient to represent the
streamwise velocity fluctuations $u(x)$ at a given height $y$ from the
wall as follows
\begin{equation}
  u(x) = \sum_{n,m} a_{nm} \Pi (\xi)
  \label{eq1}
\end{equation}
where $\Pi (\xi) = 1$ if $-1 < \xi < 1$ with $\xi = {x-x_{nm}\over
  \lambda_{n}}$ and $\Pi (\xi) = 0$ otherwise. The on-off function
$\Pi (\xi)$ is our cartoon model of a streaky structure. Streaky
structures of length $\lambda_n$ are centred at random positions
$x_{nm}$ and their intensity is given by the coefficients
$a_{nm}$. For each subscript $n$, the subscript $m$ counts the spatial
positions where cartoon structures of size $\lambda_n$ can be centred
in a given realisation. The sum in (\ref{eq1}) is over all structures
lengths $\lambda_n$ and all their positions $x_{nm}$.

The energy spectrum of $u(x)$ is $ E_{11} (k_{x}) = {(2\pi)^{2}\over
  L_{x}} \overline{\vert \hat{u}(k_{x})\vert^{2}} $ where $L_x$ is the
length of the record, $\hat{u}(k_{x})$ is the Fourier transform of
$u(x)$, and the overbar signifies an average over realisations. The
Fourier transform of $\Pi ( {x-x_{nm}\over \lambda_{n}} )$ being
$\hat{\Pi}(k_{x}, \lambda_{n}, x_{nm}) =2i k_{x}^{-1} e^{ik_{x}x_{nm}}
\sin (k_{x}\lambda_{n})$, it follows that
\begin{equation}
  \hat{u}(k_{x}) = 2i k_{x}^{-1} \sum_{nm} a_{nm} e^{ik_{x}x_{nm}} \sin
  (k_{x}\lambda_{n})
  \label{eq2}
\end{equation}
which implies that the energy spectrum is given by 
\begin{widetext}
\begin{equation}
  E_{11} (k_{x}) = 4 {(2\pi)^{2}\over L_{x}} k_{x}^{-2}
  \overline{\sum_{nm} a_{nm} e^{ik_{x}x_{nm}} \sin (k_{x}\lambda_{n})
    \sum_{pq} a_{pq} e^{-ik_{x}x_{pq}} \sin (k_{x}\lambda_{p})}.
  \label{eq3}
\end{equation}
\end{widetext}

We introduce two assumptions which were also used by
Perry \textit{et al.}
[\onlinecite{perry1986theoretical}] in their more intricate model. The first
assumption is that the positions and amplitudes of our cartoon
stuctures are uncorrelated and that different positions are not
correlated to each other either, i.e. $\overline{e^{ik_{x}x_{nm}}
  e^{ik_{x}x_{pq}}} = \delta_{pn} \delta_{qm}$. As a result, the
expression for the energy spectrum simplifies as follows:
\begin{equation}
  E_{11} (k_{x}) = 4 {(2\pi)^{2}\over L_{x}} k_{x}^{-2} \sum_{nm}
  \overline{(a_{nm})^{2}} \sin^{2} (k_{x}\lambda_{n}).
  \label{eq4}
\end{equation}

Let us say that there is an average number $N_n$ of cartoon stuctures
of size $\lambda_{n}$ centred within an integral scale along the
$x$-axis. The expression for $E_{11}(k_{x})$ simplifies even further:
\begin{equation}
  E_{11} (k_{x}) = 4 {(2\pi)^{2}\over L_{x}} k_{x}^{-2} \sum_{n}
  \overline{a_{n}^{2}} N_{n} \sin^{2} (k_{x}\lambda_{n})
  \label{eq5}
\end{equation}
where $\overline{a_{n}^{2}} \equiv \overline{(a_{nm})^{2}}$ is the
same irrespective of position $x_{nm}$.

We now consider a continuum of different structure sizes $\lambda$
rather than discrete length-scales $\lambda_n$ and the previous
expression for $E_{11}(k_{x})$ must therefore be replaced by
\begin{equation} 
E_{11} (k_{x}) = 4 {(2\pi)^{2}\over L_{x}} k_{x}^{-2} \int d\lambda
\overline{a^{2}}(\lambda) N(\lambda) \sin^{2} (k_{x}\lambda)
\label{eq6}
\end{equation}
in terms of easily understandable notation. At this point we introduce
a generalised form of the second assumption which was also used by
Perry \textit{et al.}
[\onlinecite{perry1986theoretical}]: we assume a power-law form for
$N(\lambda)$ in the range $\lambda_{i}< \lambda < \lambda_{o}$ where
$\lambda_{i} \sim y$ and $\lambda_{o} \sim \delta$, and $N(\lambda) =
0 $ outside this range for simplicity. This power law form is $N
(\lambda) = (-N_M + N_{o} (\lambda/\delta)^{-1-D})$ where $N_M$ and
$N_o$ are positive dimensionless numbers which increase propotionally
to $L_x$ so as to keep number densities constant. The number $N_M$ is
introduced to allow for the possibility of an upper bound on streaky
structure size given by $N (\lambda_{o}) = 0$, i.e. $N_{M} = N_{o}
(\lambda_{o}/\delta)^{-1-D}$ which should be small given that LSM and
VLSM streaky structures have been observed with lengths greater than
$\delta$ [see Smits \textit{et al.} \onlinecite{smits2011high}].

Vassilicos \& Hunt [\onlinecite{vassilicos1991fractal}] proved that, if $0 \le D \le 1$, then
the set of points defining the edges of the on-off functions $\Pi
(\xi)$ is fractal and $D$ is effectively the fractal dimension of this
set of points. The case where this fractal dimension is $D=1$ is the
case where these points are space-filling.
The population density assumption of Perry \textit{et al.}
[\onlinecite{perry1986theoretical}]
corresponds to $D=1$ which is also the choice we make in this work. We
now show that this choice can lead to $E_{11}(k_{x}) \sim k_{x}^{-1}$
in the range $1/\lambda_{o} \ll k_{x} \ll 1/\lambda_{i}$.

We calculate the energy spectrum by carrying out the integral in
(\ref{eq6}). This requires a model for $\overline{a^{2}}(\lambda)$
which, in this section, we chose to be as simple as possible and
therefore independent of $\lambda$ in the relevant range,
i.e. $\overline{a^{2}}(\lambda) = A^{2}/\delta$ for $\lambda_{i}<
\lambda < \lambda_{o}$ where $A^2$ is a constant. Using our models for
$N(\lambda)$ and $\overline{a^{2}}(\lambda)$ and the change of
variables $\lambda k_{x} = l$, (\ref{eq6}) becomes
\begin{equation} 
E_{11} (k_{x}) = A^{2}\delta (C_{o} (k_{x}\delta)^{-2+D} - C_{M}
(k_{x}\delta)^{-2})
\label{eq7}
\end{equation}
where 
$$ 
C_{o} = 4(2\pi)^{2} N_{o} {\delta\over L_{x}}
\int_{\lambda_{i}k_{x}}^{\lambda_{o}k_{x}} dl \sin^{2} (l) l^{-1-D}
$$ 
and
$$ 
C_{M} = 4(2\pi)^{2} N_{M} {\delta\over L_{x}} (k_{x}\delta)^{-1}
\int_{\lambda_{i}k_{x}}^{\lambda_{o}k_{x}} dl \sin^{2} (l)
$$
which is bounded from above by ${N_{M}\over L_{x}}{\lambda_{o}
  -\lambda_{i}\over \delta}$. In the attached eddy range
$1/\lambda_{o} \ll k_{x} \ll 1/\lambda_{i}$, $C_{o} \approx
4(2\pi)^{2} {N_{o}\over L_{x}} \int_{0}^{\infty} dl \sin^{2} (l)
l^{-1-D}$ which means that $C_o$ is approximately independent of
$k_{x}$ in this range.

Substituting the value $D=1$ in equation (\ref{eq7}), we get $E_{11}
(k_{x}) = A^{2} (C_{o} k_{x}^{-1} - C_{M} \delta^{-1}k_{x}^{-2})$
which is well approximated by
\begin{equation} 
E_{11} (k_{x}) \approx C_{o} A^{2} k_{x}^{-1}
\label{eq8}
\end{equation}
for wavenumbers $k_{x}\delta \gg C_{M}/C_{o}$ (i.e. $C_{o}
k_{x}^{-1}\gg C_{m} \delta^{-1} k_{x}^{-2}$). Note that $C_{M}/C_{o}$
is much smaller than 1 because $N_M$ is much smaller than $N_o$ and
that (\ref{eq8}) is valid in the range $1/\lambda_{o} \ll k_{x} \ll
1/\lambda_{i}$ where $\lambda_{o}$ scales with but is much larger than
$\delta$. For a good correspondence with the scalings of the
Townsend-Perry attached eddy model one needs to take $\lambda_{i} \sim
y$ and $A^{2} \sim U_{\tau}^{2}$.

\section{A straightforward generalisation}\label{sec:Modified TP range}

It is worth generalising the previous section's model by assuming that
$\overline{a^{2}}(\lambda)$ is not constant but varies with $\lambda$
in the range $\lambda_{i} < \lambda < \lambda_{o}$, for example as
$\overline{a^{2}}(\lambda)=(A^{2}/\delta) (\lambda/\delta)^{p}$ where
$p$ is a real number with bounds which we determine below.  The
arguments of the previous section can be reproduced till equation
(\ref{eq6}) which now becomes
\begin{eqnarray}
 E_{11} (k_{x}) & = & A^{2} \delta [c_{o} (k_{x}\delta)^{-2+D-p} -
   c_{M} (k_{x}\delta)^{-2}]
 \label{eq9}
\end{eqnarray}
where 
$$ 
c_{o} = 4(2\pi)^{2} N_{o}{\delta \over L_{x}}
\int_{\lambda_{i}k_{x}}^{\lambda_{o}k_{x}} dl \sin^{2} (l) l^{-1-D+p}
$$ 
and
$$
c_{M} = 4(2\pi)^{2} N_{M}{\delta \over L_{x}} (k_{x}\delta)^{-1-p}
\int_{\lambda_{i}k_{x}}^{\lambda_{o}k_{x}} dl \; l^{+p}\sin^{2} (l)
$$ 
which is bounded from above by ${N_{m}\over
  (1+p)L_{x}}[({\lambda_{o}\over \delta})^{1+p} -({\lambda_{i}\over
    \delta})^{1+p}]$. In the attached eddy range $1/\lambda_{o} \ll
k_{x} \ll 1/\lambda_{i}$, $c_{o} \approx 4(2\pi)^{2} {N_{o}\over
  L_{x}} \int_{0}^{\infty} dl \sin^{2} (l) l^{-1-D+p}$ which means
that $c_o$ is approximately independent of $k_{x}$ in this range if
$0<D-p<2$.

Substituting the value $D=1$ in (\ref{eq9}), we obtain the following
leading order approximation in the parameter range $-1<p<1$:
\begin{eqnarray}
 E_{11} (k_{x}) \approx c_{0} A^{2}\delta (k_{x}\delta)^{q}
 \label{eq10}
\end{eqnarray}
where
\begin{eqnarray}
 p+q=-1
 \label{eq10bis}
\end{eqnarray}
for wavenumbers $k_{x}\delta \gg (c_{M}/c_{o})^{{1\over 1-p}}$. Note
that $c_{M}/c_{o}$ is much smaller than 1 if $p$ is not too close to
$1$ because $N_M$ is much smaller than $N_o$.

The spectral shape (\ref{eq10}) is potentially significantly different
from what the classical Townsend-Perry attached eddy model predicts.
We emphasize that in this and the previous sections we have developed
a simple model based on on-off functions representing long streaky
structures which returns a wavenumber dependency of $E_{11} (k_{x})$
which is either identical to the Townsend-Perry spectral shape if
$p=0$, or different but in some ways comparable if $p\not = 0$. In the
remainder of this paper we present experimental evidence in support of
$D=1$ and (\ref{eq10})-(\ref{eq10bis}) rather than (\ref{eq8}), with
$p$ as function of $y^{+}$.

\section{Experimental set-up} \label{sec:Exp setup}

An experiment was performed in the boundary layer wind tunnel at the
Lille Mechanics Laboratory (LML) having a test section $2$m wide, 1m
high and $20.6$m long. The tests were conducted at two free stream
velocities of $3$m/s and $10$m/s corresponding to Reynolds numbers
$Re_{\theta} = 8100$ \textcolor{black}{($Re_{\tau}=2700$)} and
$Re_{\theta} = 20600$ \textcolor{black}{($Re_{\tau}=7200$)}
respectively. To capture the large streamwise wall-normal field, four
$12$ bits Hamamatsu cameras having a resolution of $2048$x$2048$
pixels were installed in series to observe a region between $19.26$m
and $20.42$m from inlet which is $1.16$m long ($\approx 3.36\delta$
and $3.85\delta$, for $Re_{\theta} = 8100$ and $20600$ respectively)
and 0.3m high ($\approx 0.86\delta$ and $1\delta)$ for $Re_{\theta} =
8100$ and $20600$ respectively). Nikon lenses of 50mm focal length
were set on the cameras and the magnification obtained was 0.05. The
Software HIRIS was used to acquire the images of the four cameras
simultaneously. A total of $22 500$ and $29 500$ samples were recorded
at the highest and lowest Reynolds numbers respectively. The flow was
seeded with $1\mu m$ Poly-Ethylene glycol and illuminated by a double
pulsed NdYAG laser at $400$mJ/pulse. The modified version by LML of
MatPIV toolbox, was used under Matlab to process the acquired images
from the 2D2C PIV. A multipass software was used with a final pass of
28x28 pixels (with a mean overlap of $65\%$) corresponding to 4mm x
4mm i.e. 33x33 wall units for $Re_{\theta} = 8100$ and 100x100 wall
units for $Re_{\theta} = 20600$.  Image deformation was applied at the final pass. The final grid had $766$ points along
the wall and 199 points in the wall-normal direction with a grid
spacing of $1.5$mm corresponding to 11 wall units and $35$ wall units
for the test cases at $Re_{\theta} = 8100$ and $Re_{\theta}
= 20600$ respectively.

\section{Results and discussion} \label{sec:Resuls and discussion}
\subsection{Validation of experimental data} \label{sec:Validation of experimental data}

\begin{figure}

	   {\includegraphics[width=0.85\textwidth]{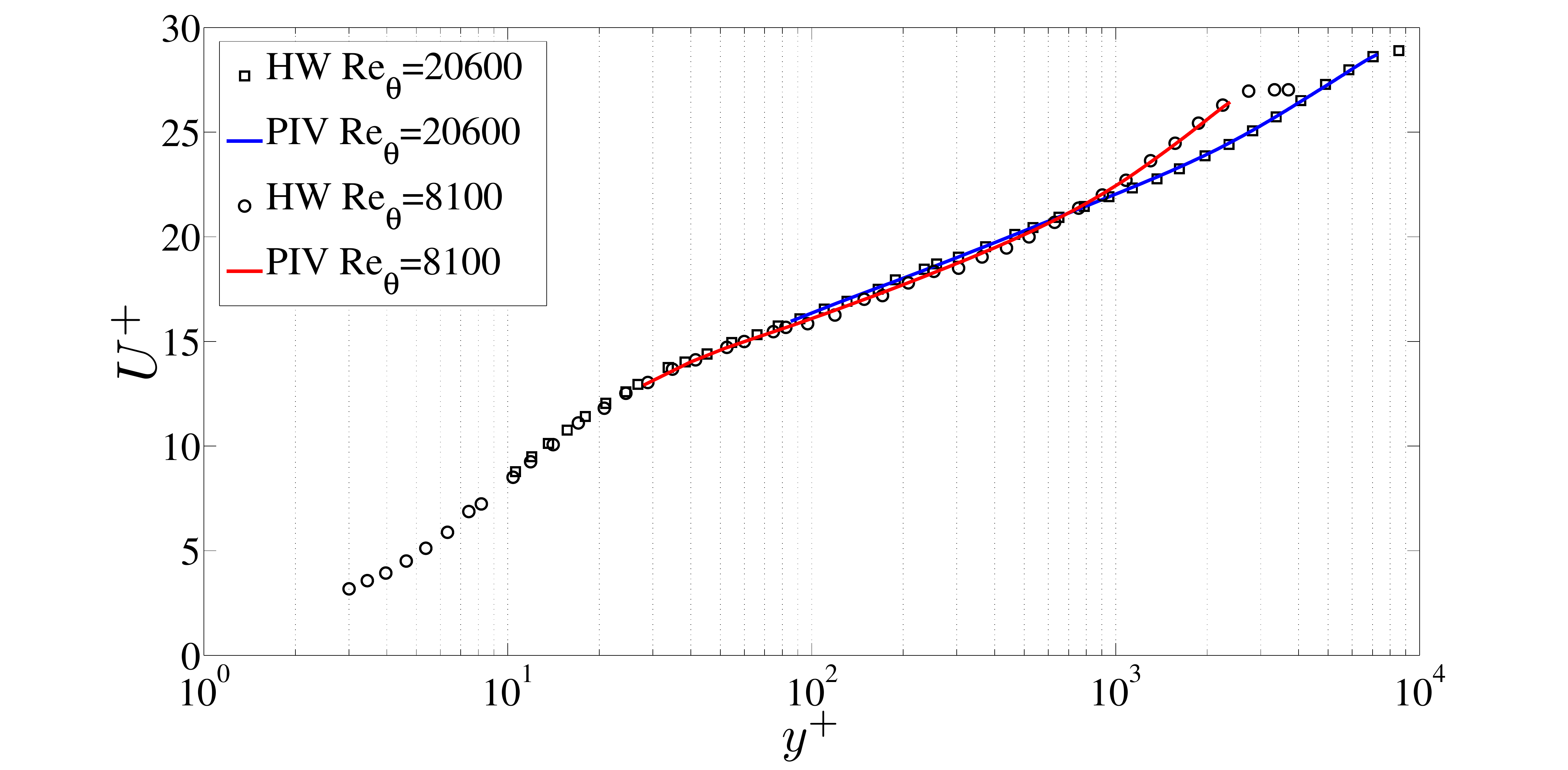}}
		{\includegraphics[width=0.85\textwidth]{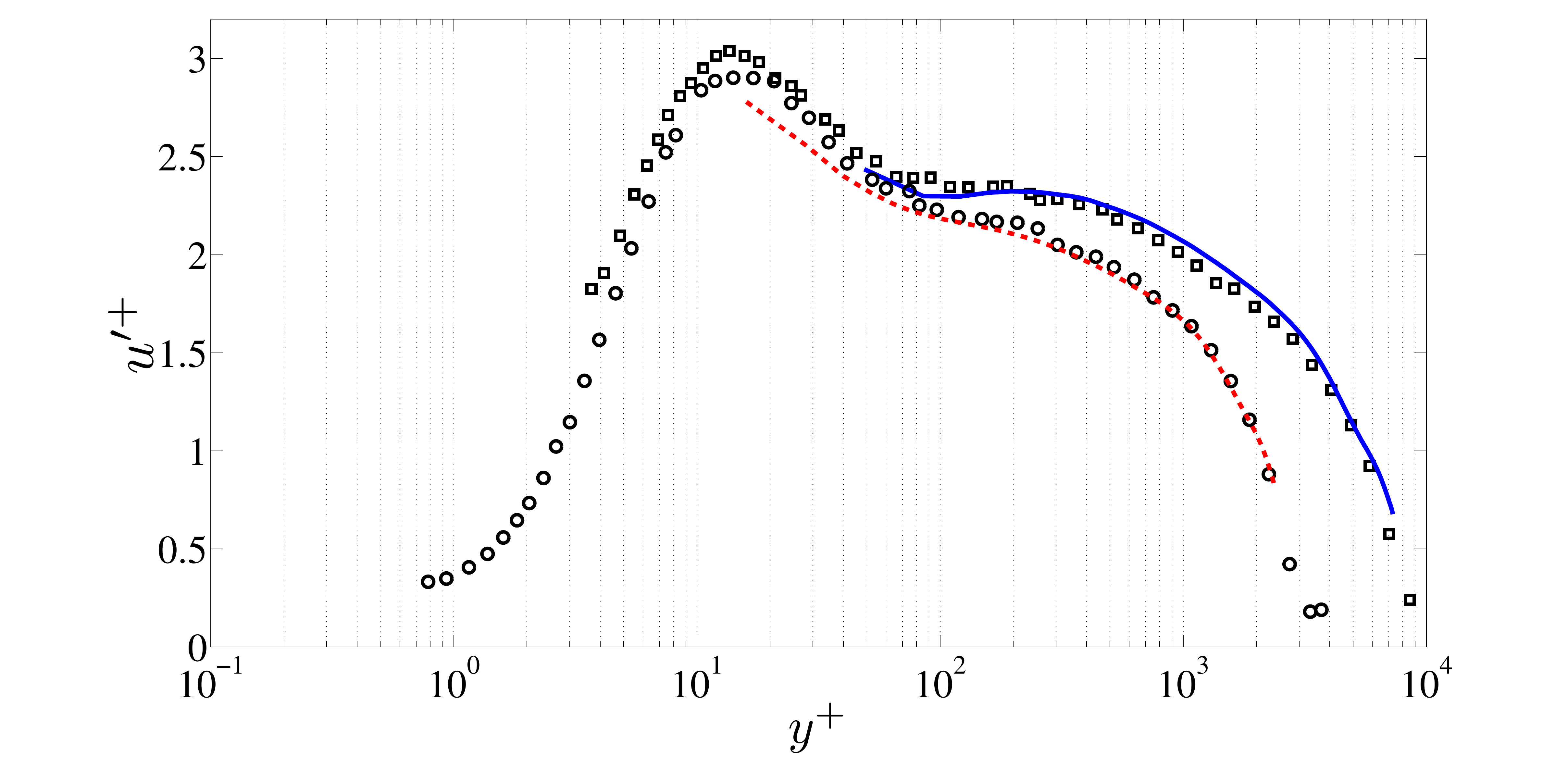}}
		\caption{Mean streamwise velocity profiles (top) and
                  rms streamwise fluctuating velocity profiles
                  (bottom) at $Re_{\theta} = 8100$ ($U_\infty
                  =3m/s$) and $Re_{\theta} = 20600$ ($U_\infty
                  =10m/s$) obtained with PIV and compared with the hot
                  wire anemometry results of Carlier \& Stanislas
                  [\onlinecite{carlier2005experimental}].}
		\label{fig:figure1}
\end{figure}

\begin{figure}
\centering

	   \centerline{\includegraphics[width=1\textwidth]{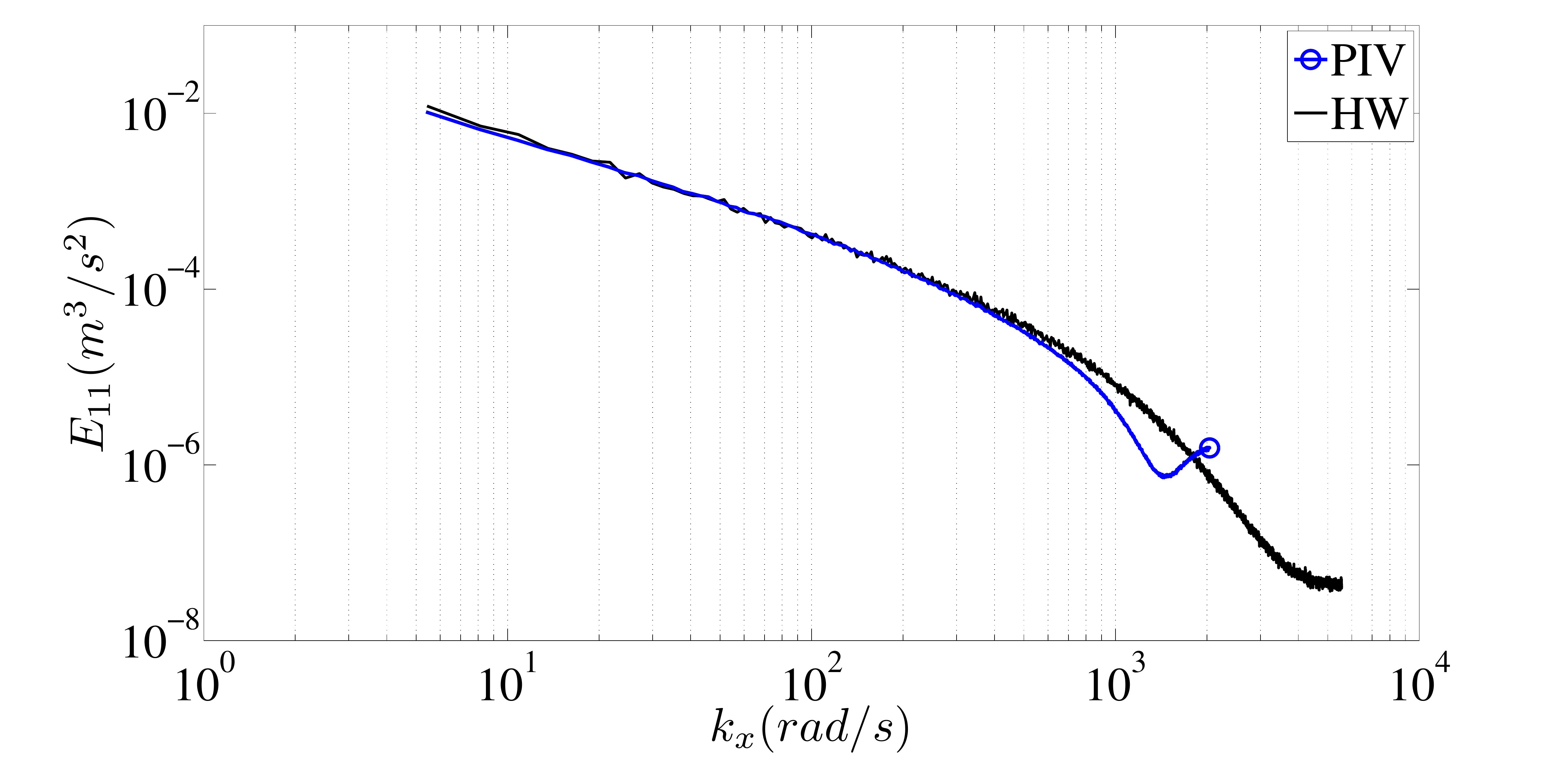}}
		\caption{Comparison of the streamwise energy spectra
                  obtained from PIV and hot-wire anemometry at
                  $y^+=200$ for $Re_{\theta} = 20600$. $E_{11}(k_{x})$
                  is in $m^{3}/s^2$ and $k_{x}$ is in $rad/m$. The
                  hot-wire anemometry was made by Carlier \& Stanislas
                  [\onlinecite{carlier2005experimental}] at $19.6$m from wind tunnel inlet in the same
                  wind tunnel.}
	\quad
	\label{fig:figure2}
\end{figure}

Figure \ref{fig:figure1} shows profiles of the mean streamwise
velocity $U$ and the rms streamwise fluctuating velocity $u'$ obtained
from PIV at $Re_{\theta} = 8100$ and $Re_{\theta} = 20600$ and
compared with the hot-wire anemometry results of Carlier \& Stanislas
[\onlinecite{carlier2005experimental}]. The mean velocity profiles are in
good agreement with the hot-wire data and are well resolved from $y^+
\approx30$ and $y^+ \approx90$ upwards for $Re_{\theta}=8100$ and
$20600$ respectively. Comparisons of the profiles of $u'^{+}$ ($u'$
scaled with inner variables) show a fairly good match with the
hot-wire data. A plateau of $u'^{+}$ is present in the range
$100<y^+<300$ for our higher Reynolds number case. Close to the wall,
the $u'^+$ values obtained from our PIV are slightly underestimated,
in particular for $Re_{\theta} = 20600$, demonstrating some filtering
of the PIV at this resolution (Foucaut \textit{et al.} [\onlinecite{foucaut2004piv}]). To compute from
PIV the energy spectra used in this paper, we used the method of
Foucaut \textit{et al.} [\onlinecite{foucaut2004piv}]. As seen in figure \ref{fig:figure2} for the
particular case of wall distance $y^+=200$ at $Re_{\theta} = 20600$,
the agreement between the spatial spectrum from the PIV and the
temporal spectrum from the hot-wire anemometry of
Carlier \& Stanislas [\onlinecite{carlier2005experimental}] is good up to wavenumbers $k_{x} \approx 500$ corresponding to length-scales of $2$mm.

\begin{figure}
\centering

	\begin{subfigure}{\textwidth}
	   \centerline{\includegraphics[width=1.1\textwidth]{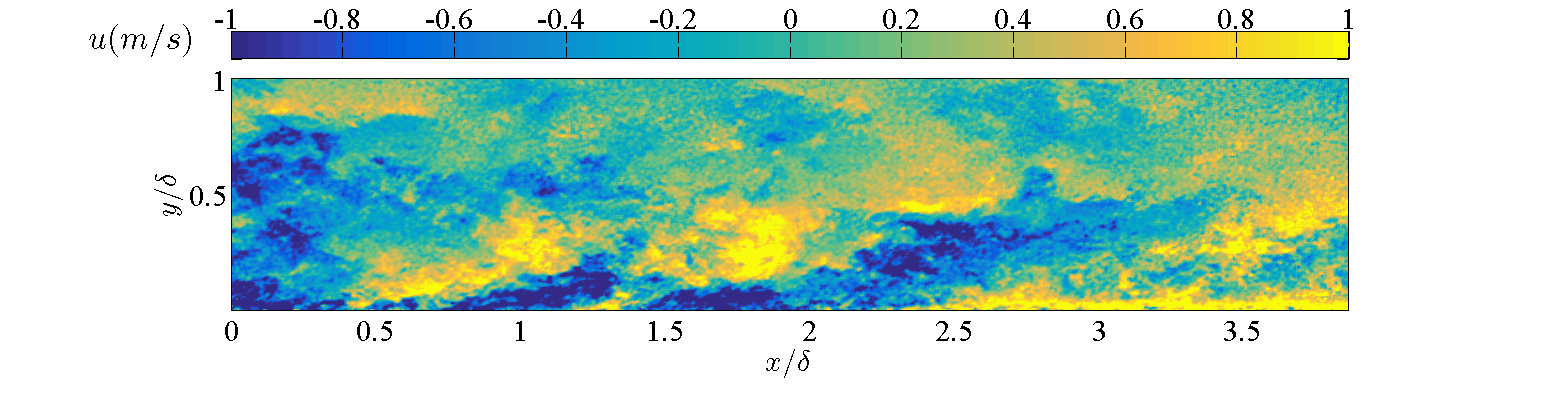}}
	\end{subfigure}   
	\quad
	\begin{subfigure}{\textwidth}
		\centerline{\includegraphics[width=1.1\textwidth]{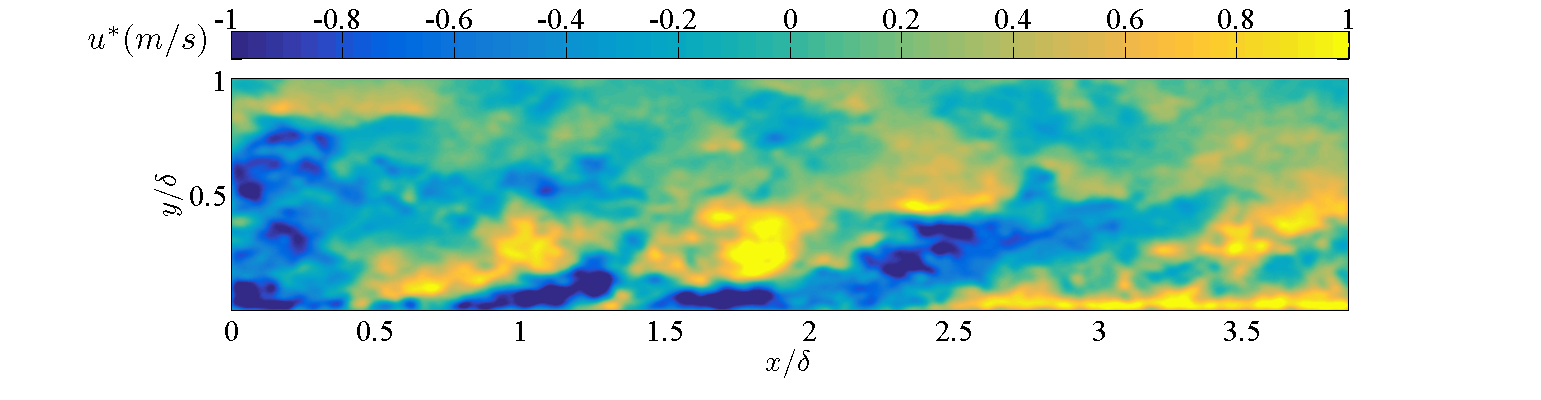}}
		\end{subfigure}
	
	\begin{subfigure}{\textwidth}
			\centerline{\includegraphics[width=1.1\textwidth]{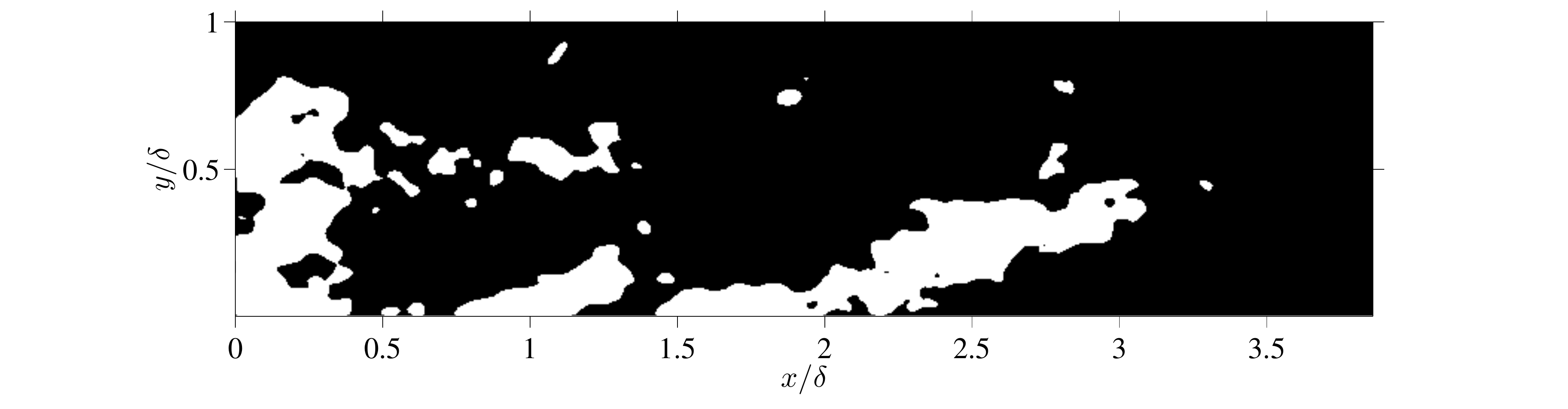}}
			\end{subfigure}
	
	\begin{subfigure}{\textwidth}
			\centerline{\includegraphics[width=1.1\textwidth]{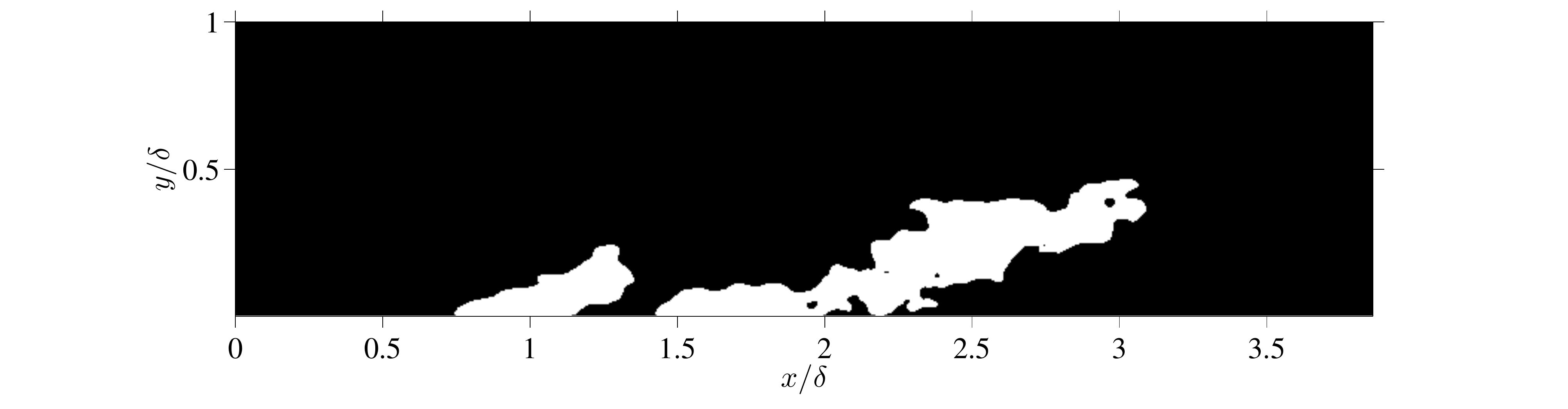}}
			\end{subfigure}
			\caption{Wall-attached elongated streaky
                          structure eduction method applied on a
                          sample instantaneous streamwise velocity
                          field at $Re_{\theta} = 20600$. From
                          top: (a) Raw instantaneous streamwise
                          fluctuating velocity component field (b) The
                          same field after application of a Gaussian
                          filter (c) Binary image obtained after
                          thresholding negative streamwise fluctuating
                          momentum regions. (d) Final image obtained
                          after cleaning as described in subsection
                          \ref{sec:Structure detection}.}
			\label{fig:figure3}
\end{figure}

\begin{figure}
\centering

	   \centerline{\includegraphics[width=0.85\textwidth]{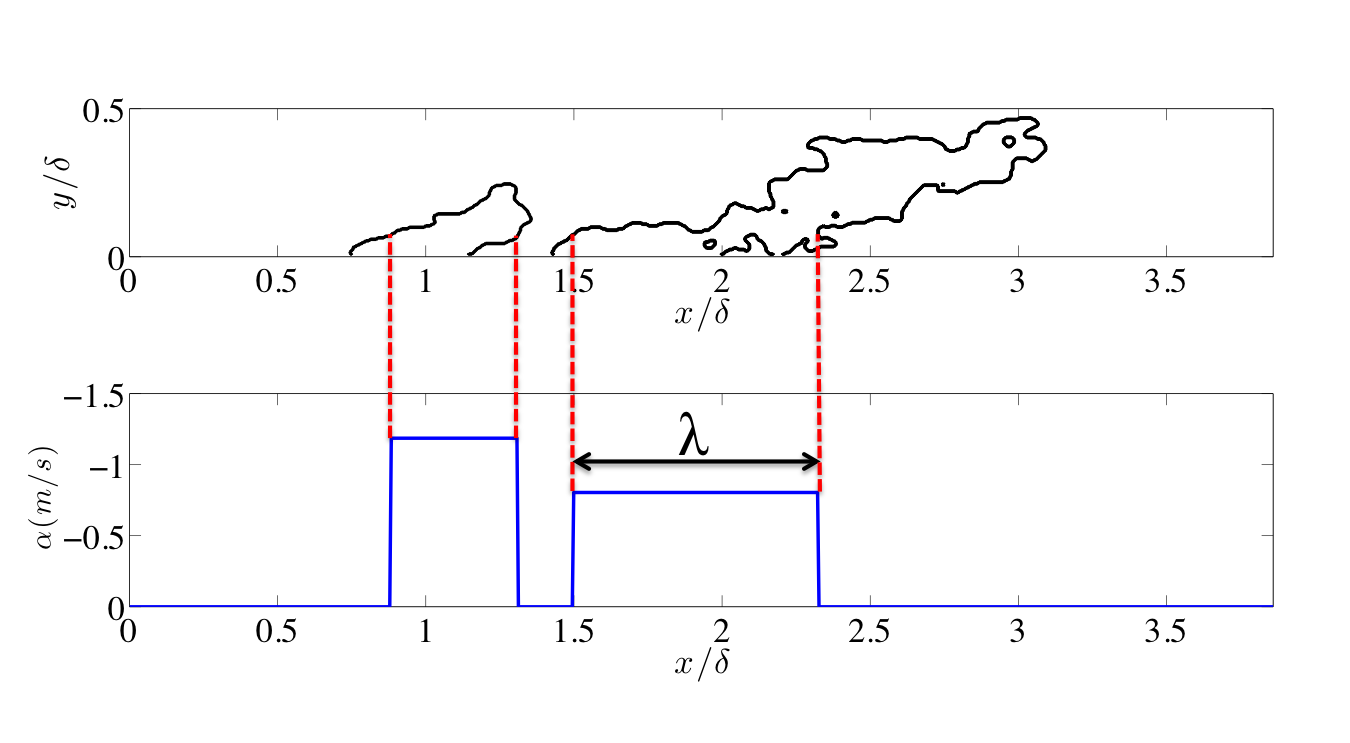}}
		\caption{Figure \ref{fig:figure3}(d) reproduced in the top plot, with,
                  in the bottom plot, the average streamwise
                  fluctuating velocity $\alpha$ and the streamwise
                  length $\lambda$ of the detected wall-attached
                  structures at $y/\delta= 0.03$.}
	\label{fig:figure4}
\end{figure}

\subsection{Structure detection} \label{sec:Structure detection}

In sections \ref{sec: simplest possible model} and \ref{sec:Modified TP range} we developed a spectral model of the streamwise
turbulence fluctuating velocity based on the concept of elongated
streaky structures which are part of attached eddies and can be
modeled as simple on-off functions. In this and the next subsections
we use our PIV data to test this concept and assess its potential as
an hypothesis for understanding near-wall streamwise energy spectra.

Figure \ref{fig:figure3}(a) shows a sample field of instantaneous streamwise
fluctuating velocity components $u$. The existence of well-defined
elongated and tilted wall-attached regions of relatively high
(positive or negative) $u$ values is clear. It is these regions that
we need to target in relation to the elongated streaky structures of
our model.

The raw instantaneous streamwise velocity fields are affected by noise
so that single structures in figure \ref{fig:figure3}(a) appear split in many little
parts. To smooth out these structures without modifying their shape
and statistics we used a two-dimensional Gaussian filter. This
filtering operation was found to be sufficient to capture and connect
the structures while retaining their overall shape.
The standard deviation of the Gaussian filter was three pixels which
corresponds to approximately $0.015\delta$ for both Reynolds numbers,
i.e. 105 wall units for $Re_{\theta} = 20600$ and 33 wall units
for $Re_{\theta} = 8100$. The result of this operation on figure
\ref{fig:figure3}(a) leads to figure \ref{fig:figure3}(b).

To educe on-off functions such as the ones required by our model we
apply a threshold $u_{th}$ on the gaussian-filetered $u^{*}$ to obtain
binary images which distinguish between $u^{*}<u_{th}$ and
$u^{*}>u_{th}$. Effects of the threshold on the statistics of educed
structures were investigated in the range $0.1 u'_{300^+}< \vert
u_{th} \vert < u'_{300^+}$ where $u'_{300^+}$ is $u'$ at $y^+ =
300$.

A threshold $u_{th}$ equal to $-0.4 u'_{300^+}$ was finally chosen to
detect low momentum structures in the present study as it corresponds
to the value that leads to least threshold-dependency of our
statistics for a negative $u_{th}$ \textcolor{black}{(for example,
  $u_{th}$ equal to $-0.2 u'_{300^+}$ or $-0.6 u'_{300^+}$ return
  results with no significant difference, see Appendix \ref{Appendix})}. This thresholding operation leads to figure \ref{fig:figure3}(c) when applied to figure \ref{fig:figure3}(b). The white structures in figure \ref{fig:figure3}(c) correspond to $u^{*}<u_{th}$.

One more step is required before comparing with our model. White
structures which cut through the vertical borders of the figure are
discarded because their streamwise extent is unknown; and white
structures which are not attached to the bottom wall (at $y=0$ but in
fact as close to $y=0$ as allowed by our PIV data,
\textcolor{black}{i.e. $y^{+}=16$ and $y^{+}=48$ for the lower and the
  higher Reynolds number cases respectively}) are also discarded
because we are concerned with wall-attached structures. With this
extra step, figure \ref{fig:figure3}(c) gives rise to figure \ref{fig:figure3}(d).

All the steps leading from raw fluctuating streamwise velocity fields
to the binary fields which we use in our statistical analysis are
depicted in figure \ref{fig:figure3}. The current study's effort is
concentrated on wall-attached elongated structures of negative
streamwise fluctuating velocity as in figure \ref{fig:figure3}(d), \textcolor{black}{but the analysis can be repeated equally well on structures of positive streamwise fluctuating velocity with results that are similar though slightly less sharp with regard to (\ref{eq10})-(\ref{eq10bis}) (see Appendix \ref{Appendix})}. The general behaviours of
positive and negative fluctuating streamwise velocity structures are
similar, the negative velocity structures being slightly longer in
agreement with Dennis \& Nickels [\onlinecite{dennis2011b}].

\subsection{Lengths of wall-attached streamwise velocity structures} \label{sec:Lengths}

We now need to obtain statistics of wall-attached elongated streaky
structures represented as on-off functions in our model and as binary
structures in the final stages of our structure eduction method. We
first label the connected components of the binary images using image
processing tools. Then we compute the streamwise length $\lambda$ of
each labelled structure at a distance $y$ from the wall, i.e. the
difference between the smallest and the largest values of streamwise
coordinate $x$ in this labelled structure at height $y$. Finally we
compute the average value $\alpha$ of the streamwise fluctuating
velocity component $u$ inside this labelled structure at height
$y$. Thus we obtain a pair $(\lambda, \alpha)$ for each labelled
structure at each height $y$ considered. This procedure is illustrated
in figure \ref{fig:figure4} where the streamwise extent $\lambda$ and
the corresponding amplitude $\alpha$ of two labelled structures at
wall distance $y/\delta= 0.03$ is shown. A total of 14493 and 19576
wall-attached binary structures were detected at $Re_{\theta} = 20600$
and $Re_{\theta} = 8100$ respectively. (As mentioned in section \ref{sec:Exp setup},
22500 and 29500 samples were recorded at the highest and lowest of our
two values of $Re_{\theta}$ respectively, and the ratios $14930/22500$
and $19576/29500$ are about the same.)

\begin{figure}

	   \centerline{\includegraphics[width=0.8\textwidth]{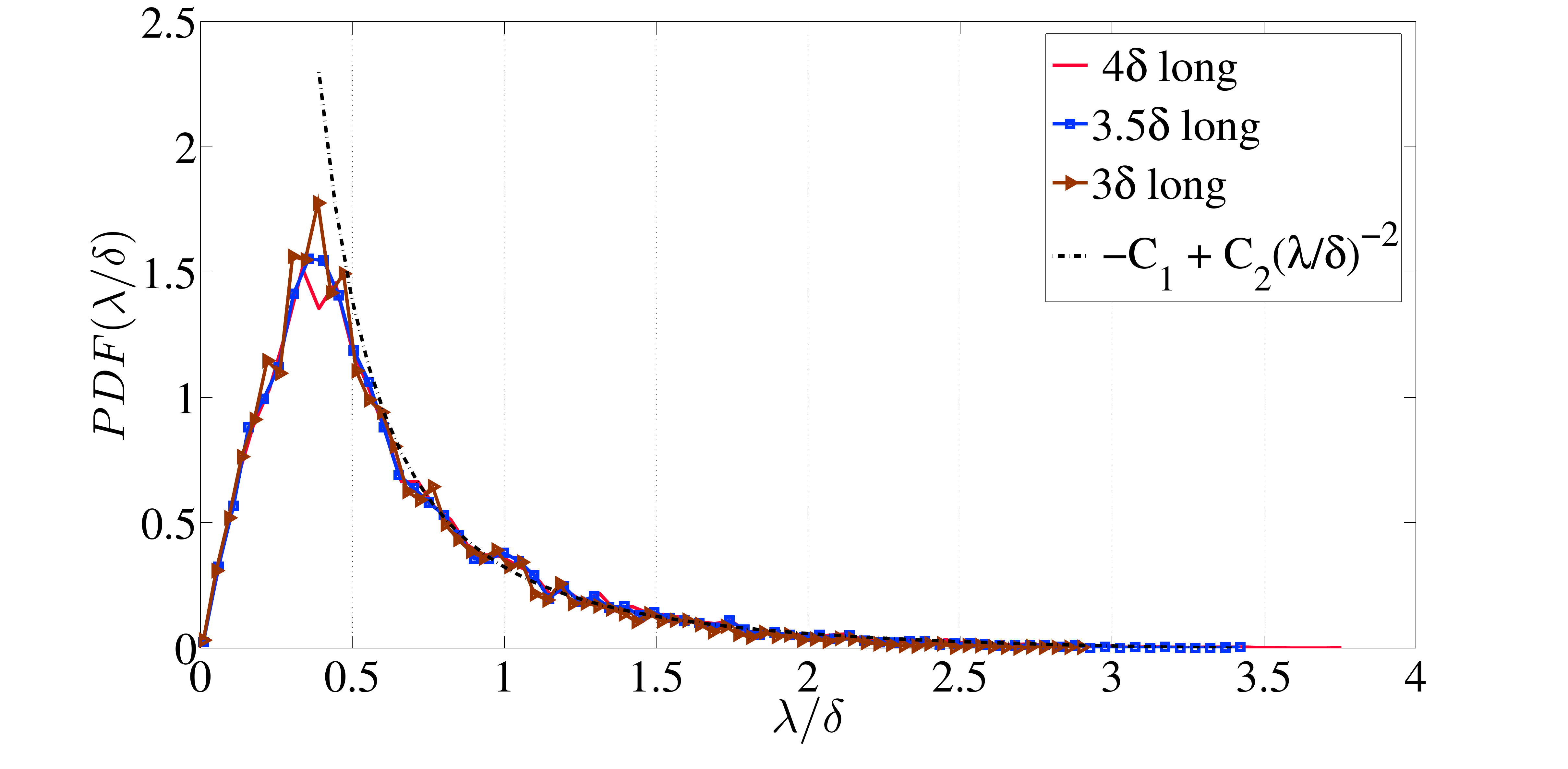}}
		\centerline{\includegraphics[width=0.8\textwidth]{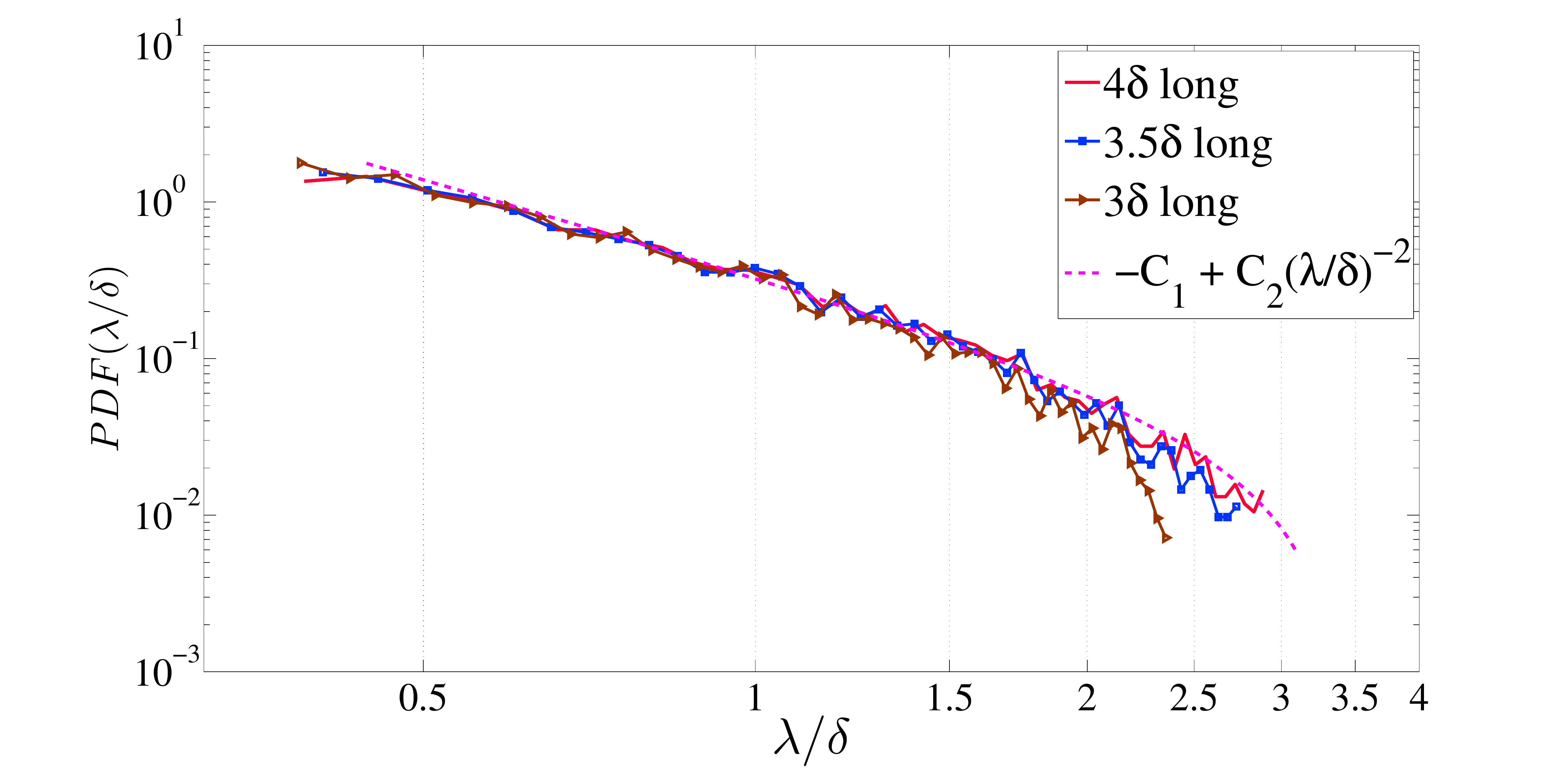}}
	\caption{PDFs of streamwise lengths $\lambda$ (see
                  figure \ref{fig:figure4}) for varying domain lengths at wall
                  distance $y^+= 195$ for $Re_{\theta} =
                  20600$. Lin-lin plot (top) and premultiplied log-log
                  plot (bottom)}
	\label{fig:figure5}
\end{figure}

\begin{figure}
	   \centerline{\includegraphics[width=0.8\textwidth]{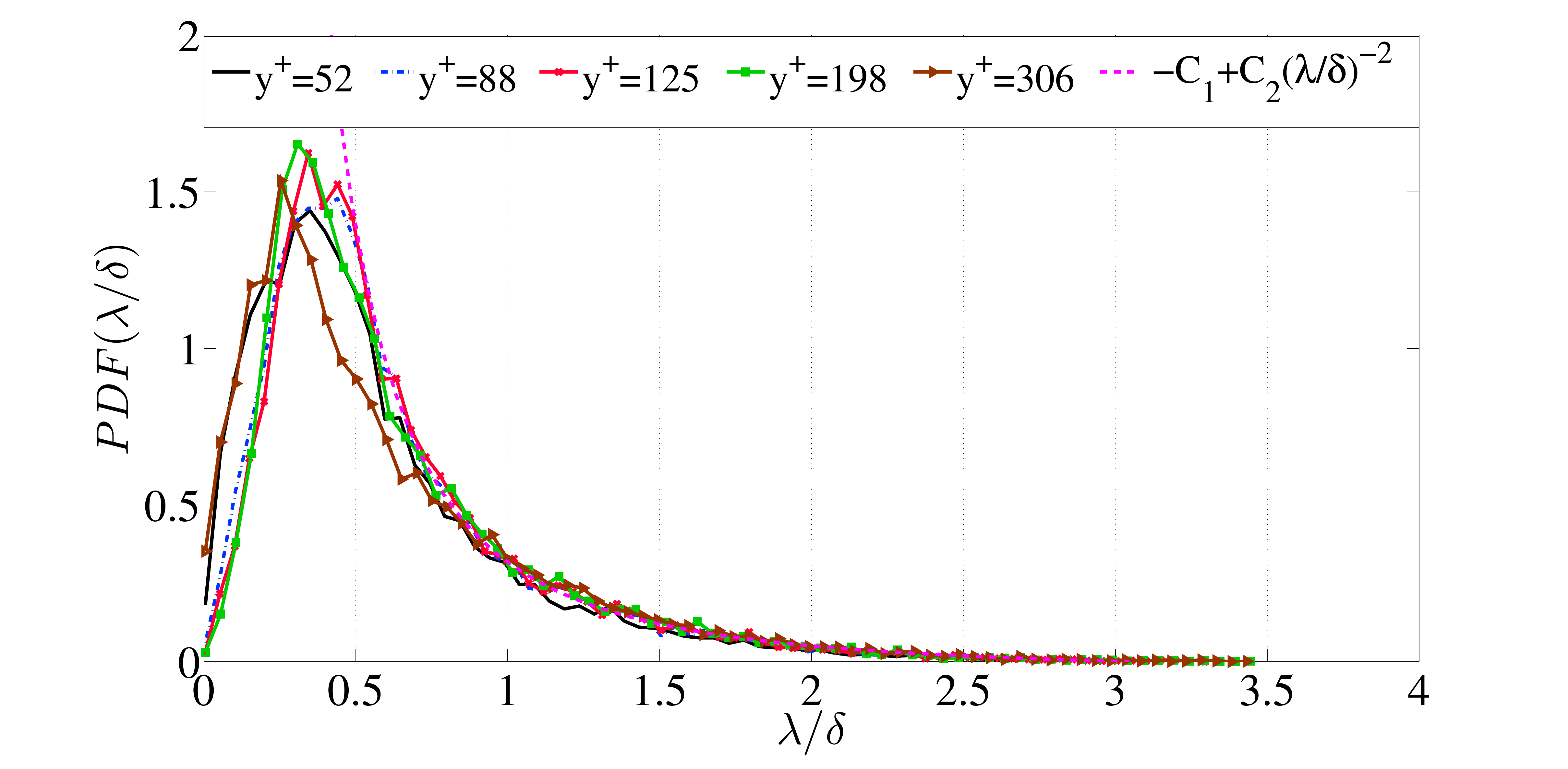}}
		\centerline{\includegraphics[width=0.8\textwidth]{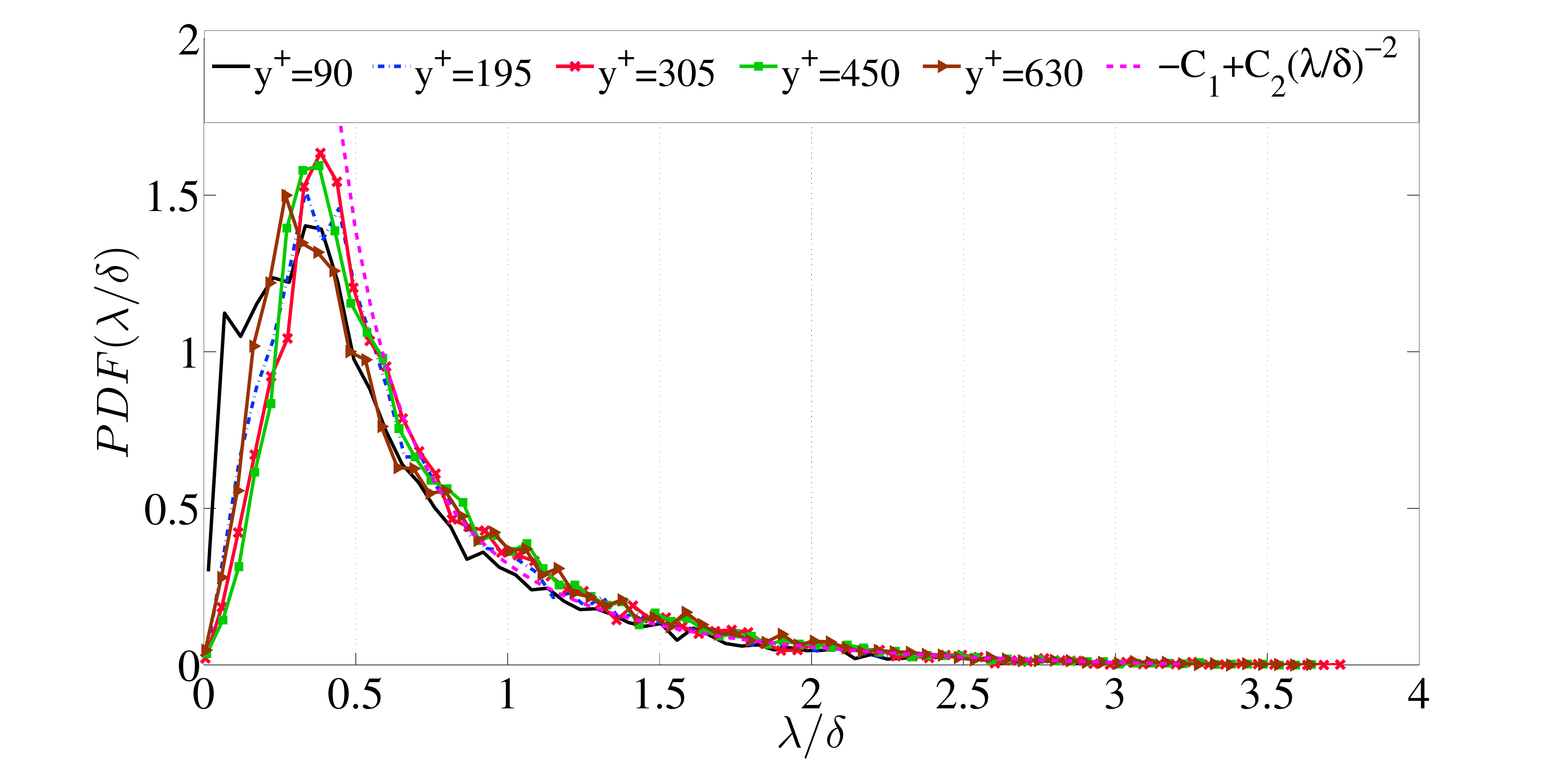}}
		\centerline{\includegraphics[width=0.8\textwidth]{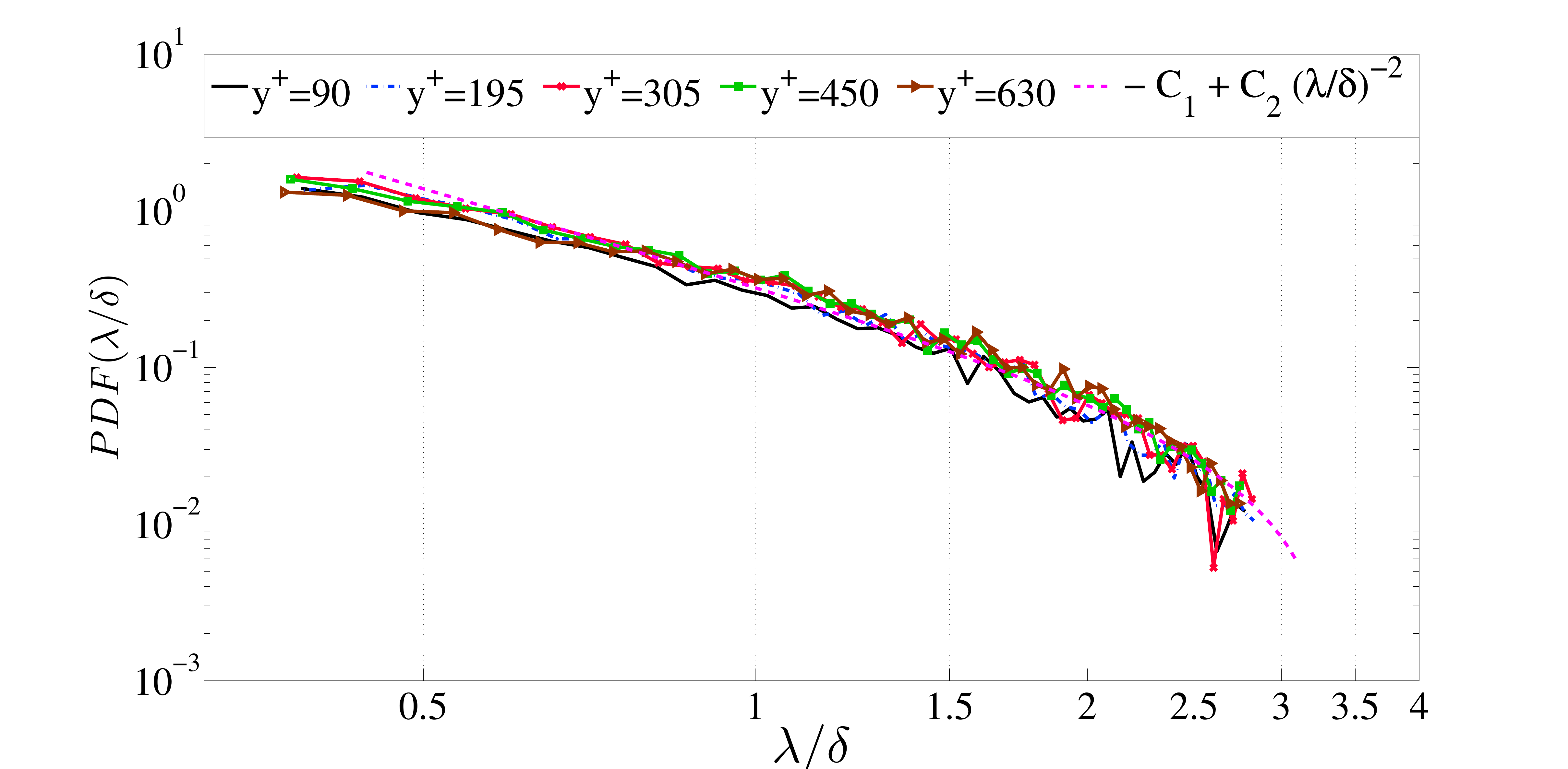}}
\caption{PDFs of streamwise lengths $\lambda$ of wall-attached
  structures (see figure \ref{fig:figure4}) at selected wall distances for
  $Re_{\theta} = 8100$ (top) and $Re_{\theta} = 20600$ (middle). The
  fits shown in these top and middle plots are for $y^+ = 198$ at
  $Re_{\theta} = 8100$ and $y^+ = 195$ at $Re_{\theta} = 20600$. The
  bottom plot is a log-log reproduction of the middle plot's data.}
		\label{fig:figure6}
\end{figure}

The model in sections \ref{sec: simplest possible model} and \ref{sec:Modified TP range} assumes that the number of wall-attached
elongated streaky structures of size $\lambda$ has a decreasing
power-law dependence on $\lambda$ in a certain range of $\lambda$
values. Following Perry \textit{et al.} [\onlinecite{perry1986theoretical}], we expect the spatial
distribution of such structures to be space-filling, which implies
(see Vassilicos \& Hunt [\onlinecite{vassilicos1991fractal}]) that the exponent of this power
law should be -2. Figures \ref{fig:figure5} and \ref{fig:figure6} show
the probability distribution function (PDF) of lengths $\lambda$ at
various wall distances.  The most probable length $\lambda$ lies
between $0.3\delta$ and $0.5\delta$ and lengths $\lambda$ longer than
3.5$\delta$ occur very rarely.

We tested for finite size effects of the field of view by computing
the PDF on smaller domains, namely 3.5$\delta$ and 3$\delta$ long in
the streamwise direction but same in the wall normal direction. As
shown in figure \ref{fig:figure5} there is no significant differences
caused by the three fields of view except that the smallest field
returns a slightly more noisy PDF. Indeed, a reduced field of view
leads to a smaller number of detected wall-attached elongated binary
structures and therefore to reduced statistical convergence.

Figures \ref{fig:figure5} and \ref{fig:figure6} show a power law
dependence on $\lambda$ between about $0.5\delta$ and $2\delta$ with
power law exponent -2, i.e. $D=1$, in all cases. Given the form of
$N(\lambda)$ hypothesised in sections \ref{sec: simplest possible model} and \ref{sec:Modified TP range}, we fit the PDF of
$\lambda/\delta$ with a functional form $-C_{1} + C_{2}
(\lambda/\delta)^{-2}$ (where $C_{1}/N_{M} = C_{2}/N_{o}$). The fit is
shown in figures \ref{fig:figure5} and \ref{fig:figure6} and is
effectively the same for both Reynolds numbers and all values of $y^+$
in the mean flow's approximate log region.
The constants $C_{1}$ and $C_{2}$ are reported in table
\ref{tab:t1}. They are indeed fairly constant over the range of wall
distances and for both Reynolds numbers. Identical results are
obtained for wall-attached structures with positive streamwise
fluctuating velocity except that $C_{1} \approx 0.02$ for both
Reynolds numbers and $C_{2} \approx 0.29$ for $Re_{\theta} = 8100$
(see Table \ref{tab:t3} in Appendix).  It is worth noting that the lower bound of
the range
where the PDF of $\lambda/\delta$ is well approximated by $-C_{1} +
C_{2} (\lambda/\delta)^{-2}$ seems to increase slightly with
increasing $y^+$.

\begin{table}
{\centering
\def~{\hphantom{0}}
\begin{ruledtabular}
  \begin{tabular}{l|ccccc|ccccc}
    \multirow{1}{*}{$Re_\theta$} &
      \multicolumn{5}{c|}{20600} &
      \multicolumn{5}{c}{8100} \\   
    \hline
    $y^+$ & 90 & 195 & 305 & 450 & 630 & 52 & 88 & 125 & 198 & 306 \\
    $C_1$ & 0.03 & 0.03 & 0.03 & 0.03 & 0.03 & 0.04 & 0.04 & 0.04 & 0.03 & 0.03 \\
    $C_2$ & 0.32 & 0.35 & 0.35 & 0.37 & 0.37 & 0.32 & 0.35 & 0.35 & 0.35 & 0.33 \\
  \end{tabular}
  \caption{Values of the constants $C_1$ and $C_2$ in the form $-C_{1}
    + C_{2} (\lambda/\delta)^{-2}$ of the PDF of
    $\lambda/\delta$. In the $Re_{\theta} = 20600$ case, the fit
    is over a range of $\lambda/\delta$ bounded from above by 3.8 and
    from below by 0.49 ($y^{+} = 90$), 0.55 ($y^{+} = 195$), 0.54
    ($y^{+} = 305$), 0.58 ($y^{+} = 450$) and 0.78 ($y^{+} = 630$). In
    the $Re_{\theta} = 8100$ case, the fit is over a range of
    $\lambda/\delta$ bounded from above by 3.4 and from below by 0.54
    ($y^{+} = 52$), 0.53 ($y^{+} = 88$), 0.53 ($y^{+} = 125$), 0.56
    ($y^{+} = 198$) and 0.65 ($y^{+} = 306$).}
    \label{tab:t1}
    \end{ruledtabular}
} 
\end{table}

\subsection{Energy spectra} \label{sec:spectra}

\begin{figure}

	   \centerline{\includegraphics[width=0.85\textwidth]{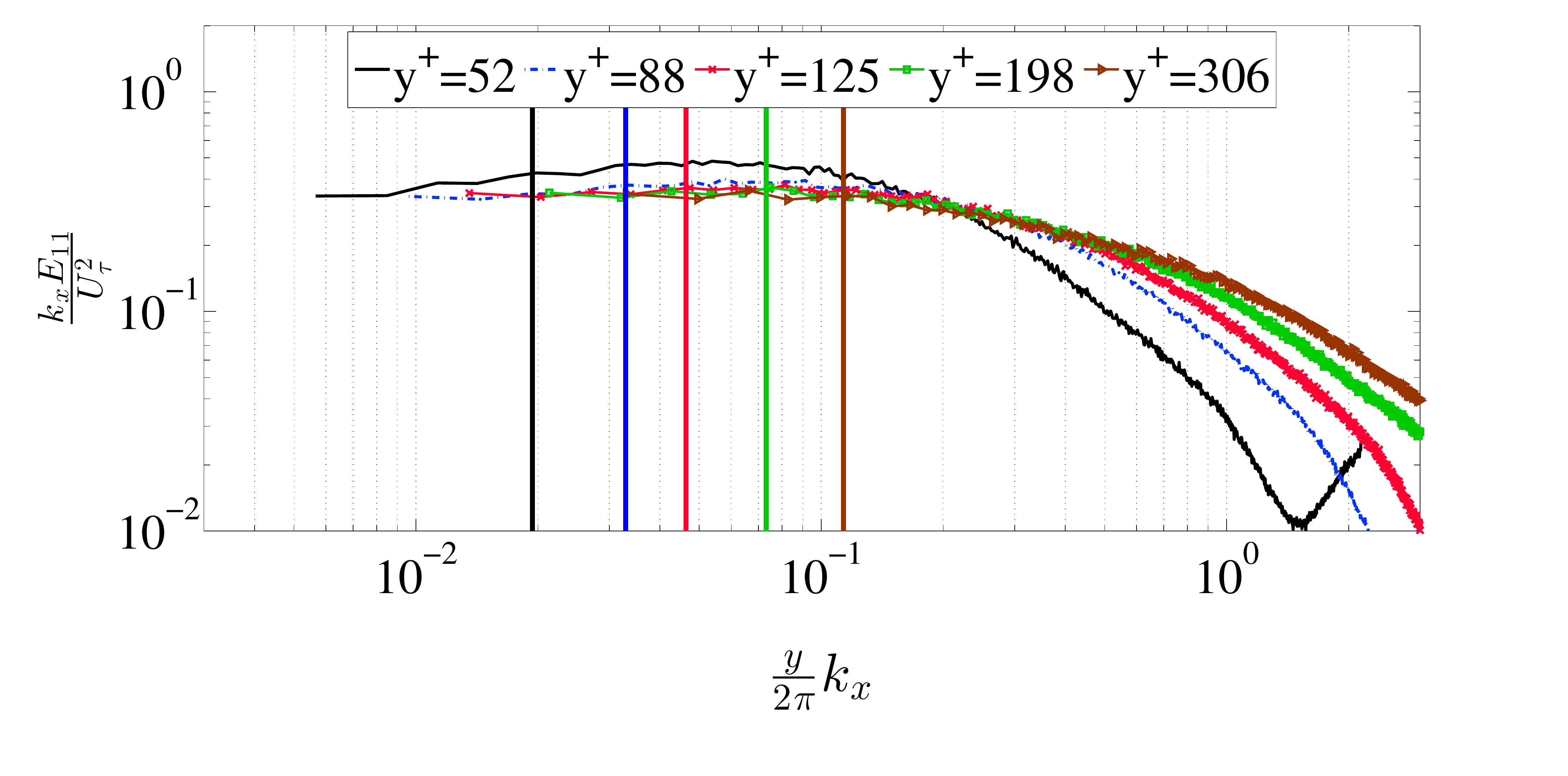}}
	   \centerline{\includegraphics[width=0.85\textwidth]{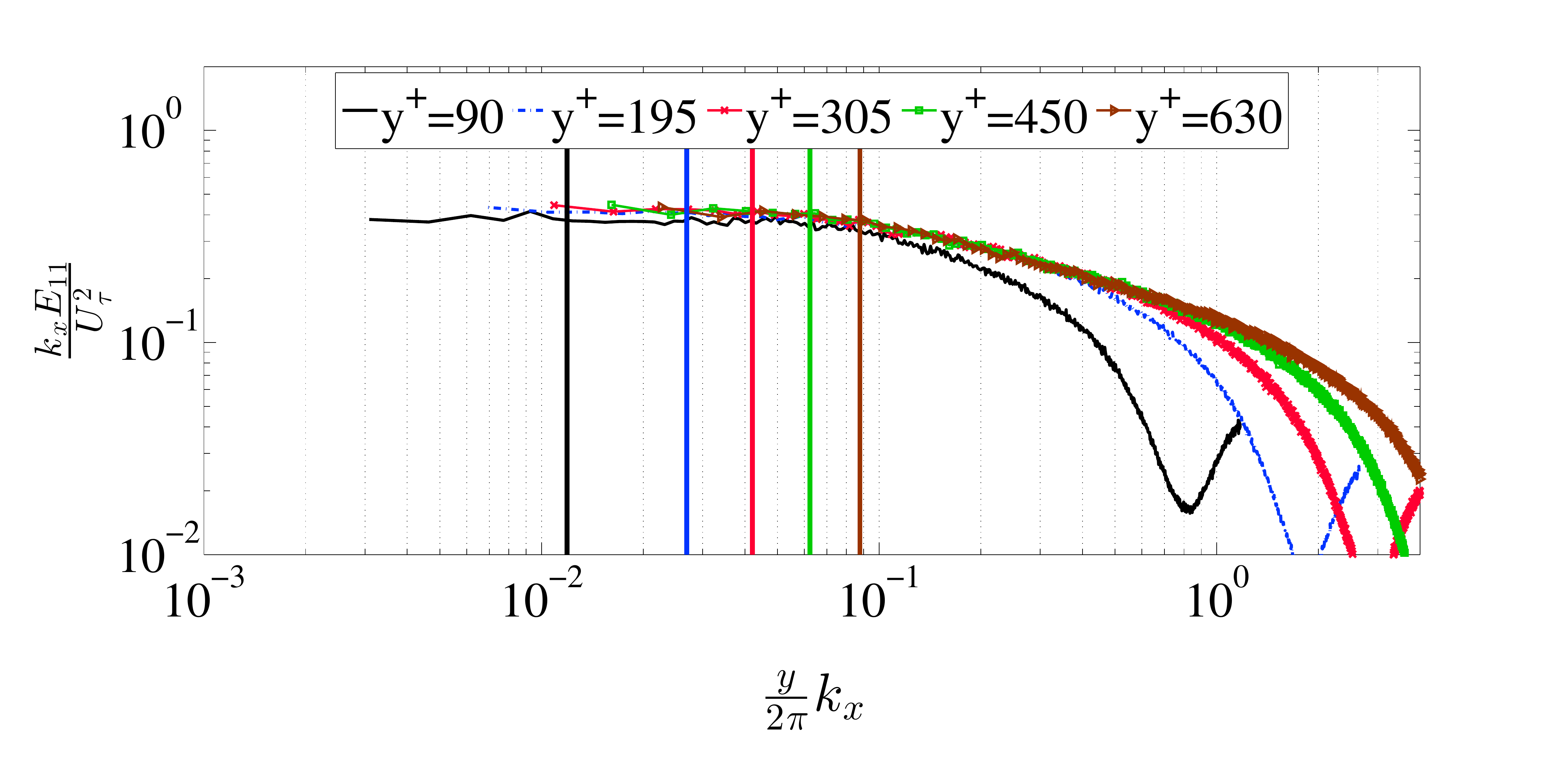}}
			\caption{Log-log plots of pre-multiplied
                          streamwise energy spectra at selected wall
                          distances for $Re_{\theta} = 8100$ above and
                          $Re_{\theta} = 20600$ below. Vertical lines
                          show \textcolor{black}{${y\over 2\pi}k_{x} =
                            (y/2\pi) (2\pi/\delta)=\frac{y}{\delta}$
                            (corresponding to $k_{x}=2\pi/\delta$)}
                          with the same color code as the legend of
                          the figure.}
			\label{fig:figure7}
\end{figure}

\begin{figure}

			\centerline{\includegraphics[width=0.85\textwidth]{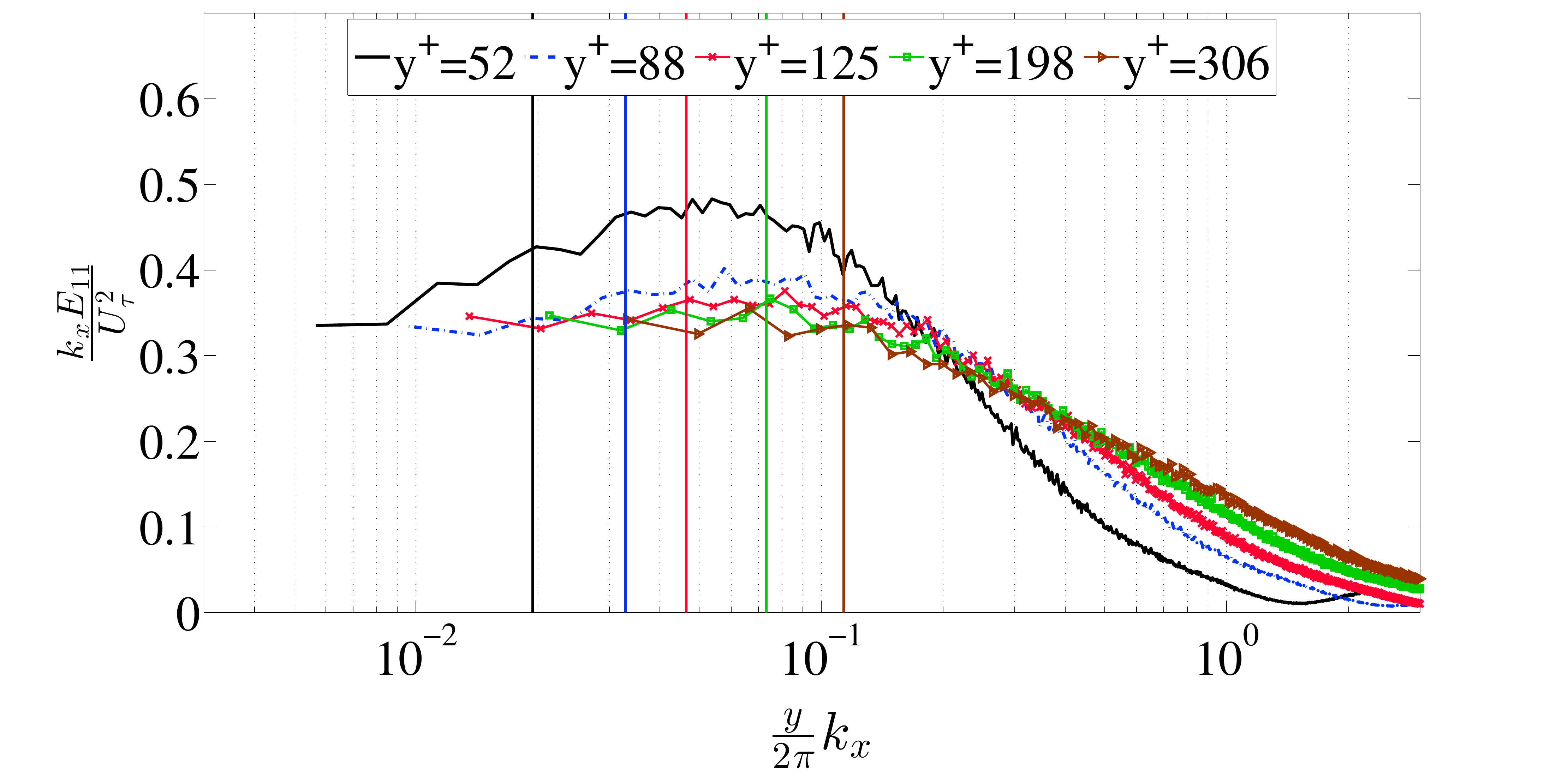}}
			\centerline{\includegraphics[width=0.85\textwidth]{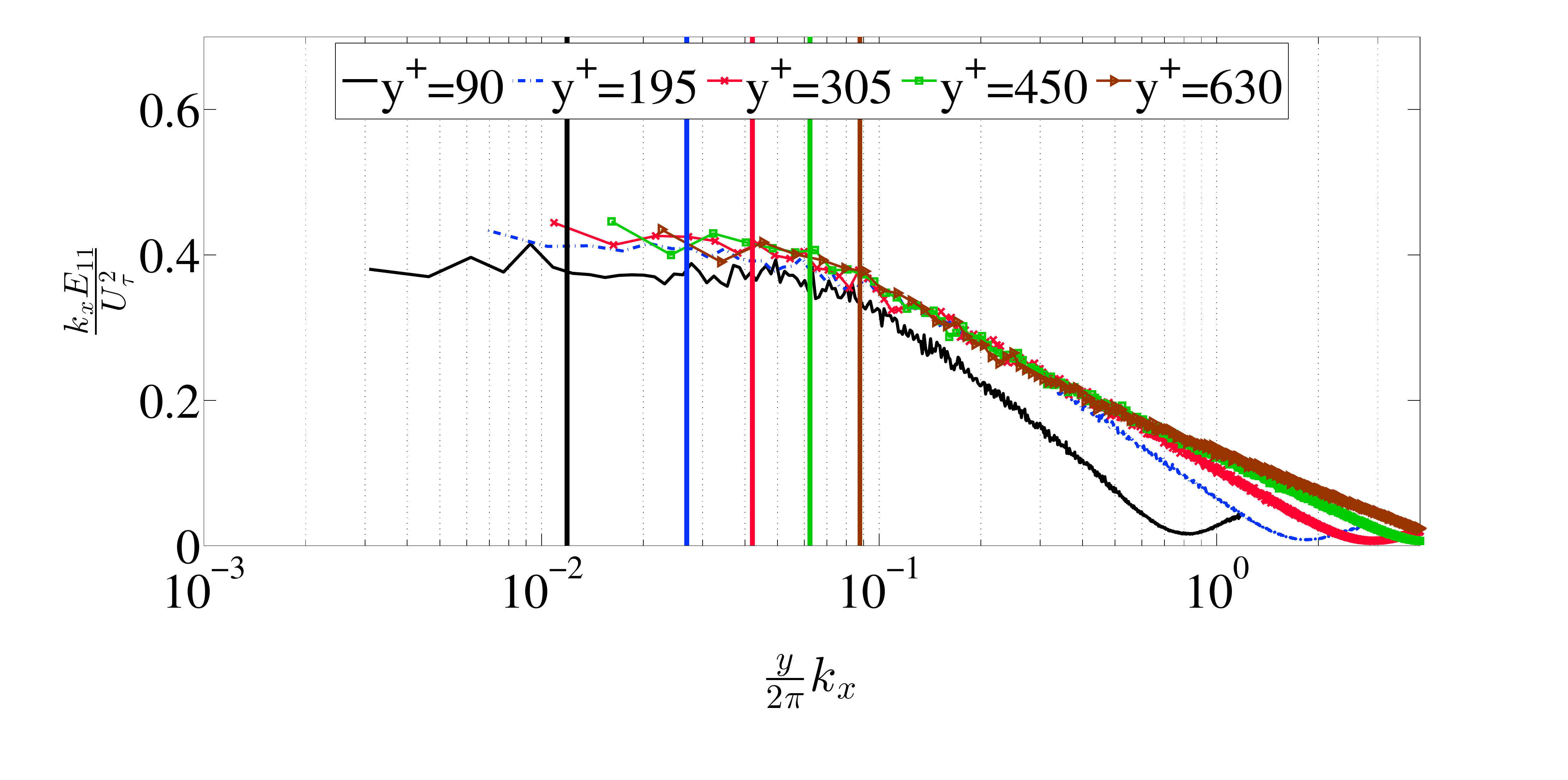}}
			\caption{Same as figure \ref{fig:figure7} in lin-log plots.} 
			\label{fig:figure8}
\end{figure}

Figure \ref{fig:figure7} shows log-log plots of premultiplied energy spectra of
streamwise fluctuating velocities $u(x)$ which have been obtained from
our PIV data at various normalised wall distances $y^+$ for both
Reynolds numbers. These plots might suggest that $E_{11} (k_{x}) \sim
U_{\tau}^{2} k_{x}^{-1}$ in a range of wavenumbers $2\pi/(4\delta)
\lesssim k_{x} \lesssim 0.63/y$
for $y^{+}$ larger than about 88 and smaller than the value of $y^+$
where this range of wavenumbers no longer exists. The apparent
$k_x^{-1}$ wavenumber range is close to a decade long at $y^+=90$ for
$Re_{\theta} = 20600$ and shorter for higher wall normal distances and
for the lower Reynolds number ($Re_{\theta} = 8100$). One would be
justified to conclude that this is indeed experimental support for the
Townsend-Perry $k_{x}^{-1}$ spectrum if the only available theoretical
glasses through which to look at these spectral plots were those of
the Townsend-Perry attached eddy model.
However the situation is subtler and, in effect, quite different.

A closer look at the spectra in the lin-log plot of figure \ref{fig:figure8} suggests
the possibility for small corrections to this conclusion, particularly
at the lower of the two $Re_{\theta}$ values, but the result
(\ref{eq10})-(\ref{eq10bis}) of our model in section \ref{sec:Modified TP range} may pave the way for a
significantly different interpretation. This model leads to $E_{11}
(k_{x}) \sim (k_{x}\delta)^{q}$ with $p+q=-1$ if $D=1$. Support for
$D=1$ has been obtained and reported in the previous subsection in the
range of lengths $\lambda$ between about $0.5\delta$ and $2\delta$. It
is therefore worth taking a closer look at our energy spectra in the
corresponding wavenumber range. For our data, this wavenumber range
turns out, in fact, to be comparable to the wavenumber range
$2\pi/(4\delta) \lesssim k_{x} \lesssim 0.63/y$ mentioned in the
previous paragraph as a candidate for Townsend-Perry
scaling. Specifically, $k_{x}/(2\pi) = 2/\delta$ corresponds to $k_{x}
y/(2\pi) = 0.25$, $0.41$, $0.58$, $0.91$ and $1.41$ in increasing
order of the $y^+$ values in figures \ref{fig:figure7} and  \ref{fig:figure8} for $Re_{\theta} = 8100$;
and to $k_{x} y/(2\pi) = 0.15$, $0.33$, $0.53$, $0.78$ and $1.1$ in
increasing order of the $y^+$ values in figures \ref{fig:figure7} and  \ref{fig:figure8} for
$Re_{\theta} = 20600$. The wavenumber range $0.5/\delta \le
k_{x}/(2\pi) \le 2/\delta$ where the analysis in the remainder of our
paper is carried out is therefore not radically different for our data
from the wavenumber range $2\pi/(4\delta) < k_{x} < 0.63/y$ where one
would interpret our spectra to have a Townsend-Perry scaling for
$y^{+} \ge 88$.

\begin{figure}

	\begin{subfigure}{0.495\textwidth}
	   {\includegraphics[width=0.65\textwidth]{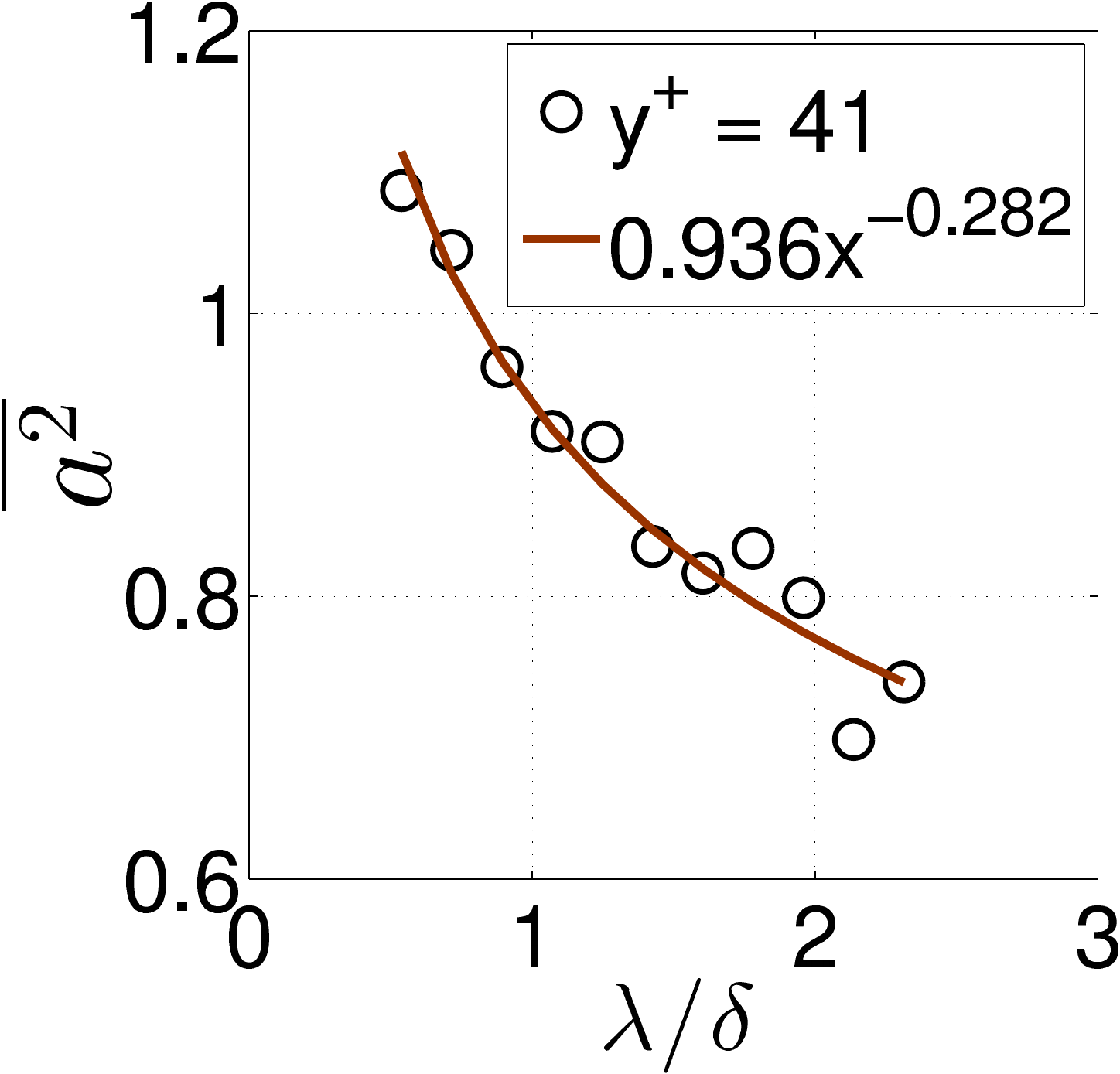}}
	\end{subfigure}   
	\begin{subfigure}{0.495\textwidth}
    {\includegraphics[width=0.65\textwidth]{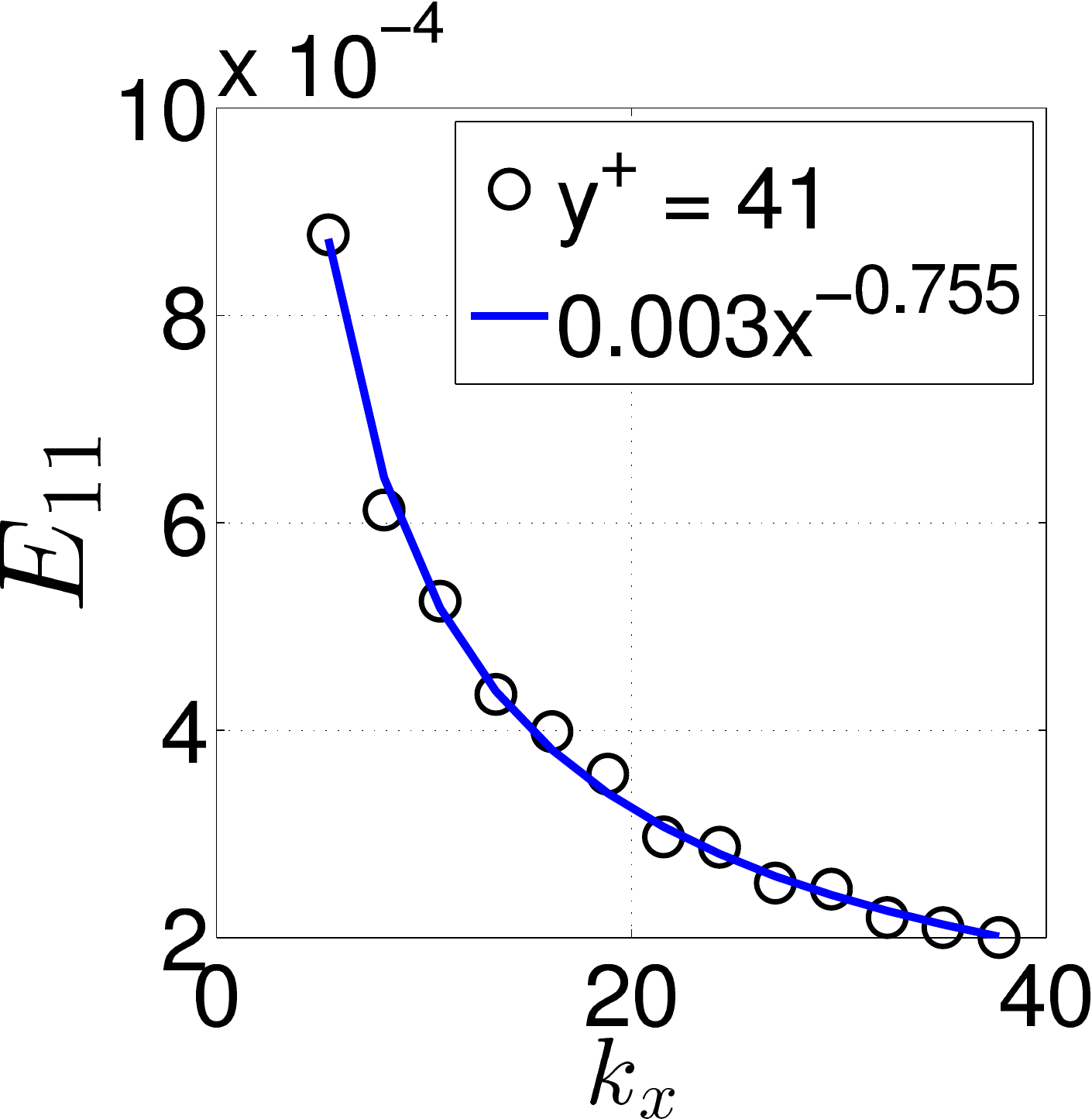}}
	\end{subfigure}
	\begin{subfigure}{0.495\textwidth}
	{\includegraphics[width=0.65\textwidth]{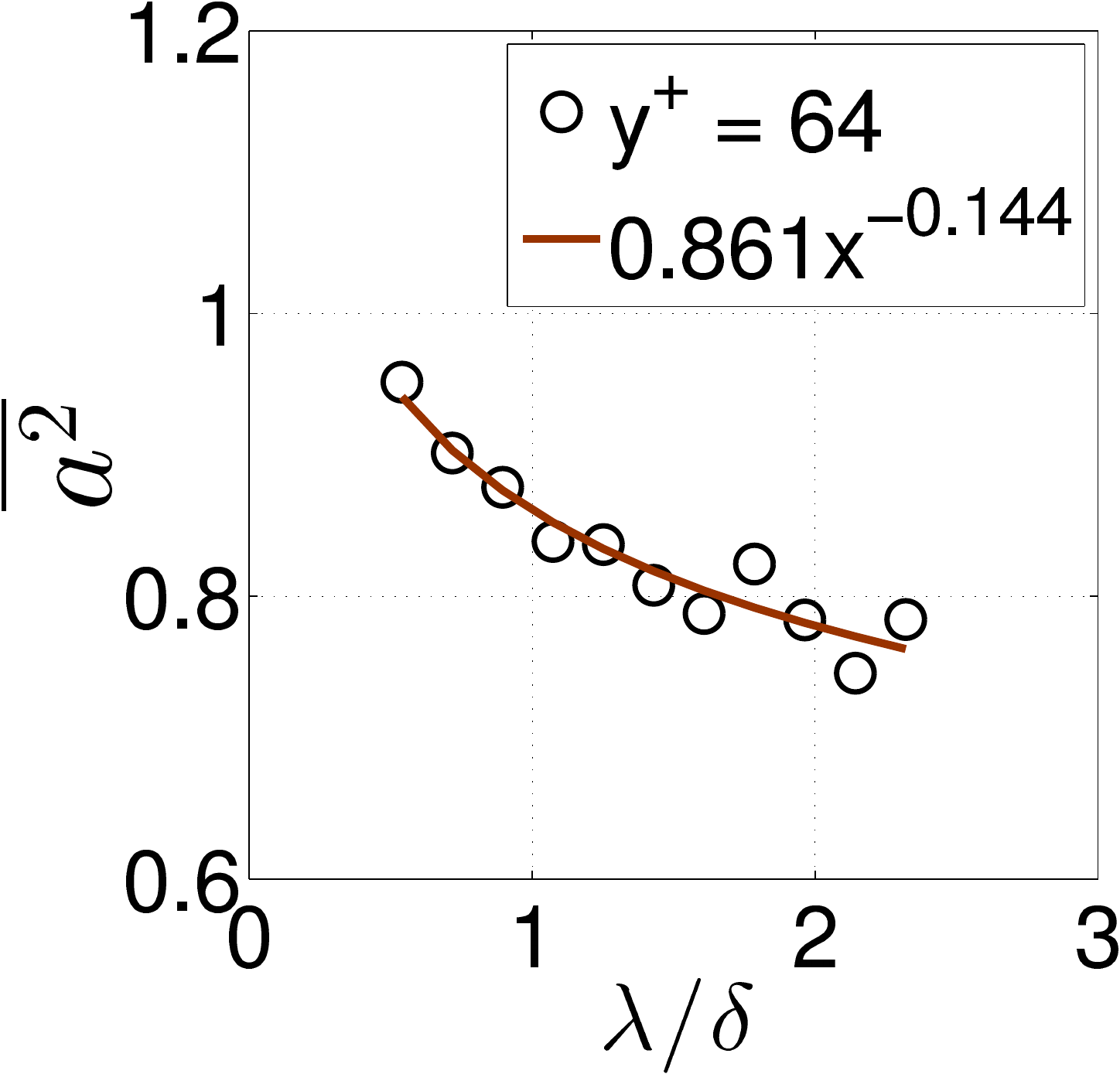}}
	\end{subfigure}   
	\begin{subfigure}{0.495\textwidth}
		    {\includegraphics[width=0.65\textwidth]{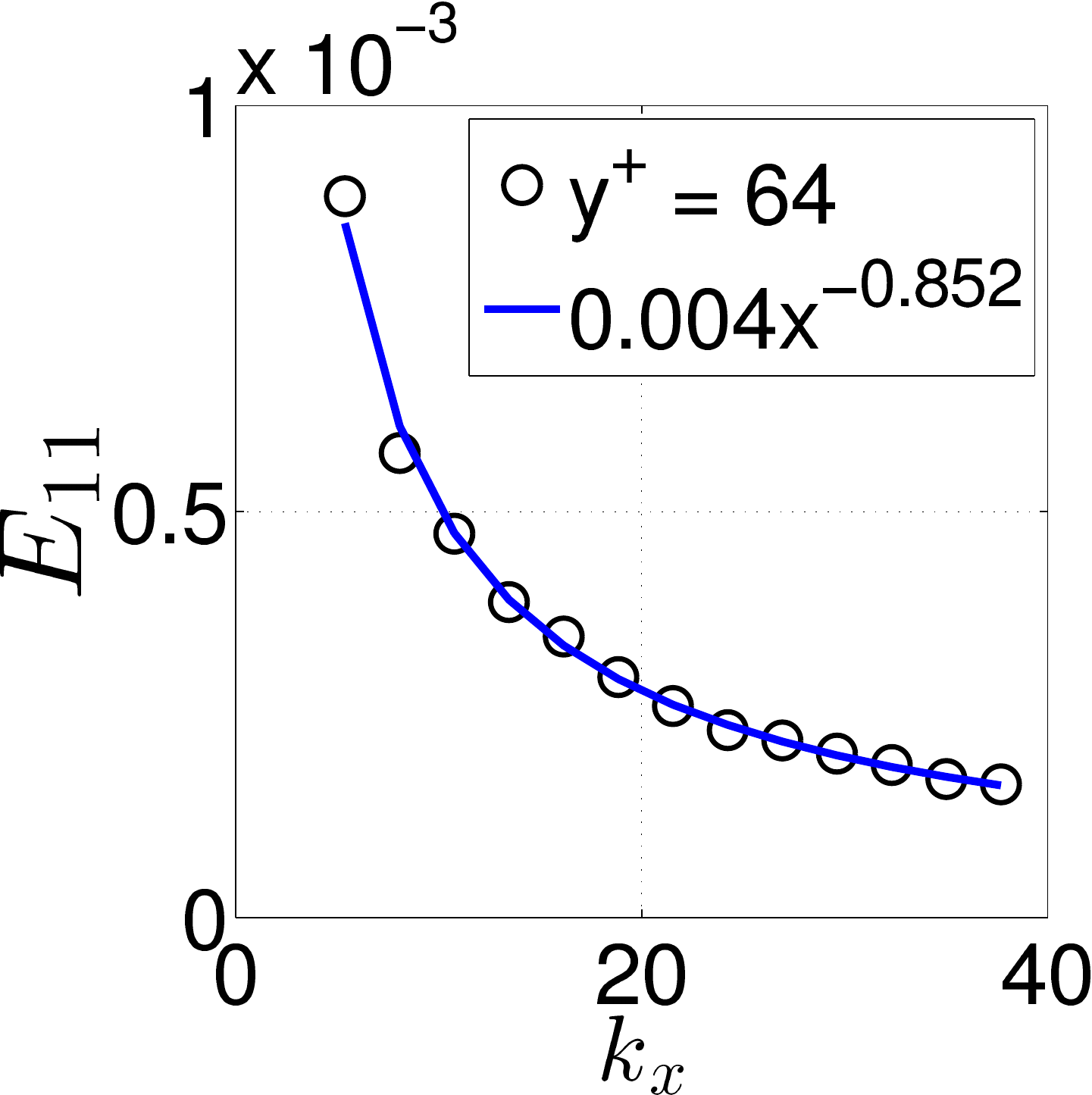}}
	\end{subfigure}
	\begin{subfigure}{0.495\textwidth}
		   {\includegraphics[width=0.65\textwidth]{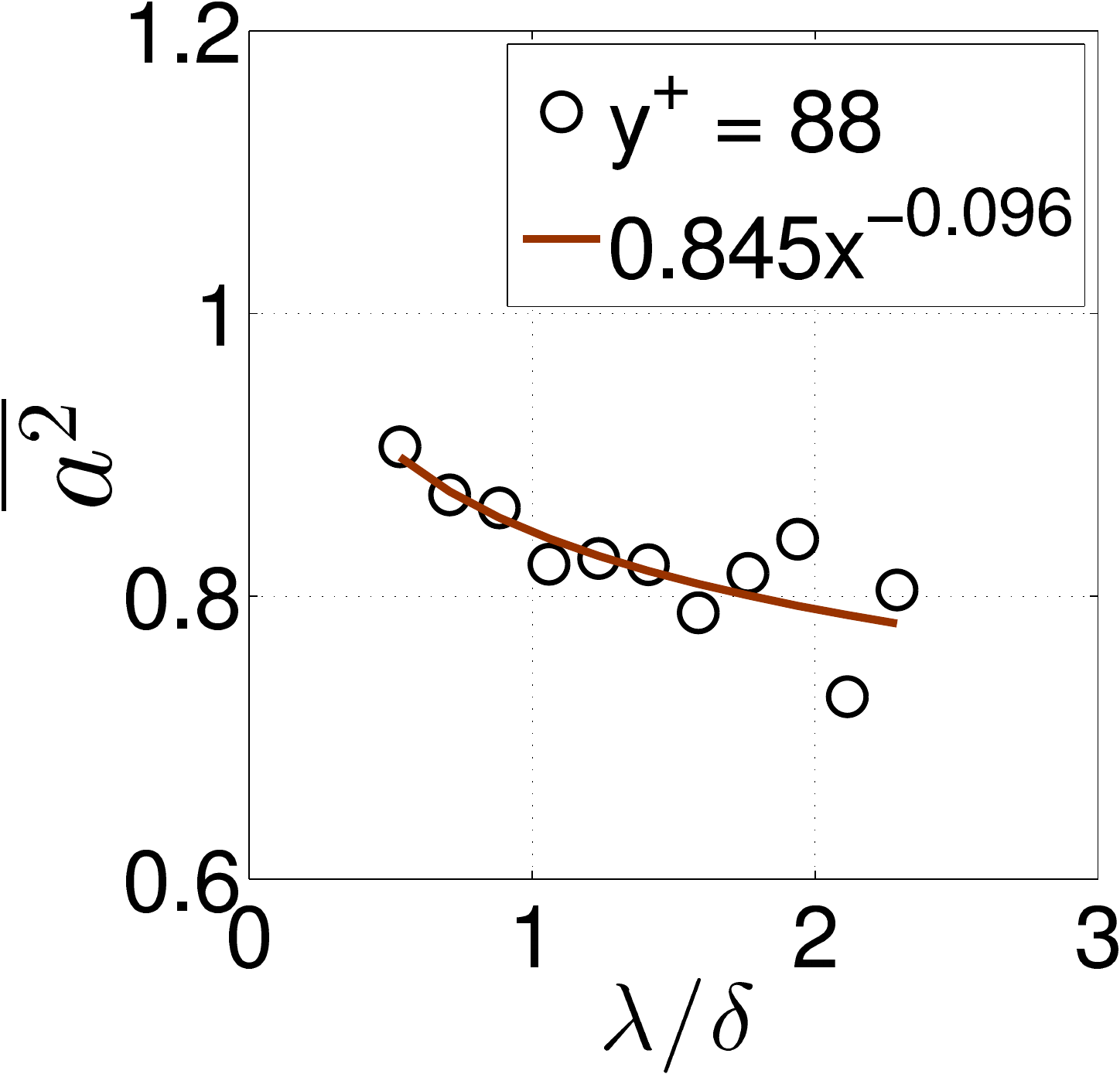}}
	\end{subfigure}   
	\begin{subfigure}{0.495\textwidth}
	{\includegraphics[width=0.65\textwidth]{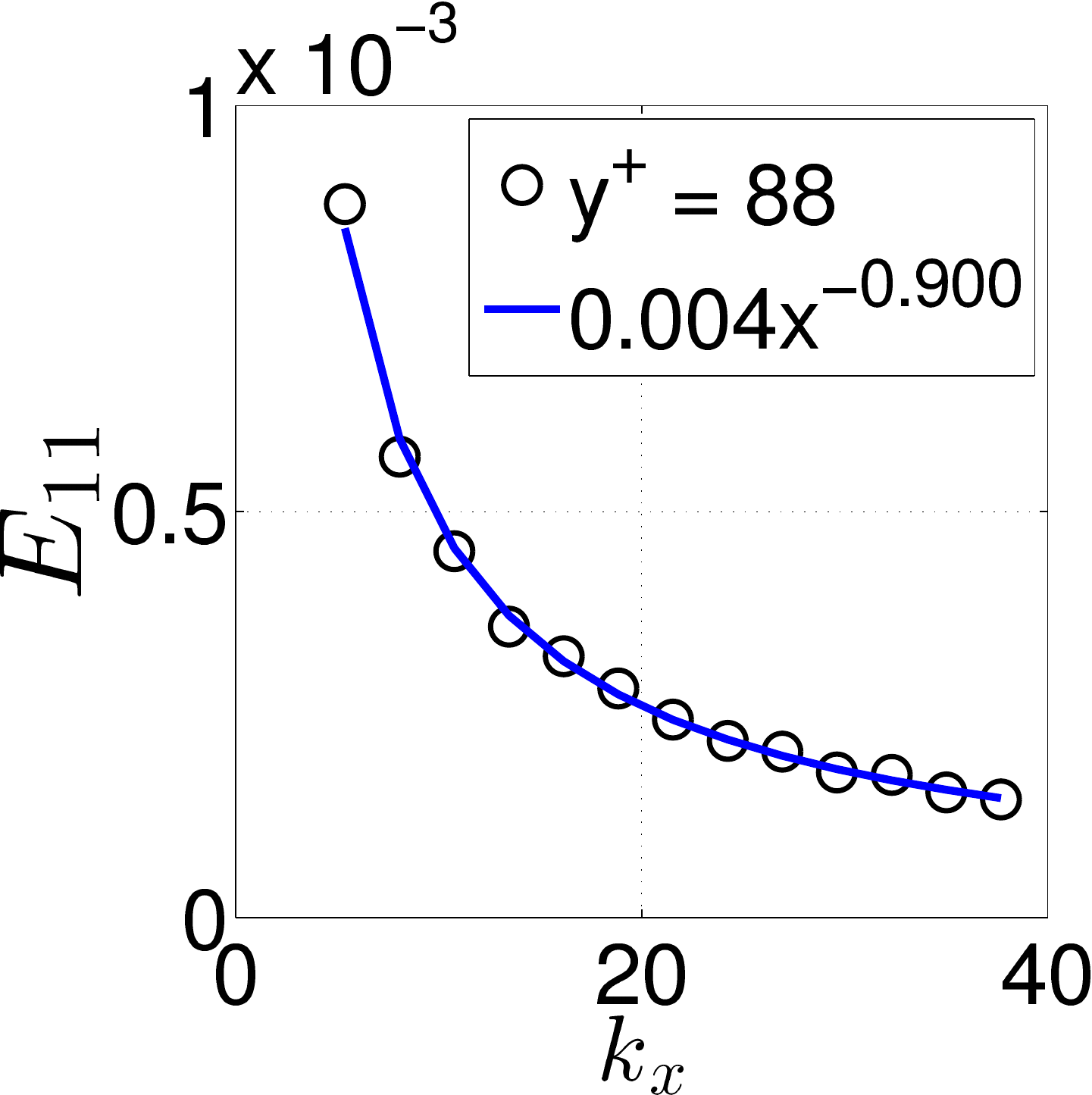}}
	\end{subfigure}
	\begin{subfigure}{0.495\textwidth}
		   {\includegraphics[width=0.65\textwidth]{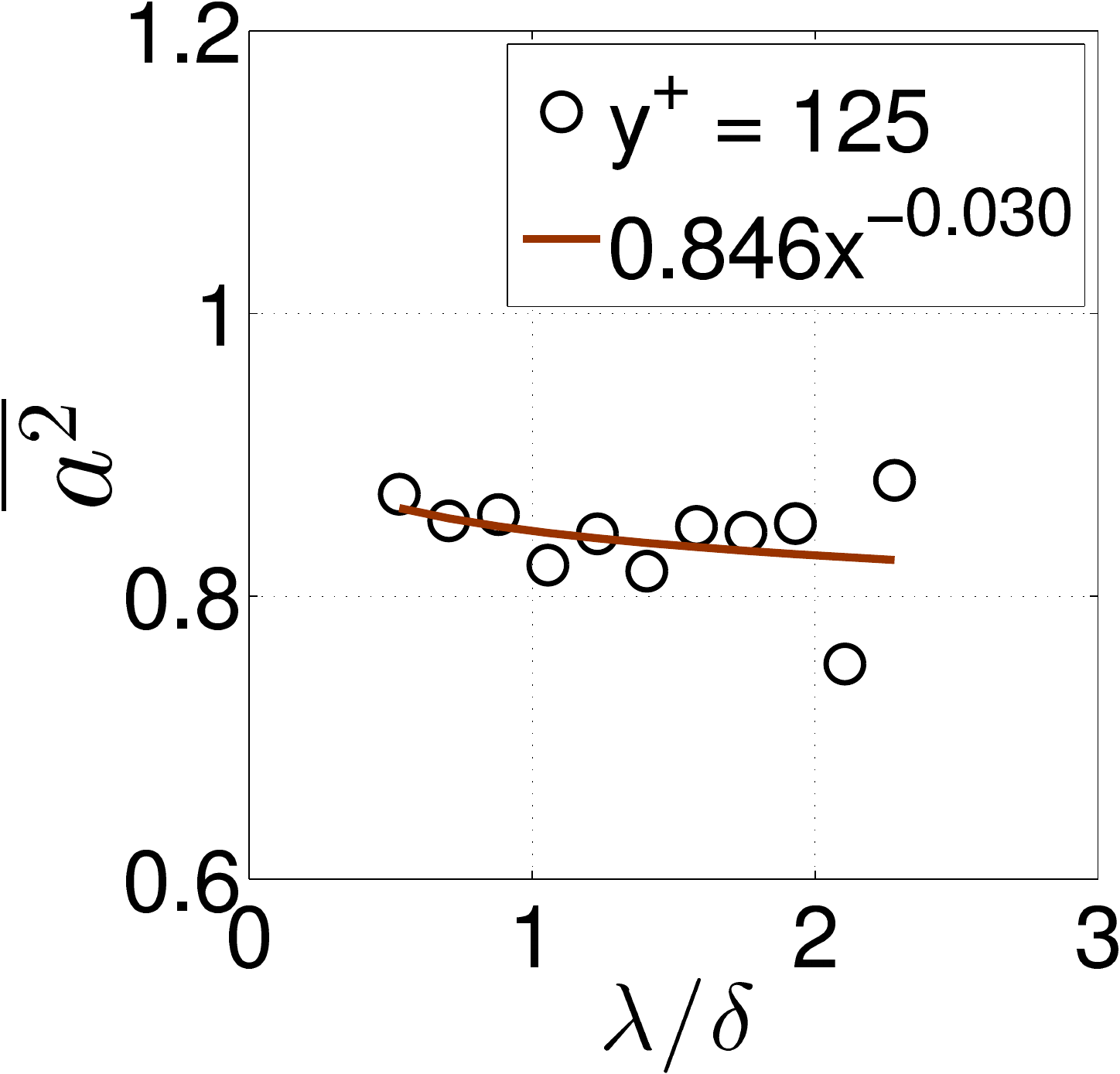}}
	\end{subfigure}   
	\begin{subfigure}{0.495\textwidth}
	    {\includegraphics[width=0.65\textwidth]{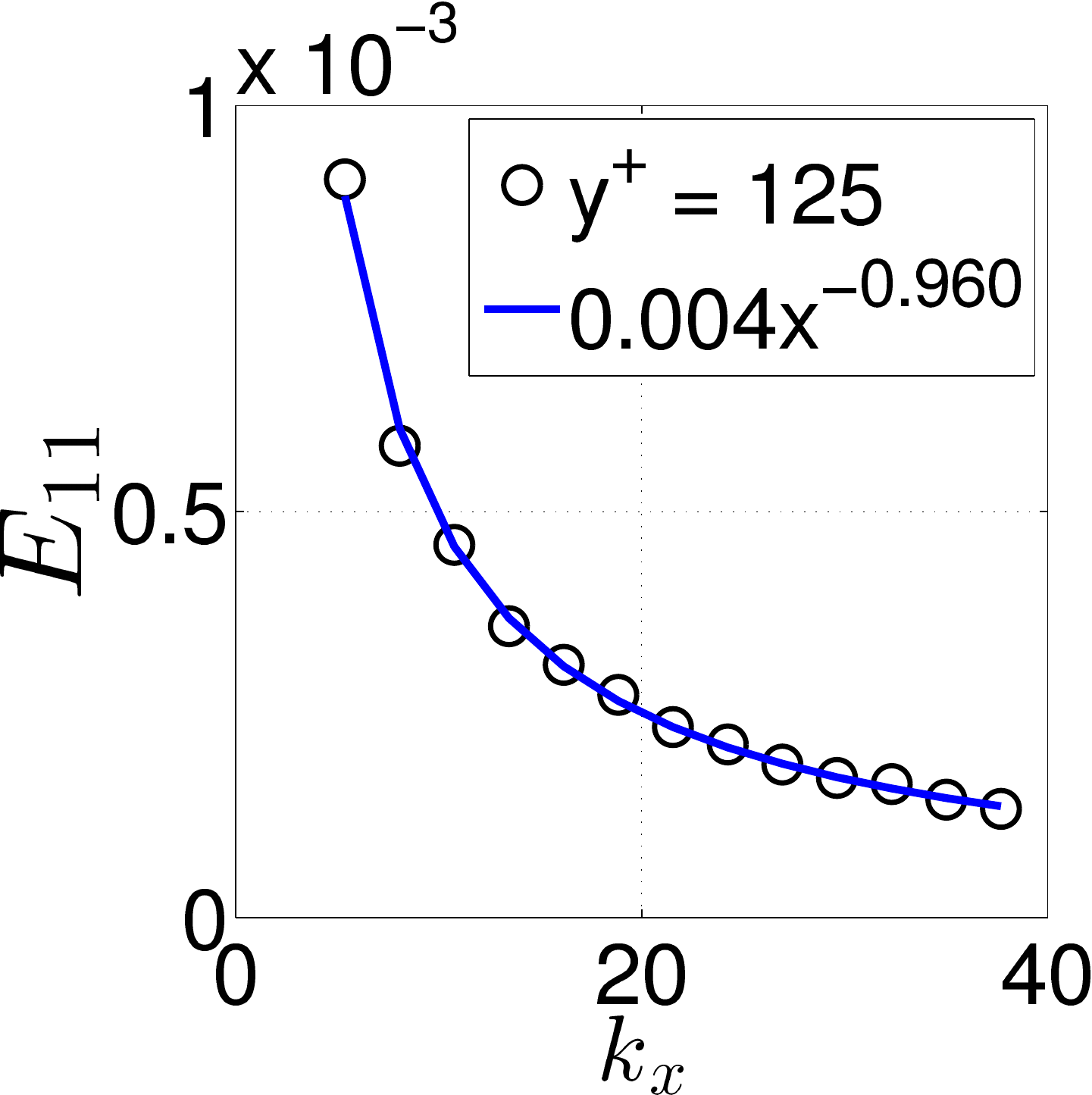}}
	\end{subfigure}
			
		\caption{Lin-lin plots of $\overline{a^{2}}$ versus
                  $\lambda/\delta$ (left) and streamwise energy
                  spectra plotted at wall distances $y^+ =$ 41, 64, 88
                  and 125 (from top to bottom) at $Re_{\theta} =
                  8100$.}
		\label{fig:figure9}
\end{figure}

\begin{figure}
\centering

	\begin{subfigure}{0.495\textwidth}
	\centering
	   {\includegraphics[width=0.65\textwidth]{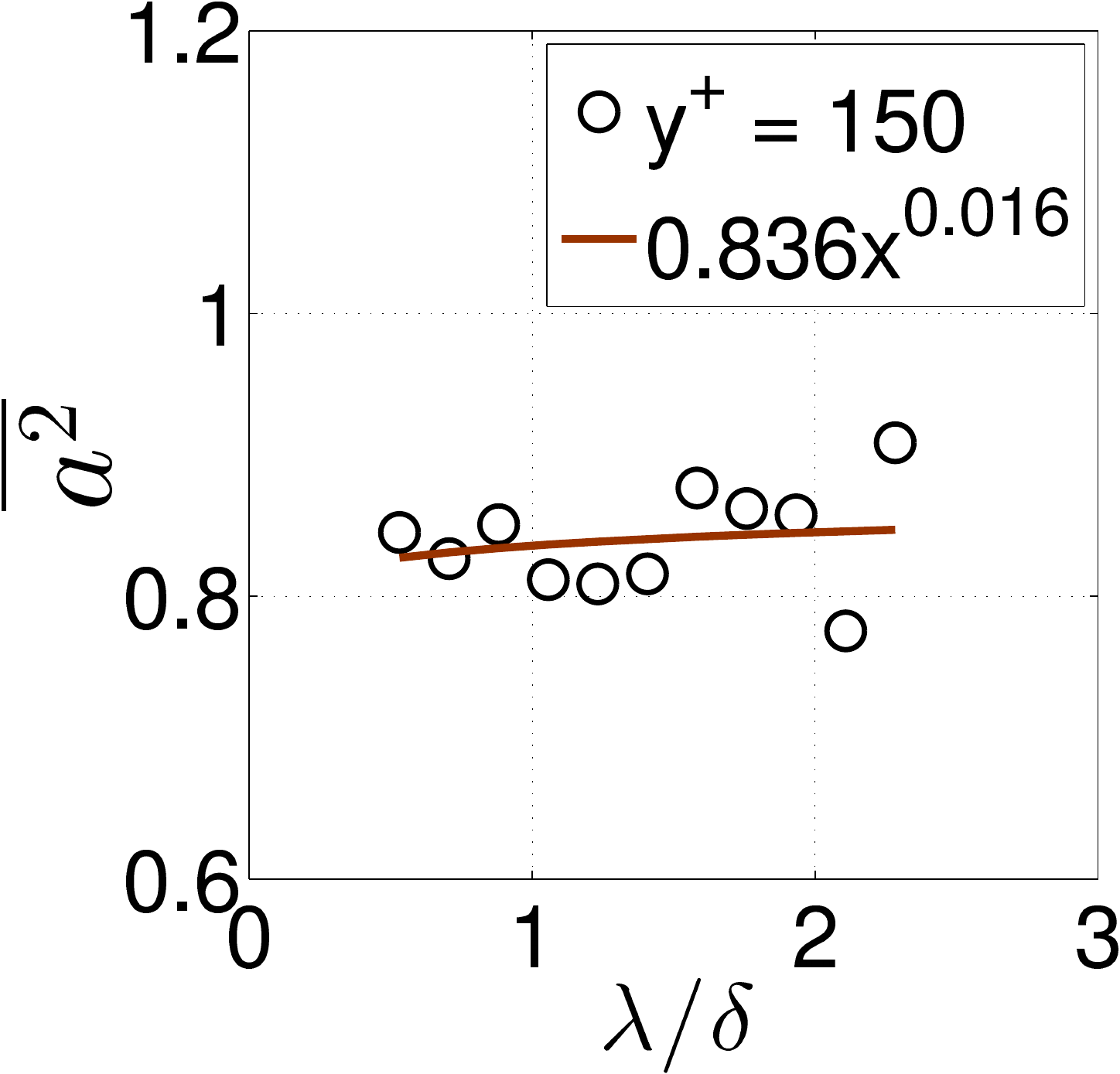}}
	\end{subfigure}   
	\begin{subfigure}{0.495\textwidth}
	\centering
    {\includegraphics[width=0.65\textwidth]{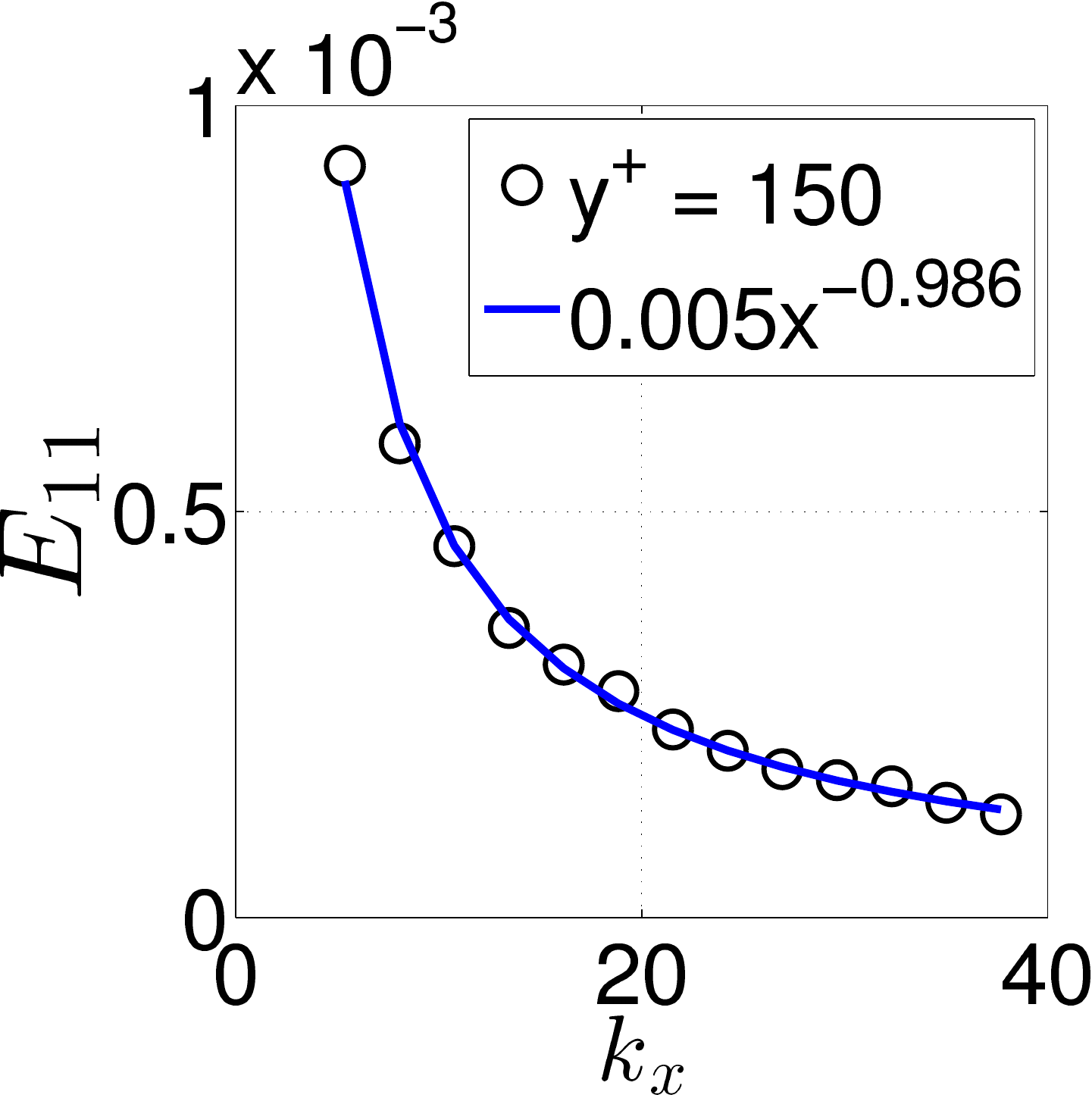}}
	\end{subfigure}
	\begin{subfigure}{0.495\textwidth}
	\centering
	{\includegraphics[width=0.65\textwidth]{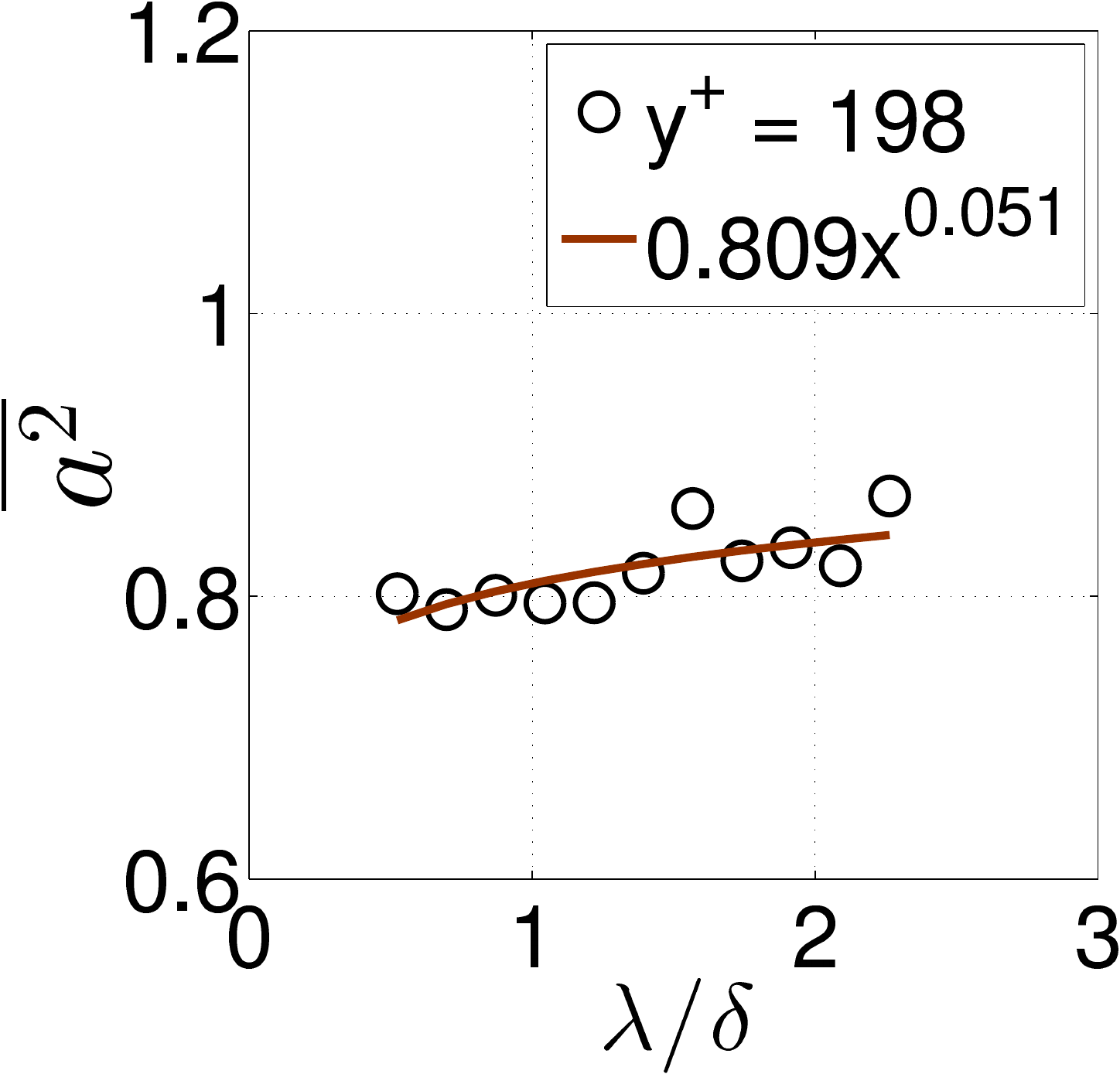}}
	\end{subfigure}   
	\begin{subfigure}{0.495\textwidth}
			\centering
		    {\includegraphics[width=0.65\textwidth]{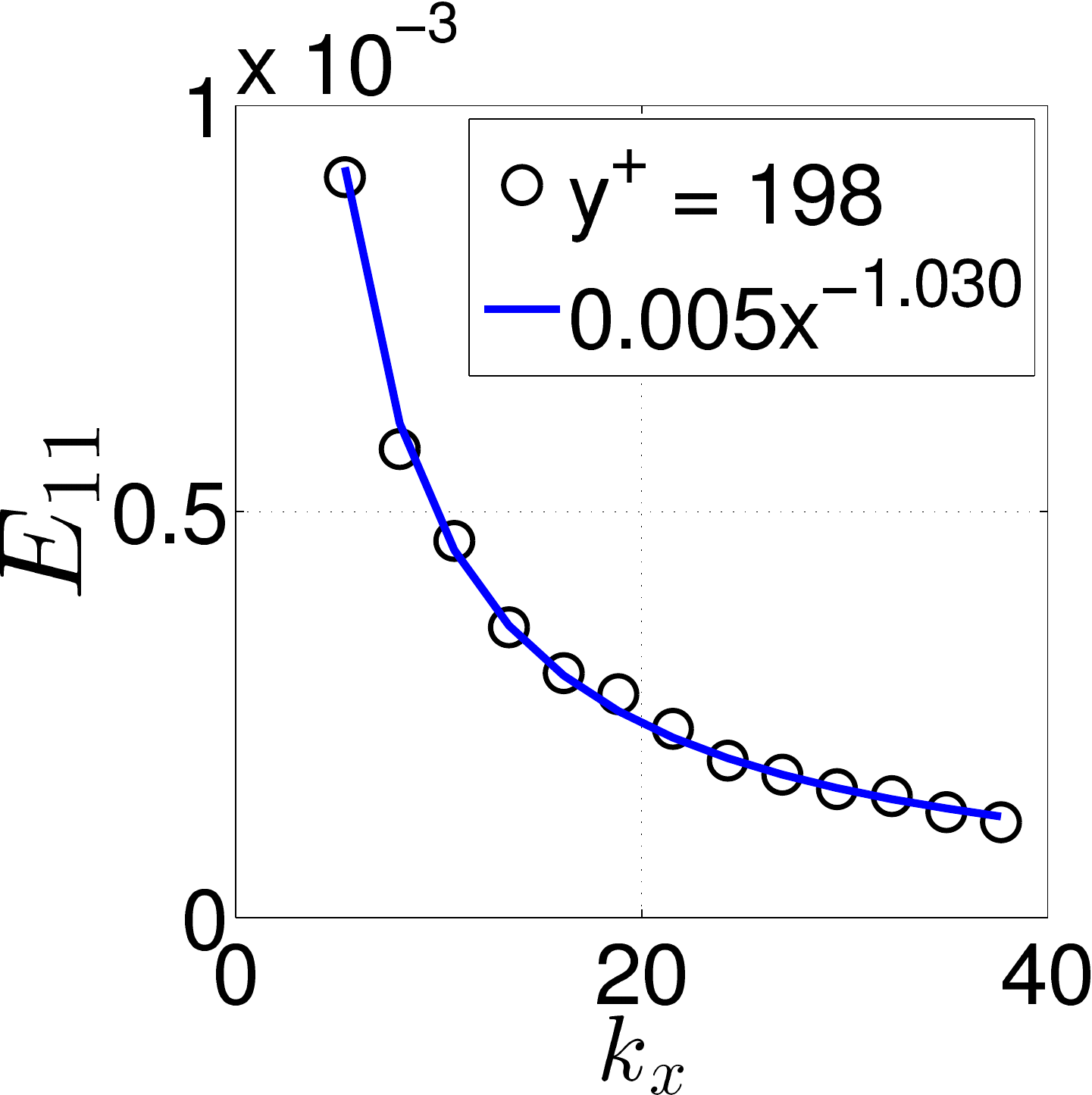}}
	\end{subfigure}
	\begin{subfigure}{0.495\textwidth}
		\centering
		   {\includegraphics[width=0.65\textwidth]{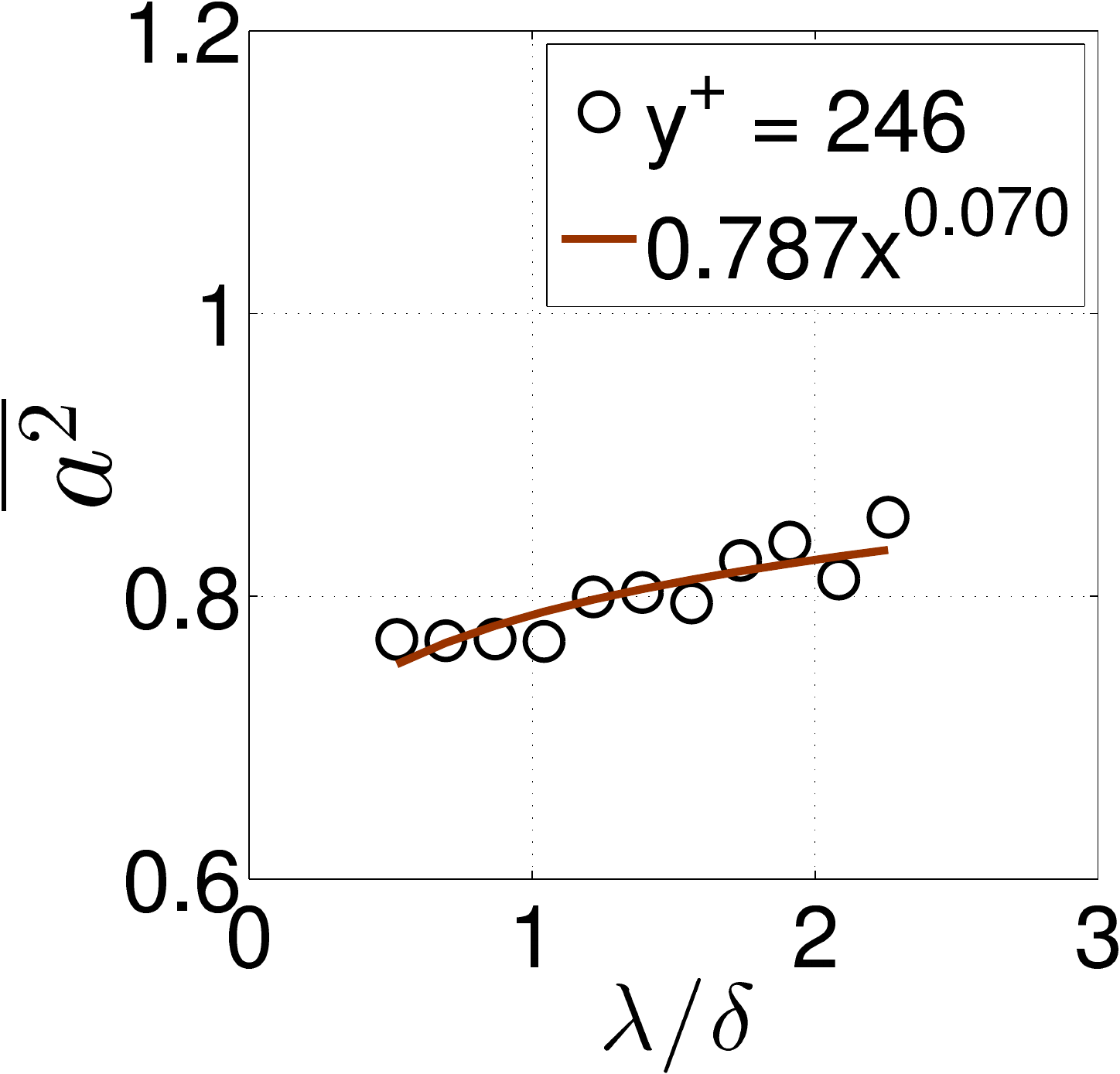}}
	\end{subfigure}   
	\begin{subfigure}{0.495\textwidth}
	\centering
	{\includegraphics[width=0.65\textwidth]{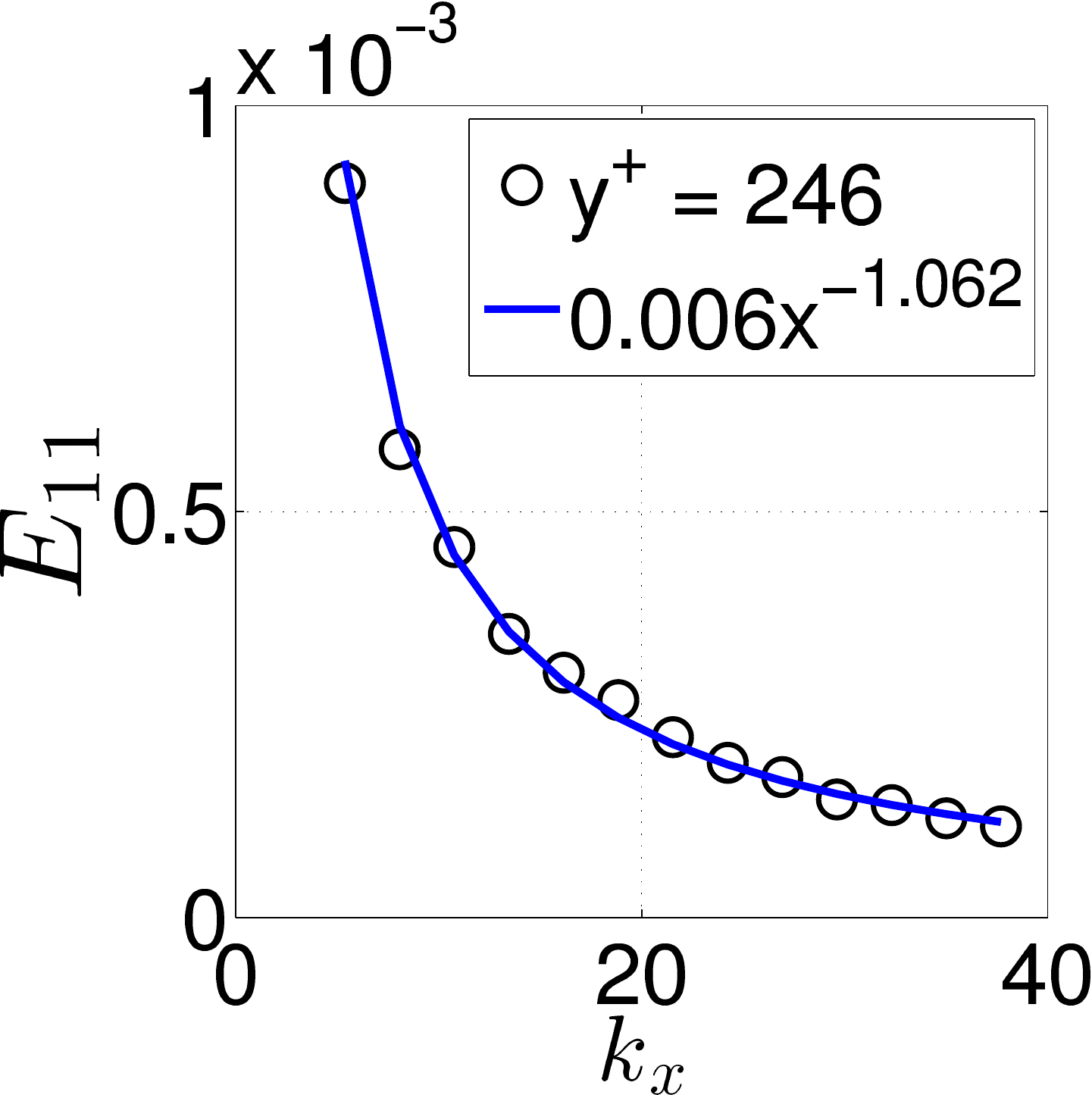}}
	\end{subfigure}
	\begin{subfigure}{0.495\textwidth}
		\centering
		   {\includegraphics[width=0.65\textwidth]{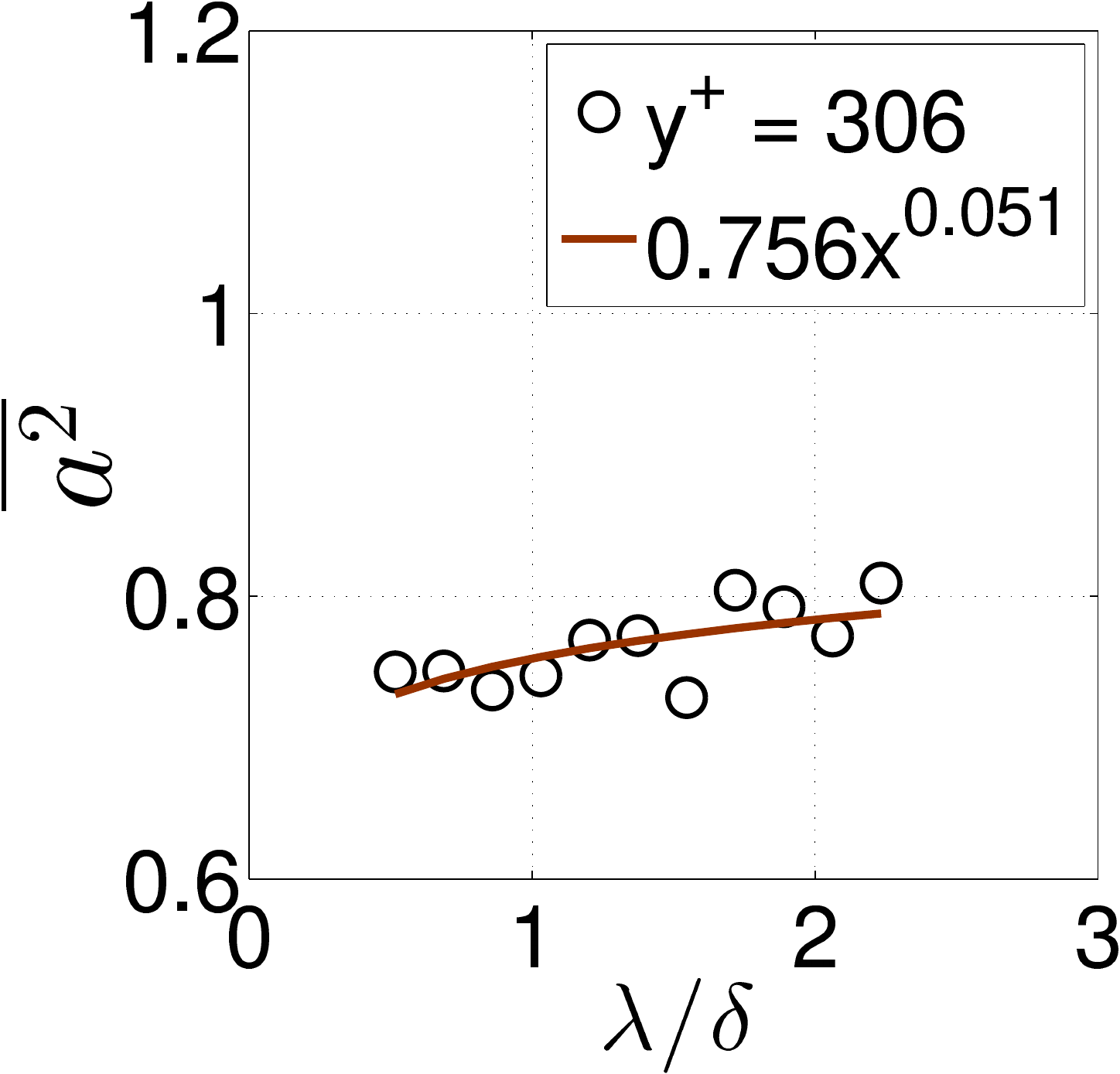}}
	\end{subfigure}   
	\begin{subfigure}{0.495\textwidth}
		\centering
	    {\includegraphics[width=0.65\textwidth]{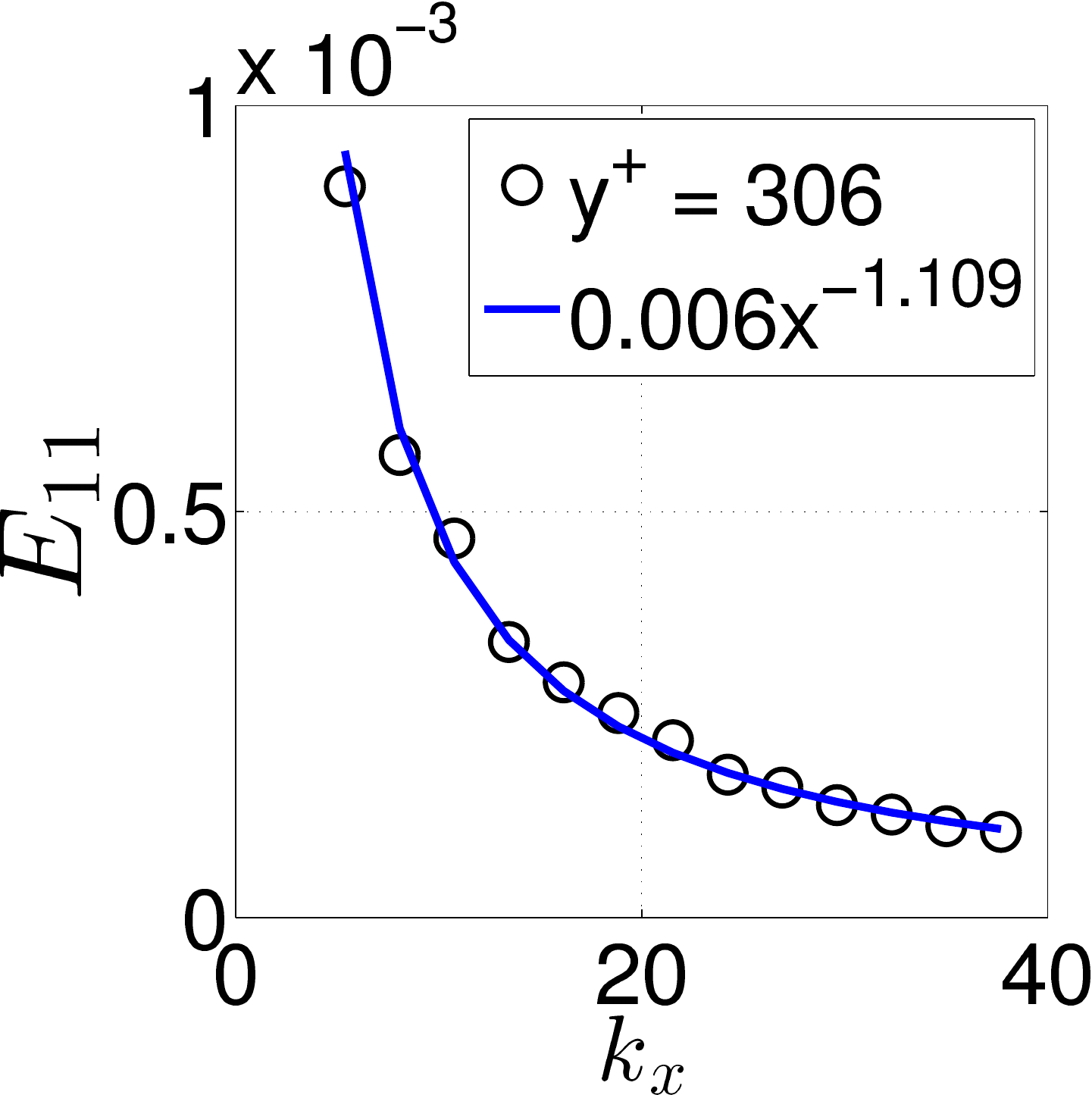}}
	\end{subfigure}

		\caption{Lin-lin plots of $\overline{a^{2}}$ versus
                  $\lambda/\delta$ (left) and streamwise energy
                  spectra plotted at wall distances $y^+ =$ 150, 198, 246
                  and 306 (from top to bottom) at $Re_{\theta} =
                  8100$.}
		\label{fig:figure10}
\end{figure}

\begin{figure}
\centering

	\begin{subfigure}{0.495\textwidth}
	\centering
	   {\includegraphics[width=0.65\textwidth]{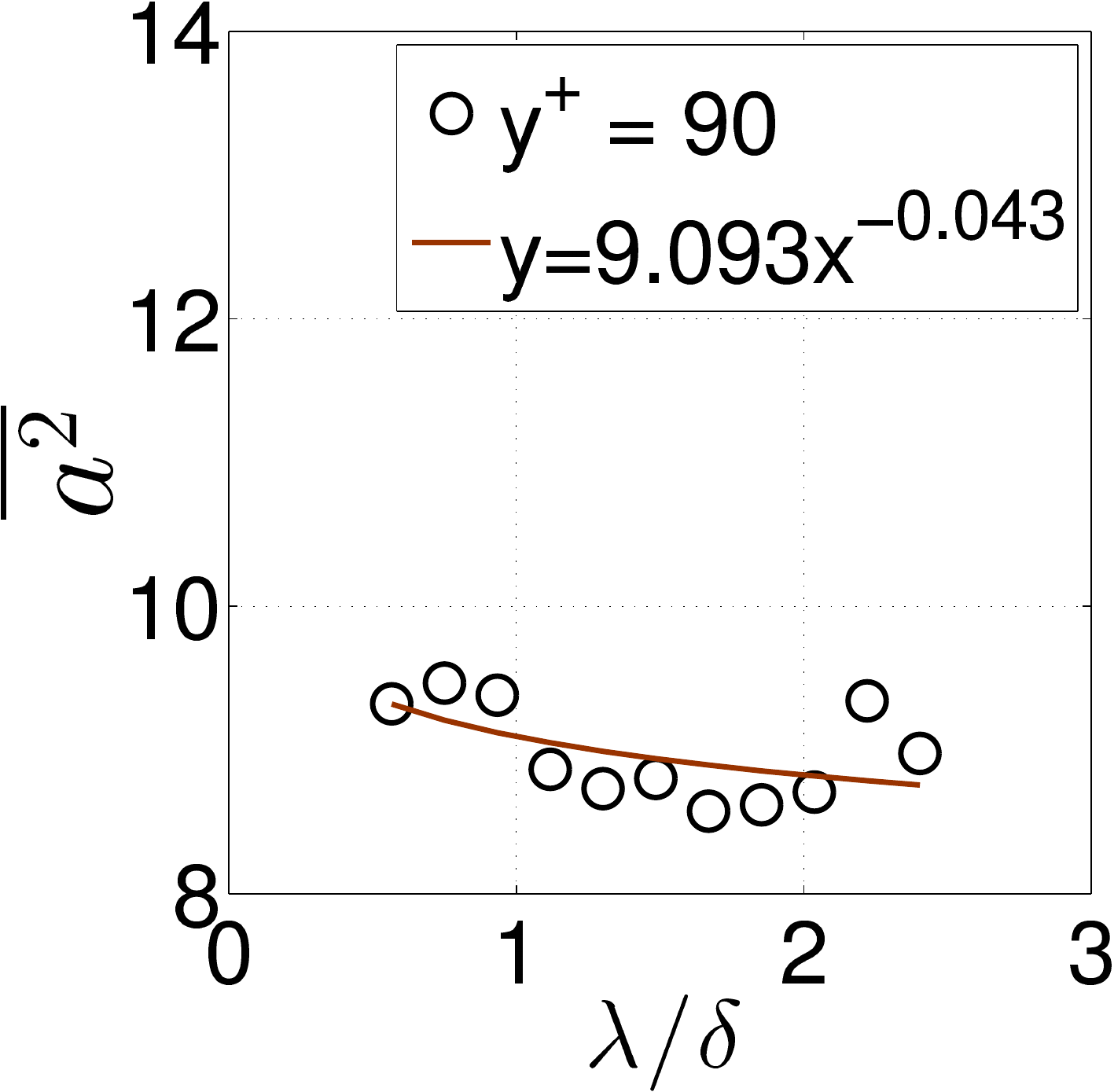}}
	\end{subfigure}   
	\begin{subfigure}{0.495\textwidth}
	\centering
    {\includegraphics[width=0.65\textwidth]{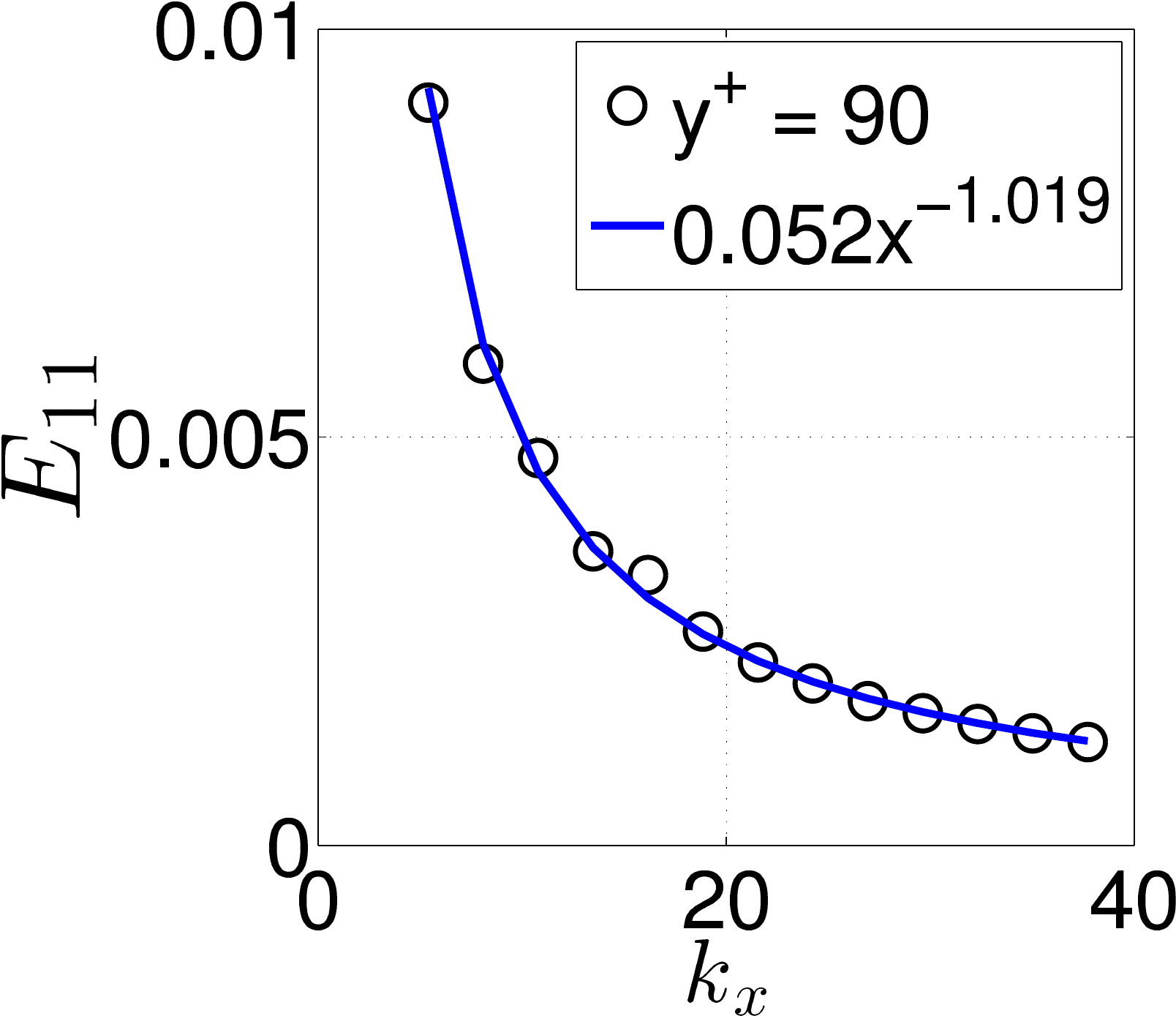}}
	\end{subfigure}
	\begin{subfigure}{0.495\textwidth}
	\centering
	{\includegraphics[width=0.65\textwidth]{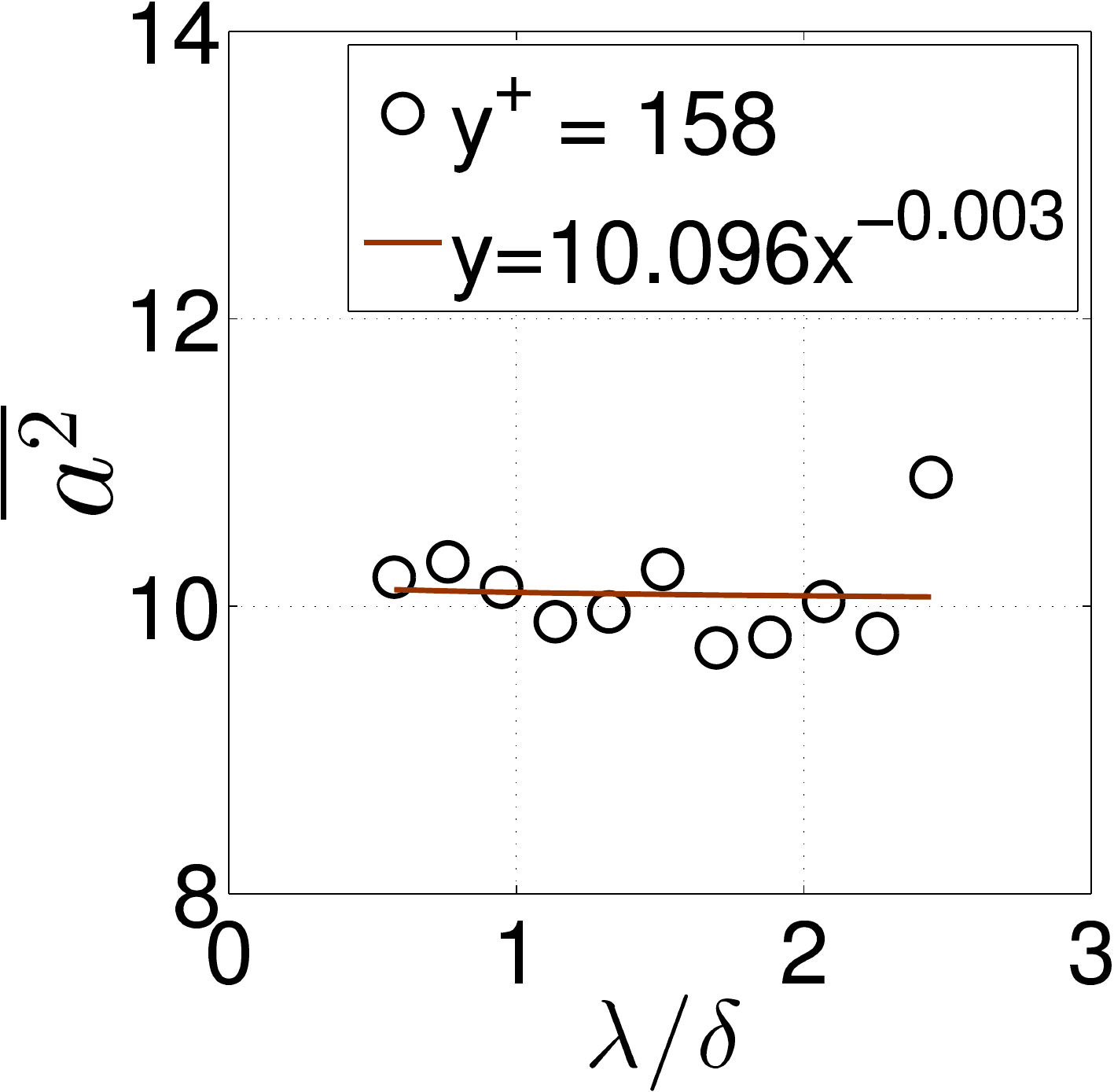}}
	\end{subfigure}   
	\begin{subfigure}{0.495\textwidth}
			\centering
		    {\includegraphics[width=0.65\textwidth]{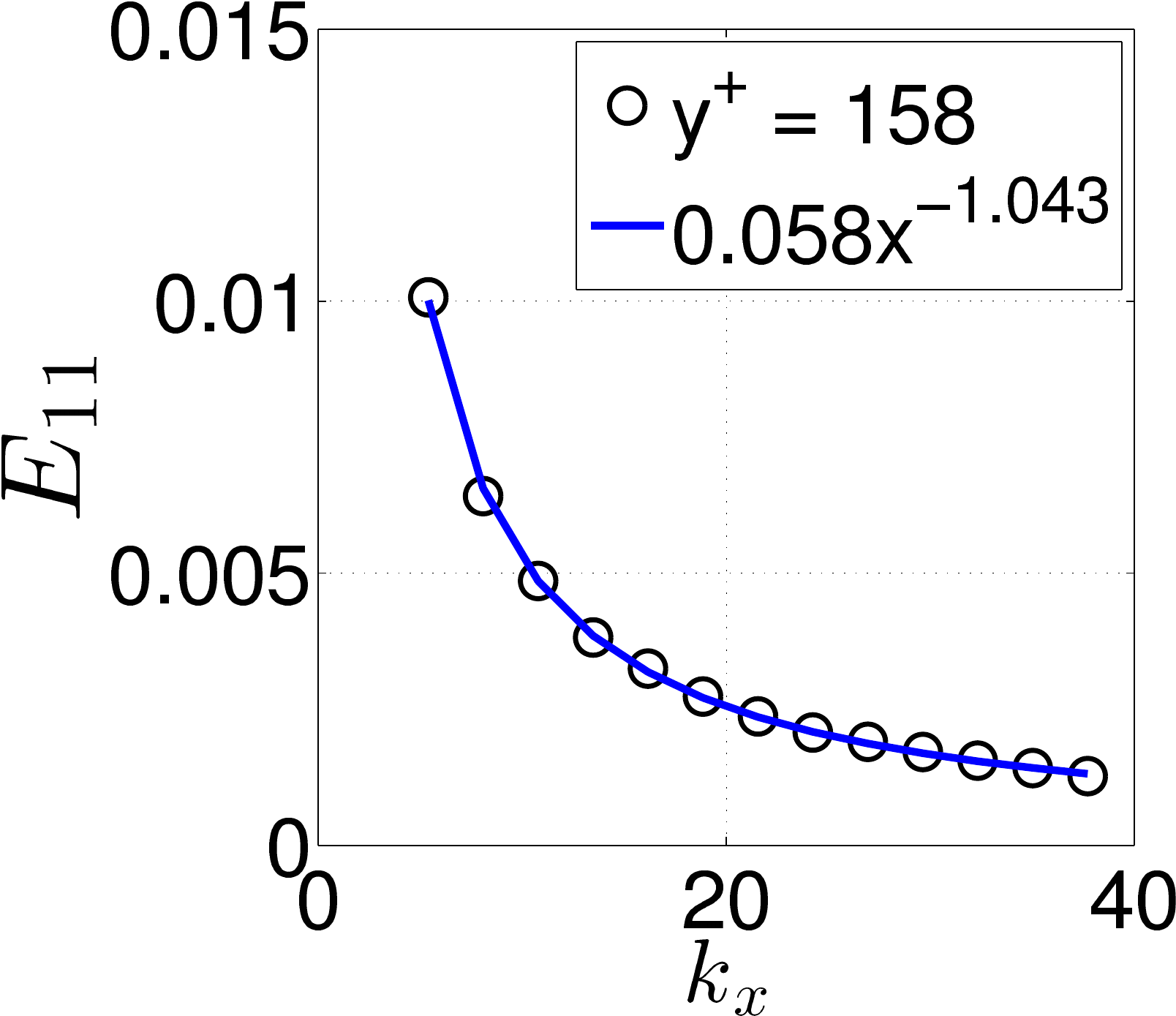}}
	\end{subfigure}
	\begin{subfigure}{0.495\textwidth}
		\centering
		   {\includegraphics[width=0.65\textwidth]{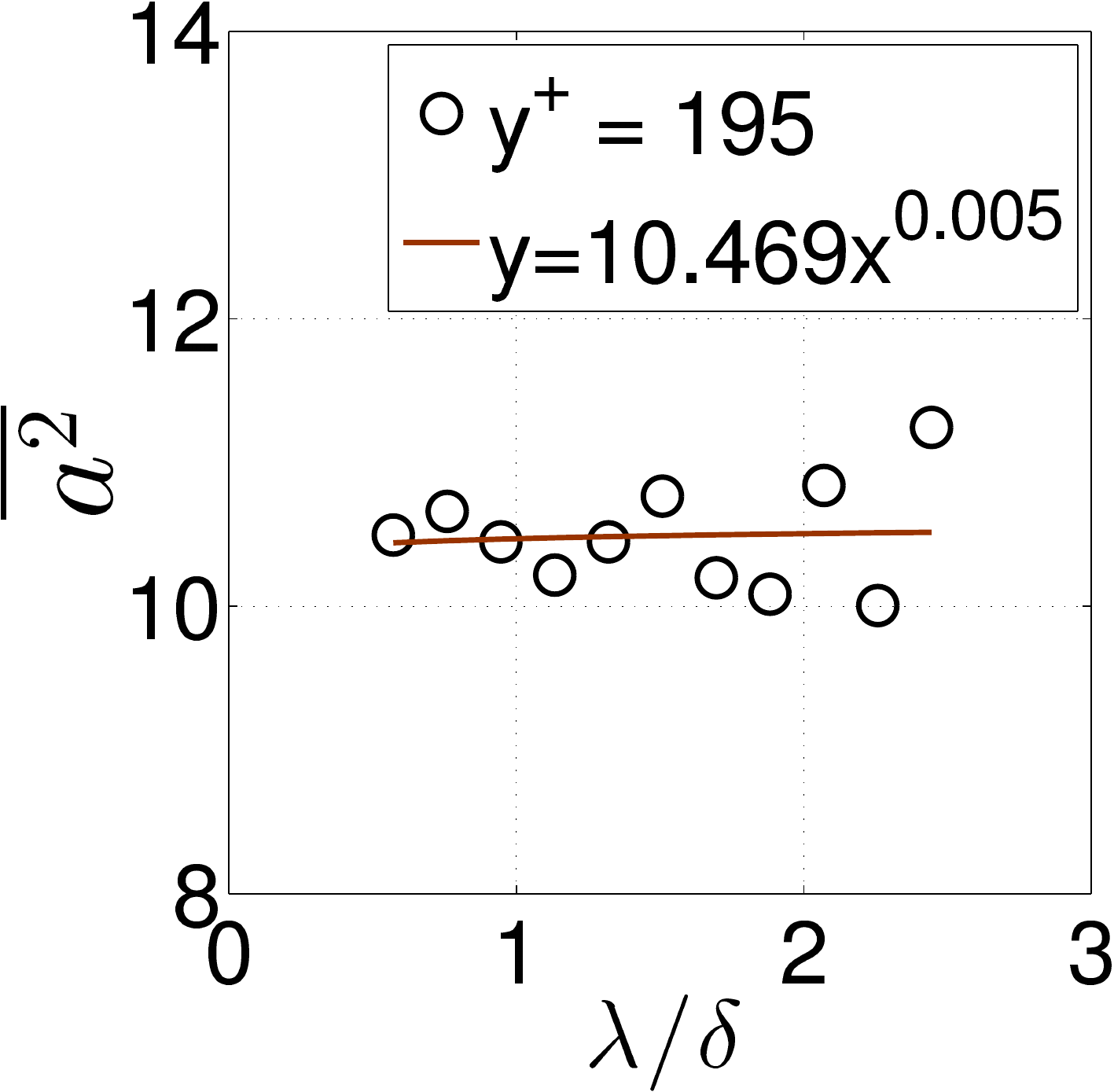}}
	\end{subfigure}   
	\begin{subfigure}{0.495\textwidth}
	\centering
	{\includegraphics[width=0.65\textwidth]{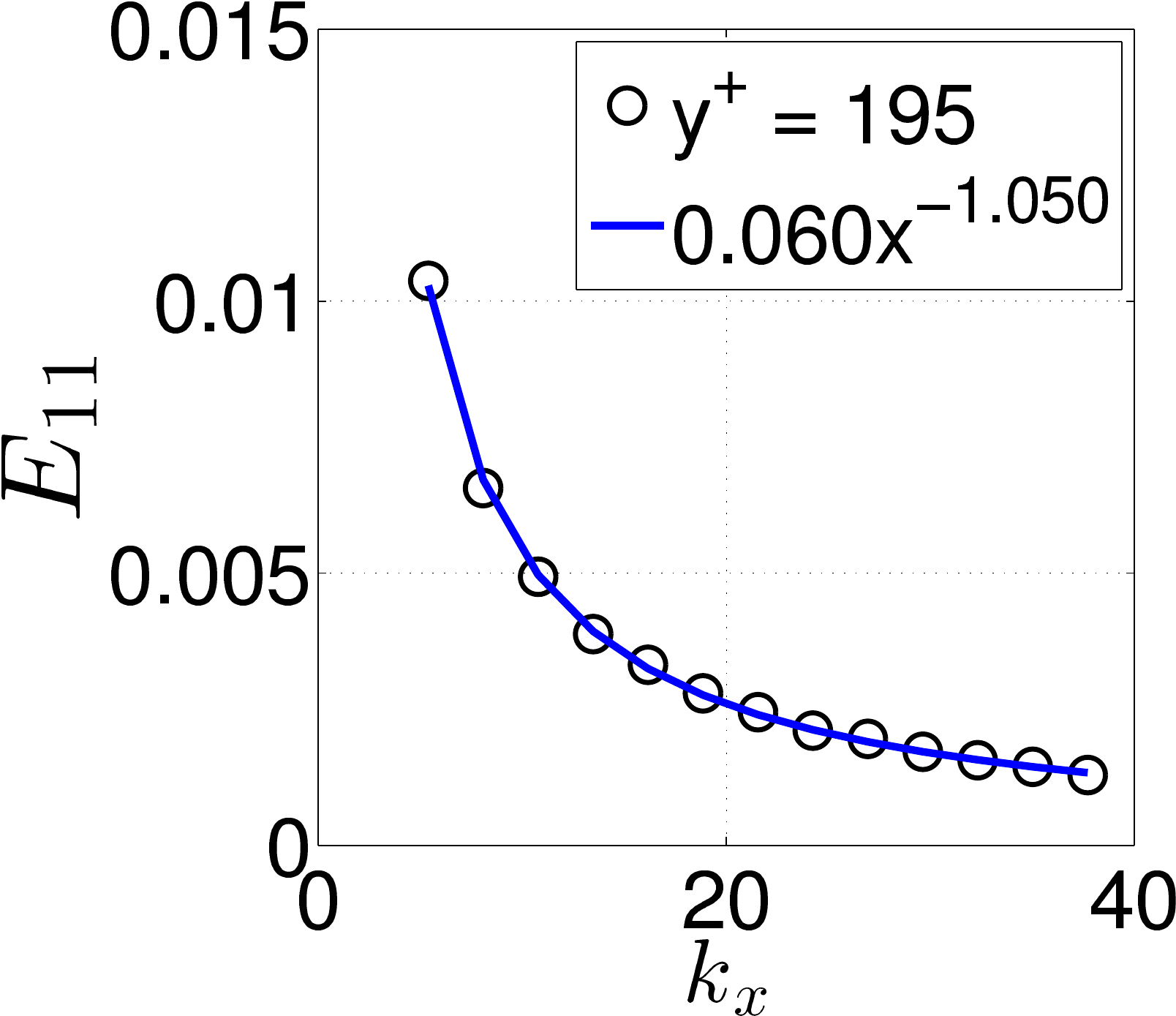}}
	\end{subfigure}
	\begin{subfigure}{0.495\textwidth}
		\centering
		   {\includegraphics[width=0.65\textwidth]{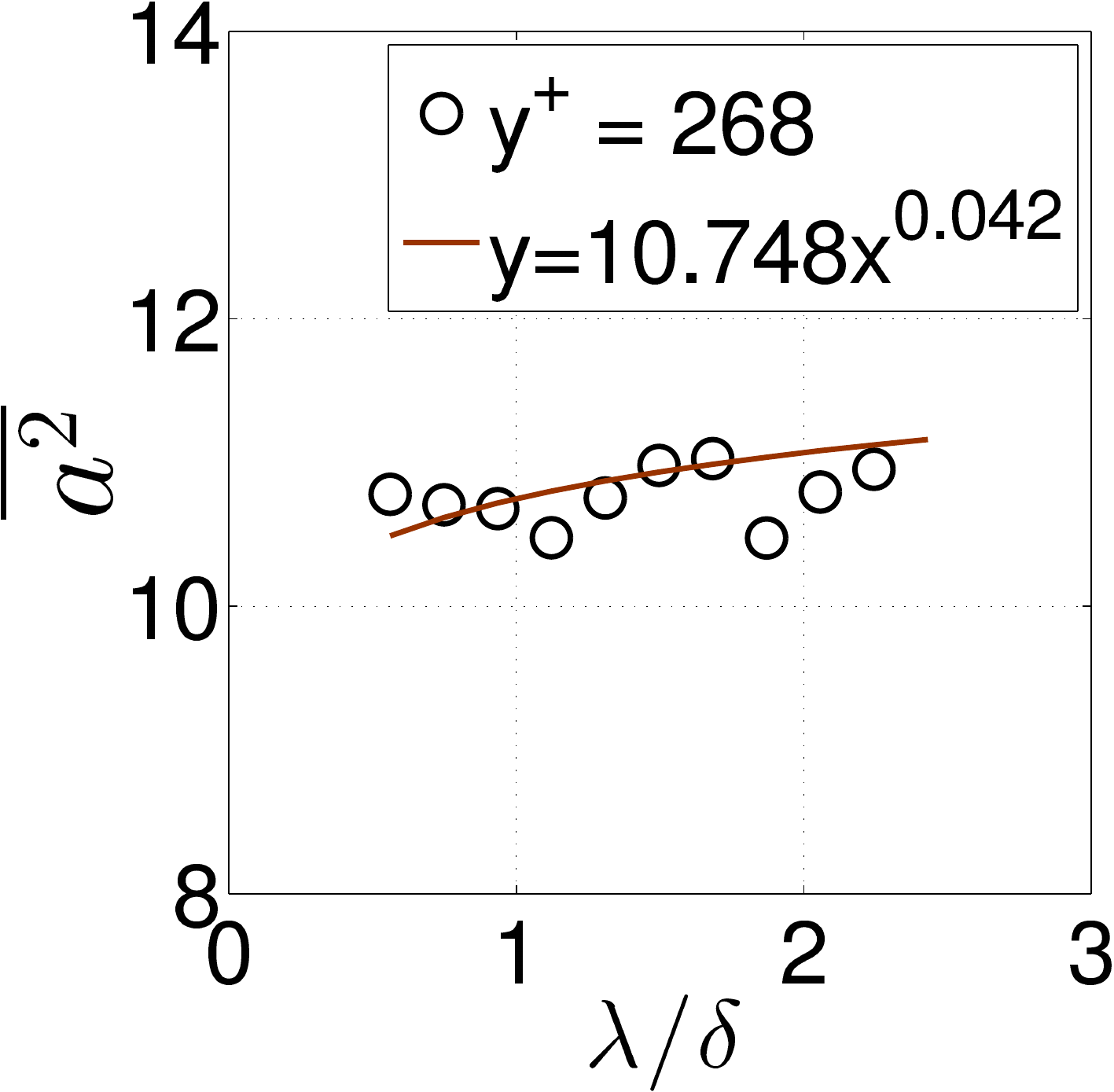}}
	\end{subfigure}   
	\begin{subfigure}{0.495\textwidth}
		\centering
	    {\includegraphics[width=0.65\textwidth]{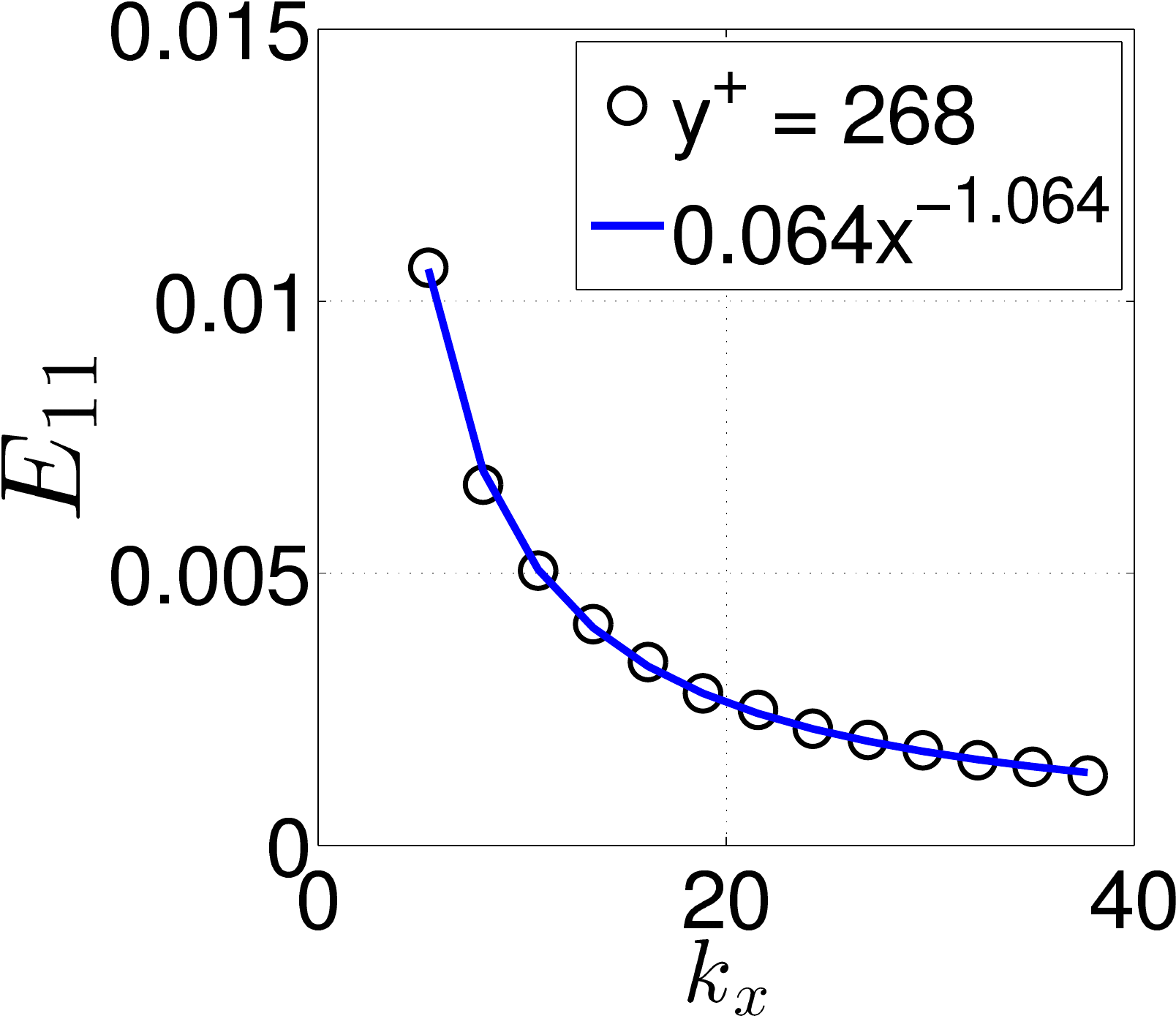}}
	\end{subfigure}

		\caption{Lin-lin plots of $\overline{a^{2}}$ versus
                  $\lambda/\delta$ (left) and streamwise energy
                  spectra plotted at wall distances $y^+ =$ 90, 158,
                  195 and 268 (from top to bottom) at $Re_{\theta}
                  = 20600$.}
		\label{fig:figure11}
\end{figure}

\begin{figure}
\centering

	\begin{subfigure}{0.495\textwidth}
	\centering
	   {\includegraphics[width=0.95\textwidth]{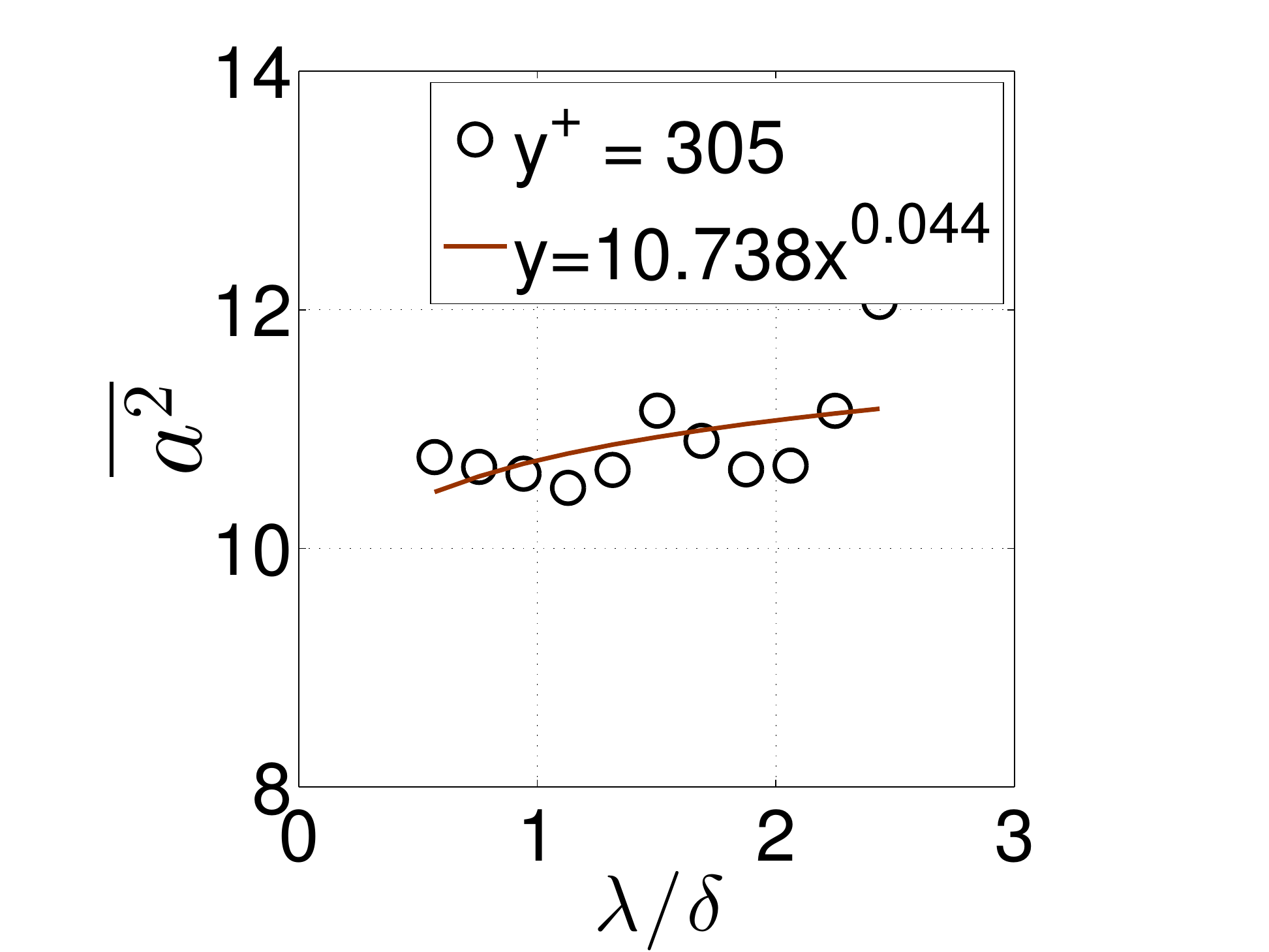}}
	\end{subfigure}   
	\begin{subfigure}{0.495\textwidth}
	\centering
    {\includegraphics[width=0.95\textwidth]{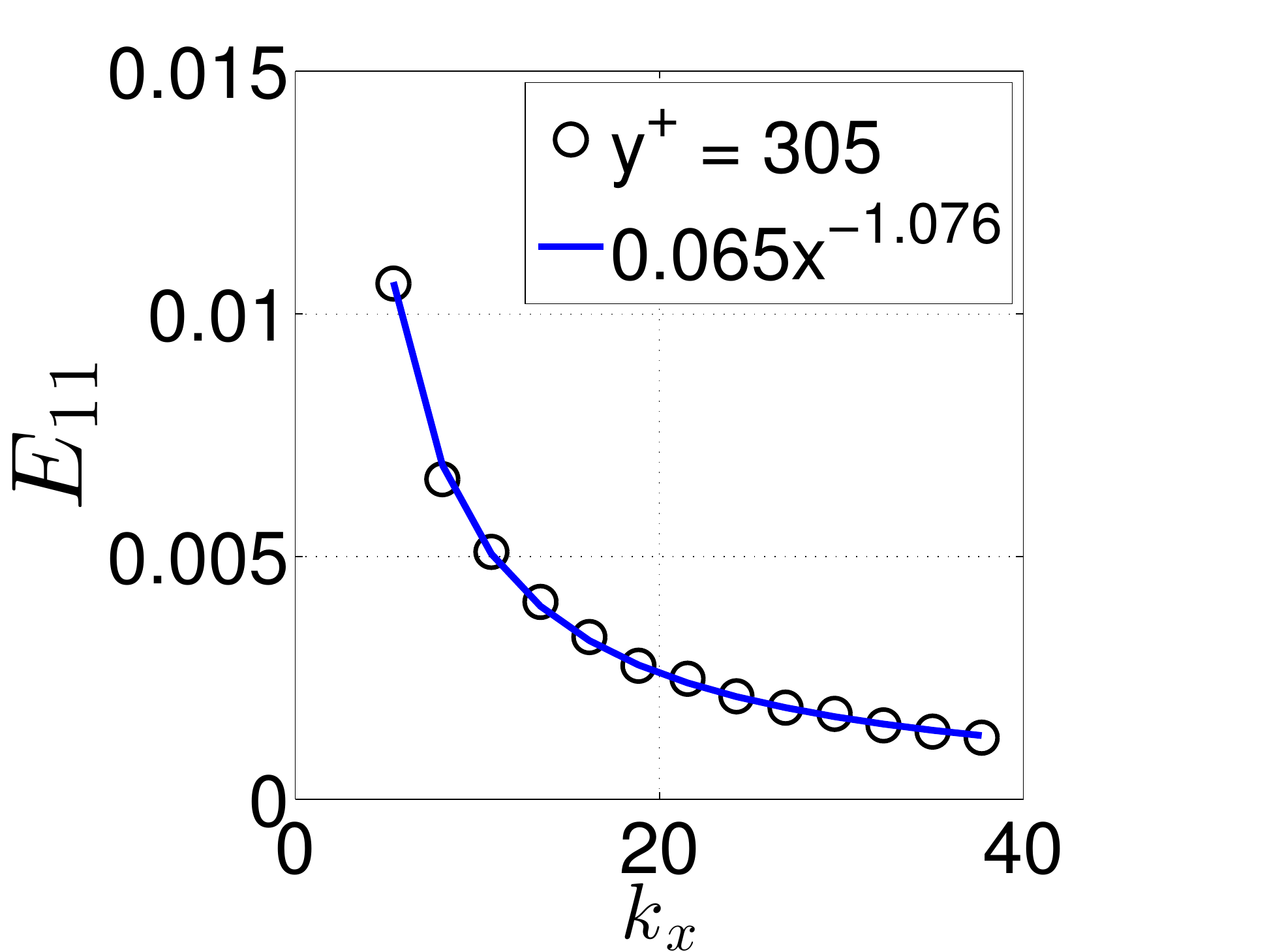}}
	\end{subfigure}
	\begin{subfigure}{0.495\textwidth}
	\centering
	{\includegraphics[width=0.95\textwidth]{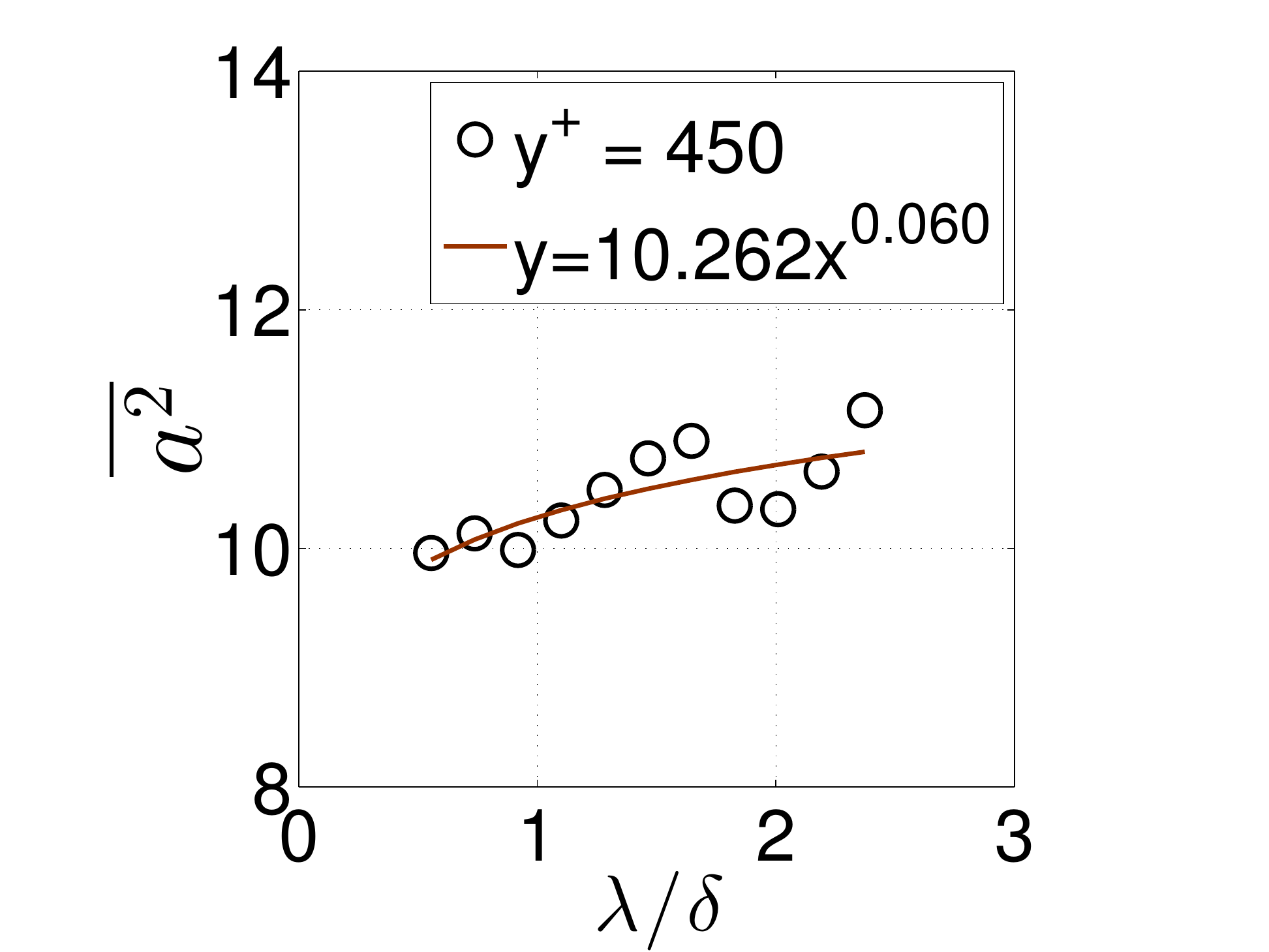}}
	\end{subfigure}   
	\begin{subfigure}{0.495\textwidth}
			\centering
		    {\includegraphics[width=0.95\textwidth]{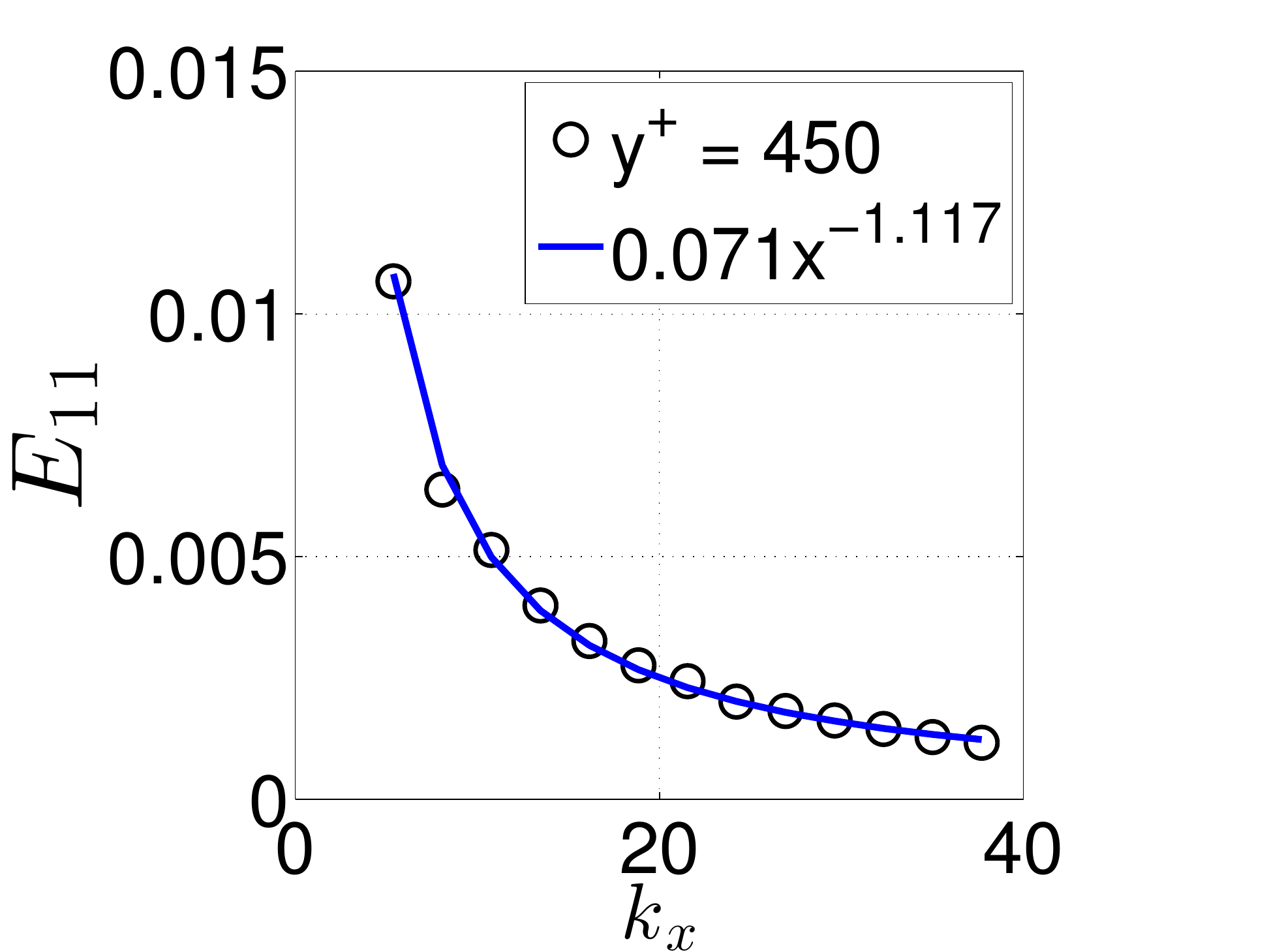}}
	\end{subfigure}
	\begin{subfigure}{0.495\textwidth}
		\centering
		   {\includegraphics[width=0.95\textwidth]{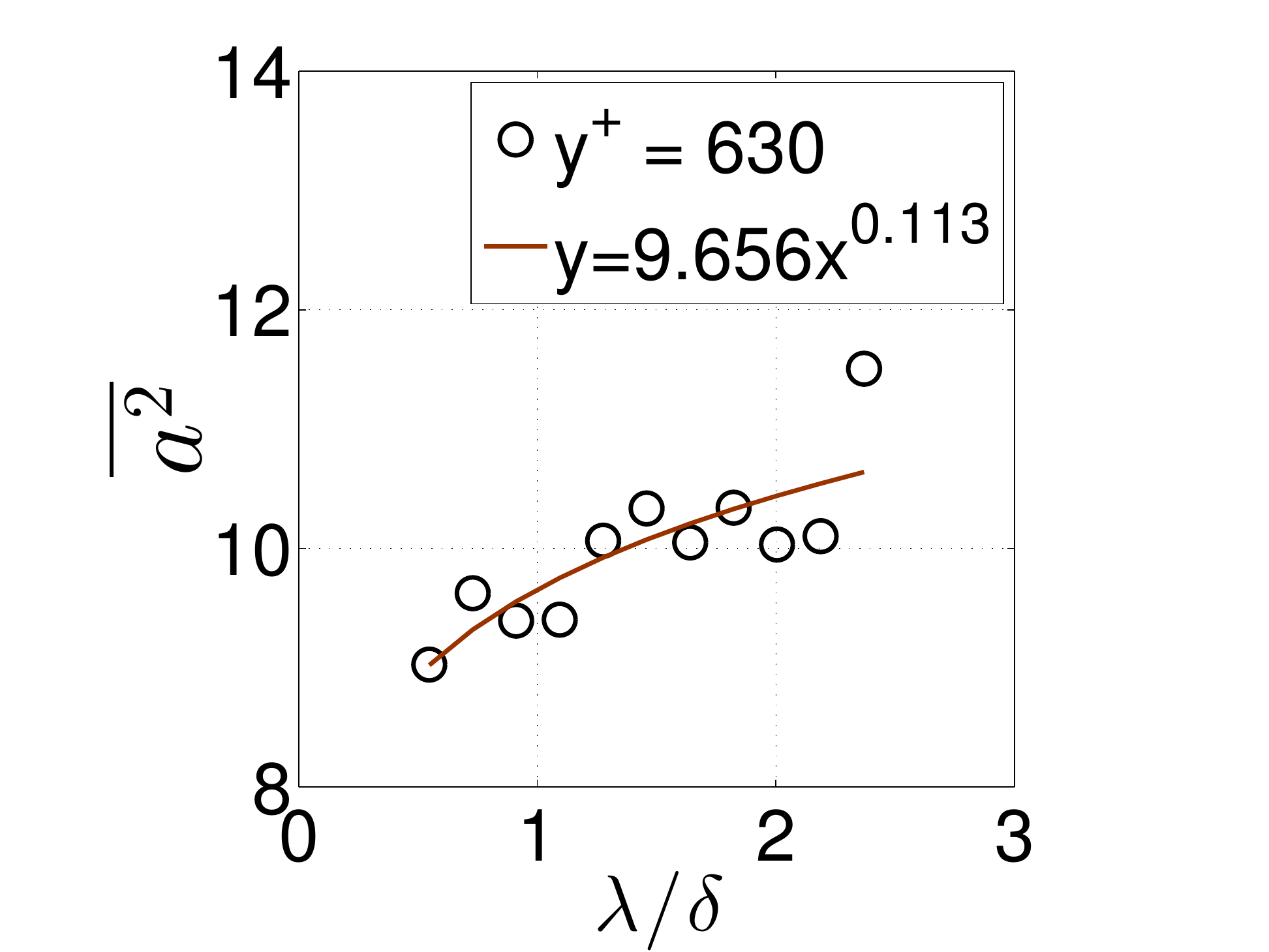}}
	\end{subfigure}   
	\begin{subfigure}{0.495\textwidth}
	\centering
	{\includegraphics[width=0.95\textwidth]{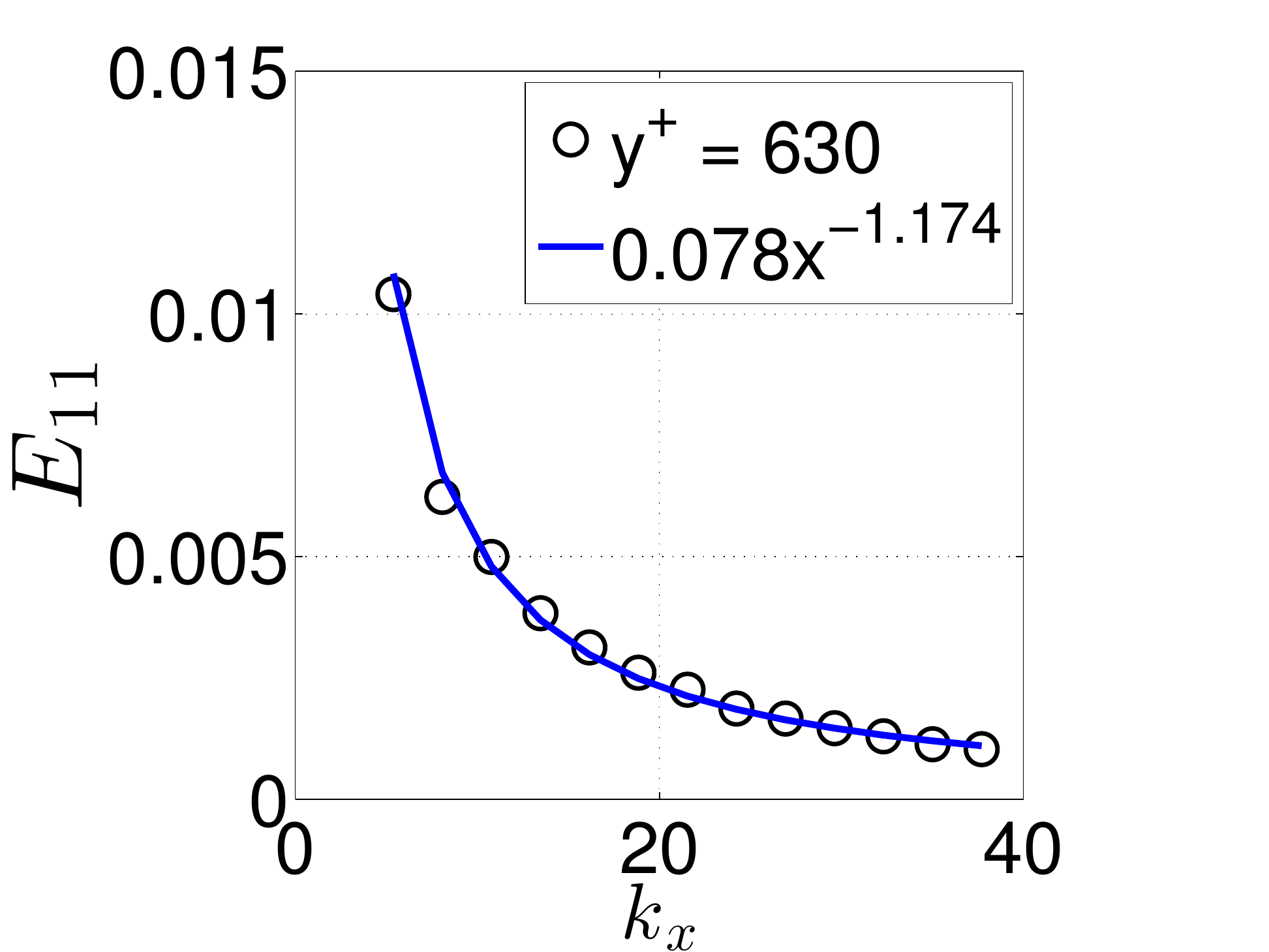}}
	\end{subfigure}
	\begin{subfigure}{0.495\textwidth}
		\centering
		   {\includegraphics[width=0.95\textwidth]{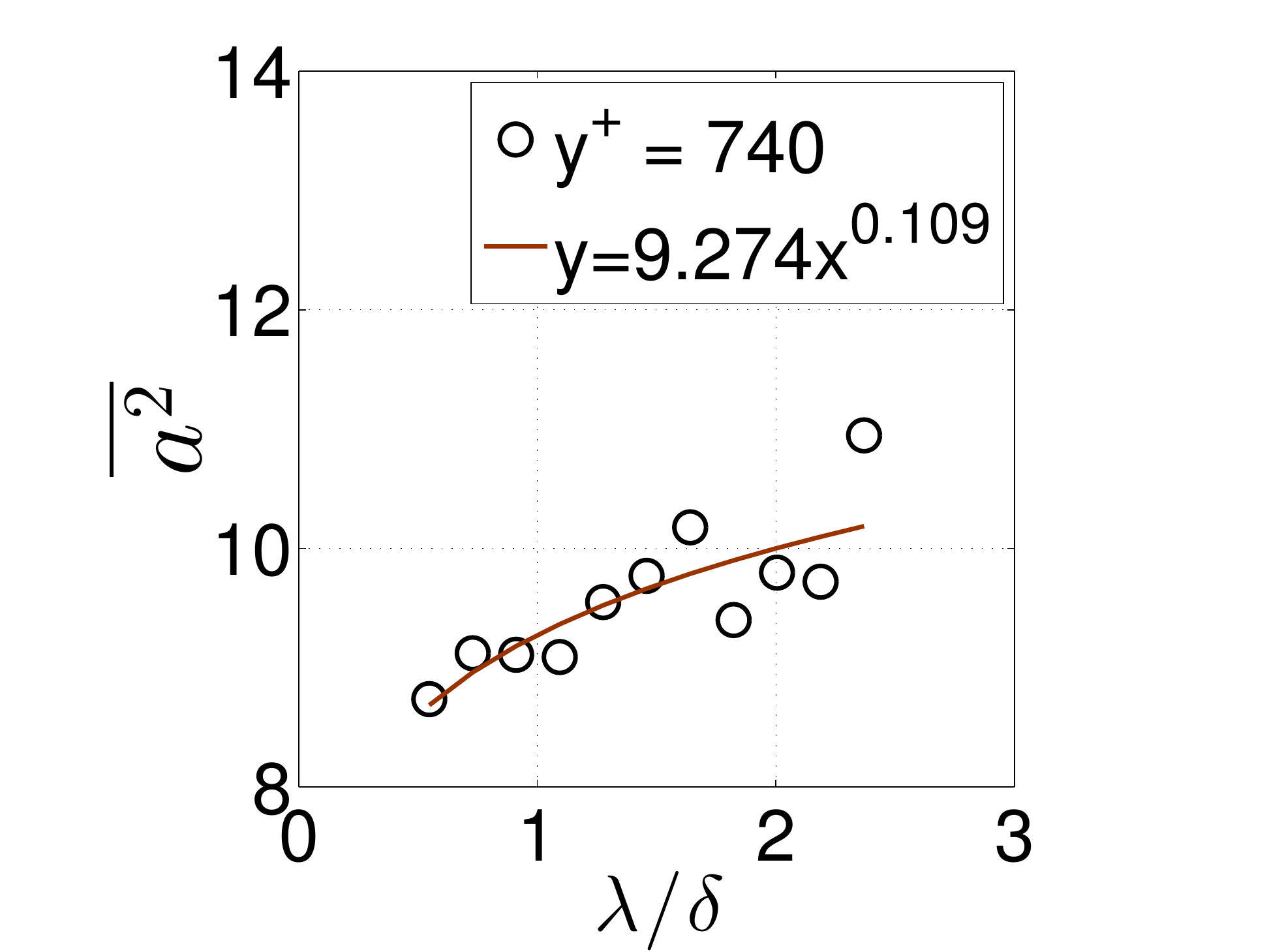}}
	\end{subfigure}   
	\begin{subfigure}{0.495\textwidth}
		\centering
	    {\includegraphics[width=0.95\textwidth]{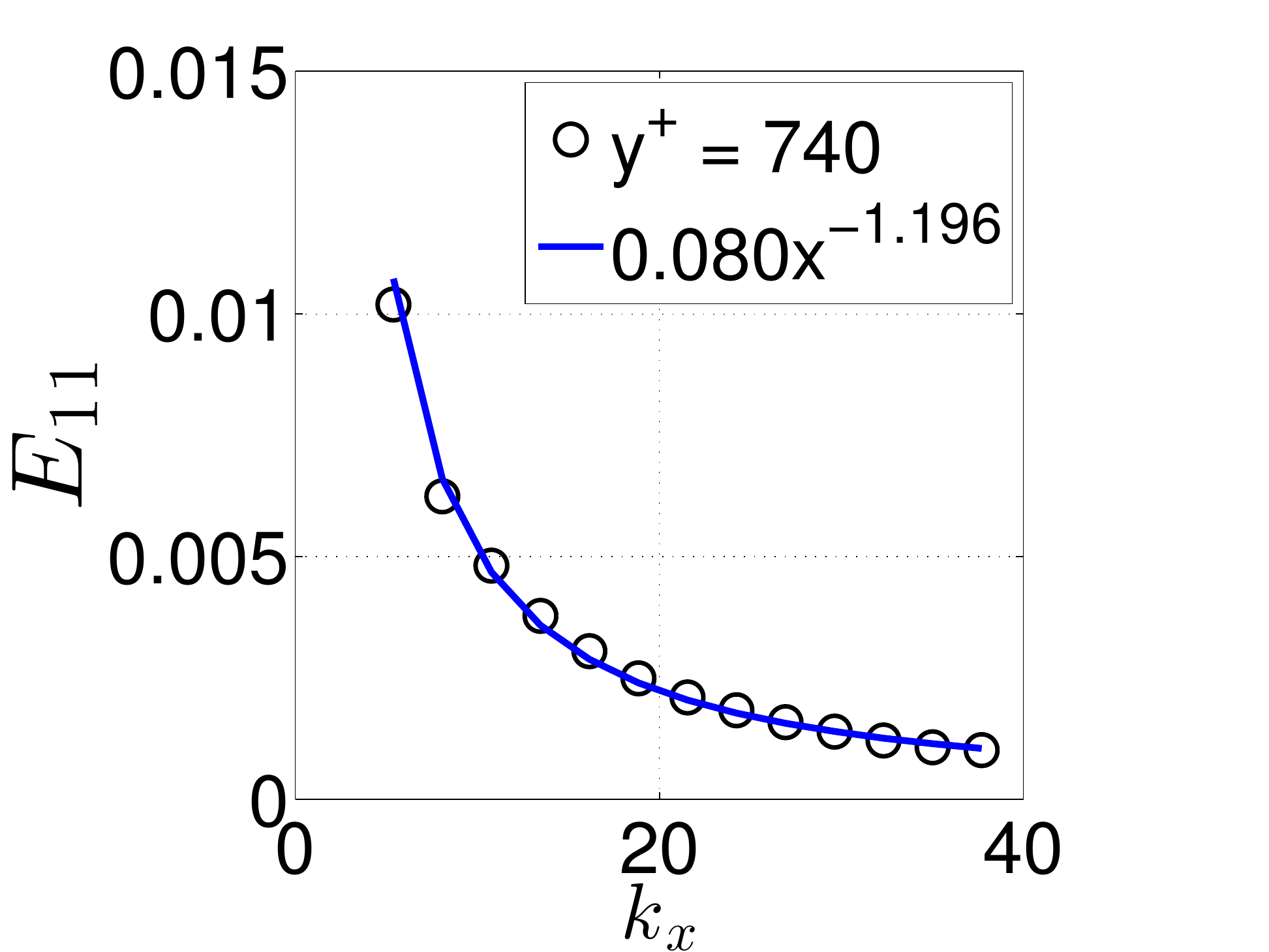}}
	\end{subfigure}

		\caption{Lin-lin plots of $\overline{a^{2}}$ versus
                  $\lambda/\delta$ (left) and streamwise energy
                  spectra plotted at wall distances $y^+ =$ 305, 450,
                  630 and 740 (from top to bottom) at $Re_{\theta}
                  = 20600$.}
		\label{fig:figure12}
\end{figure}

In figures \ref{fig:figure9} to \ref{fig:figure12} we plot $\overline{a^{2}}$ versus $\lambda/\delta$
where $\overline{a^{2}}$
is the average of $\alpha^{2} / \Delta \lambda$ conditional on the
streamwise length of a labelled structure being between $\lambda$ and
$\lambda + \Delta \lambda$ ($\alpha$ and $\lambda$ being obtained as
explained in the first paragraph of subsection \ref{sec:Lengths}). The upper values
of $\lambda/\delta$ in these plots are all below about 2.3 because we
do not have enough samples of educed structures beyond $\lambda/\delta
\approx 2.3$ to obtain values of $\overline{a^2}$ which are
statistically converged. The lower values of $\lambda/\delta$ in these
plots are all close to $1/2$ because the range where the PDF of
$\lambda/\delta$ has been found in the previous subsection to be well
approximated by $-C_{1} + C_{2}(\lambda/\delta)^{-2}$ is bounded from
below by about $1/2$ in all our $y^+$ and $Re_{\theta}$ cases. In
figures \ref{fig:figure9} to \ref{fig:figure12} we also plot $E_{11} (k_{x})$ in the corresponding
wavenumber range $0.25/\delta \le k_{x}/(2\pi) \le 2/\delta$ which, as
discussed in the previous paragraph, may be close to the wavenumber
range $2\pi/(4\delta) < k_{x} < 0.63/y$ that one could interpret as a
Townsend-Perry range. We do not have enough data and high enough
Reynolds numbers to clearly distinguish between these two ranges in
the present work.

As an aside for the moment, note that the large-scale motions (LSMs)
and very large-scale motions (VLSMs), which have been found to exist
in the logarithmic and lower wake regions of a turbulent boundary
layer 
(see Kovasznay \textit{et al.} [\onlinecite{kovasznay1970large}], Brown \& Thomas [\onlinecite{brown1977large}], Hutchins \& Marusic [\onlinecite{hutchins2007evidence}], Dennis \& Nickels [\onlinecite{ dennis2011a}] and Lee \& Sung [\onlinecite{lee2011very}])
generally refer to elongated regions of streamwise velocity
fluctuations having a streamwise extent from about $2\delta$ to
$3\delta$ for LSMs and larger than $3\delta$ for VLSMs
(see Kim \& Adrian [\onlinecite{kim1999very}], Guala \textit{et al.} [\onlinecite{guala2006large}] and Balakumar \& Adrian [\onlinecite{balakumar2007large}]). The LSMs near
the wall and the VLSMs have been interpreted as being responsible for the $k_{x}^{-1}$ scaling range of the
turbulence spectrum (Smits \textit{et al.} [\onlinecite{smits2011high}]). The range of scales we
concentrate on, in figures \ref{fig:figure9} to \ref{fig:figure12}, just about includes some LSMs at
its upper range.

\begin{figure}
\centering

    \begin{subfigure}{\textwidth}
       \centerline{\includegraphics[width=0.65\textwidth]{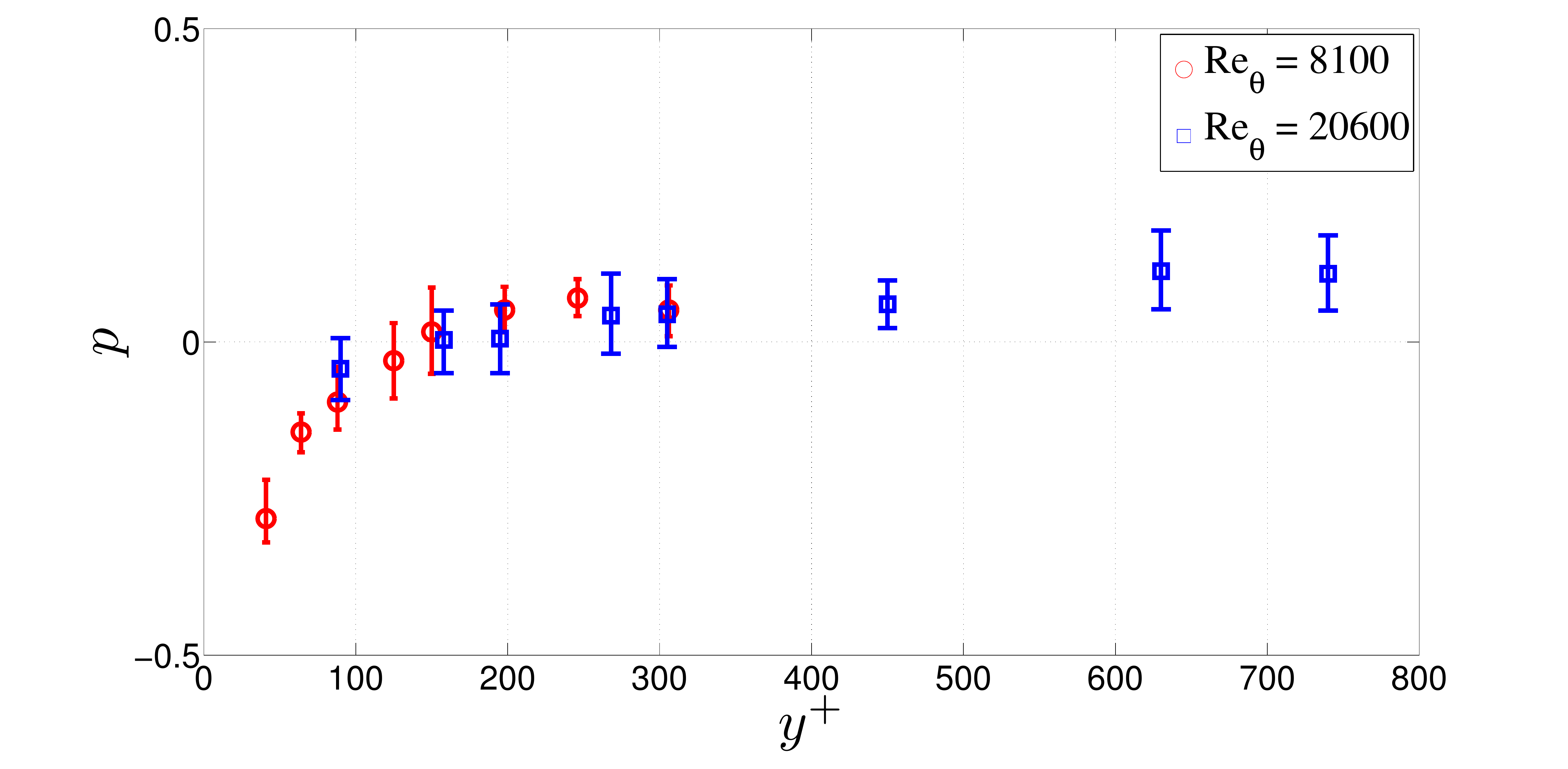}}
    \end{subfigure}  
(a)\\
    \begin{subfigure}{\textwidth}
        \centerline{\includegraphics[width=0.65\textwidth]{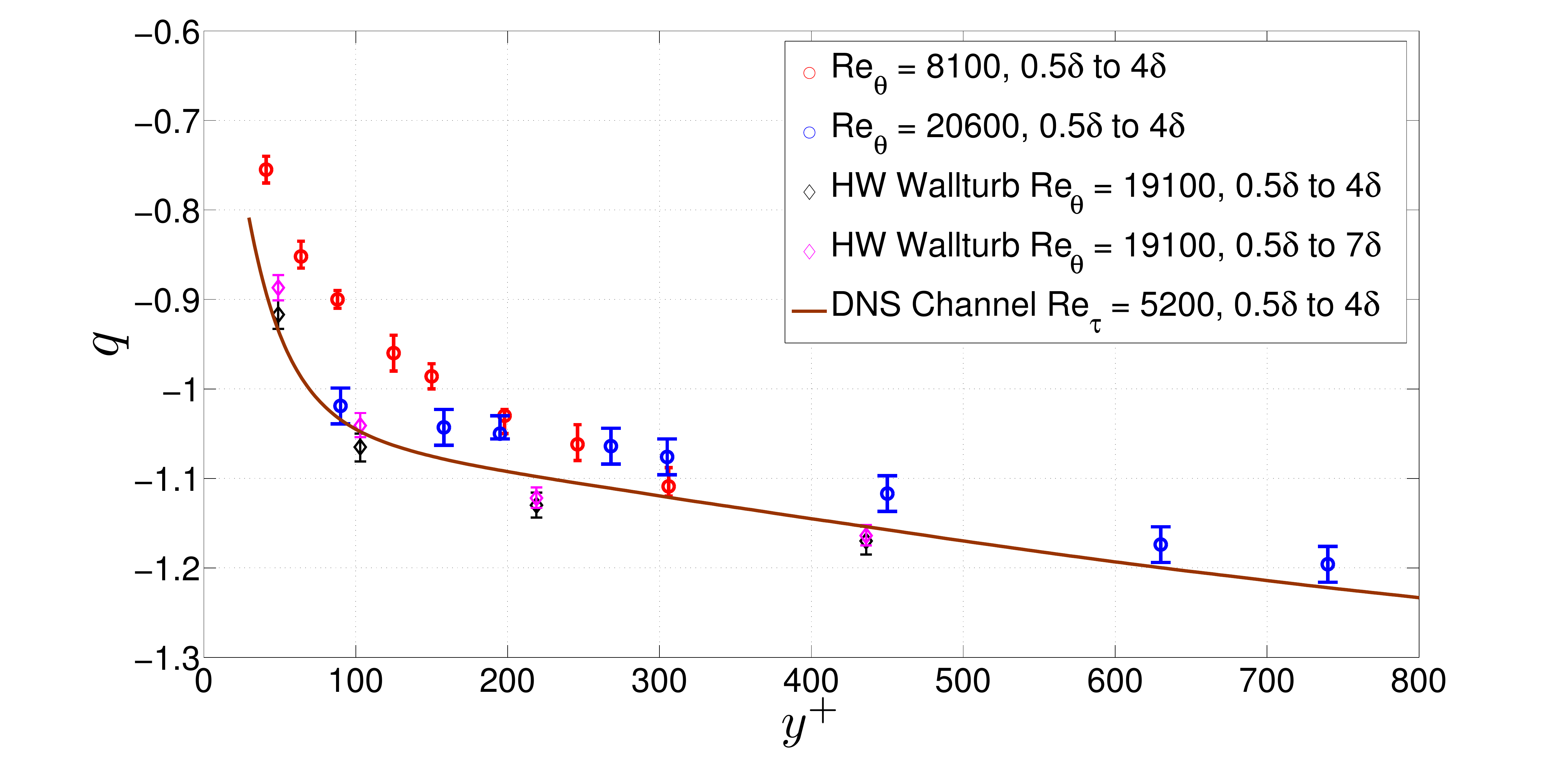}}
        \end{subfigure}
(b)\\   
    \begin{subfigure}{\textwidth}
            \centerline{\includegraphics[width=0.65\textwidth]{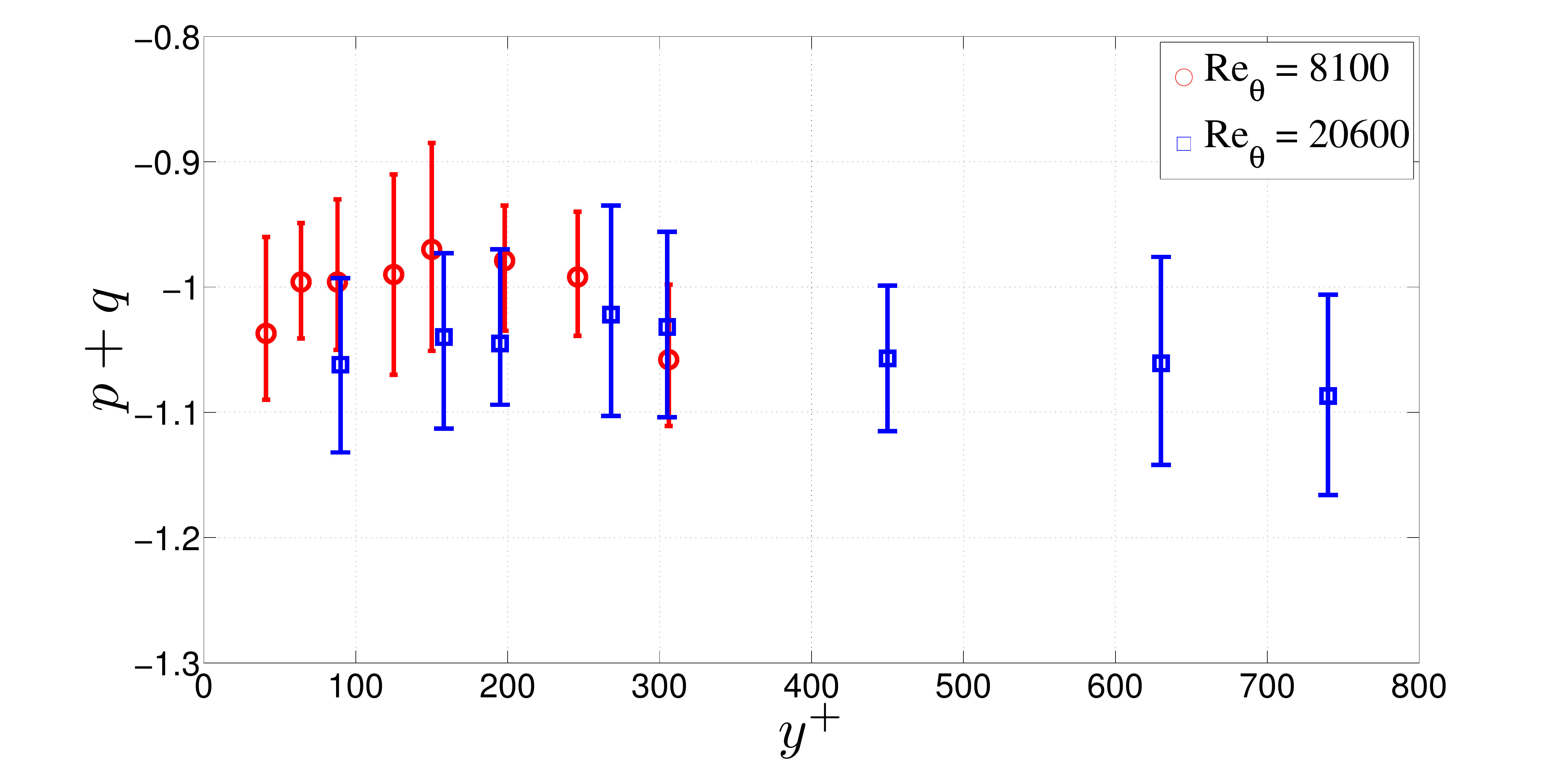}}
            \end{subfigure}
(c)\\   
            \caption{From top to bottom: (a) Exponents $p$ obtained
              from the best power-law fit of $\overline{a^{2}} \sim
              (\lambda/\delta)^{p}$. (b) Exponents $q$ obtained from
              the best power-law fit of $E_{11} \sim k_{x}^{q}$ for
              the present PIV data, the HWA turbulent boundary layer
              data of Tutkun \textit{et al.} [\onlinecite{tutkun09}] (see figure \ref{fig:figureadd}(a) and the DNS of
              turbulent channel flow data of Lee \& Moser [\onlinecite{leemoser15}] (see
              figure \ref{fig:figureadd}(b). (c) $p+ q$ versus $y^+$. All these fits are
              obtained over the range of scales investigated in
              figures \ref{fig:figure9} to \ref{fig:figure12} (except for the HWA case in (b) where we
              have also included a fit over a range of length-scales
              extended up to $7\delta$). The resulting exponents are
              plotted with the $95$\% confidence intervals for these
              fits.}
            \label{fig:figure13}
\end{figure}

\begin{figure}
	   \centerline{\includegraphics[width=0.8\textwidth]{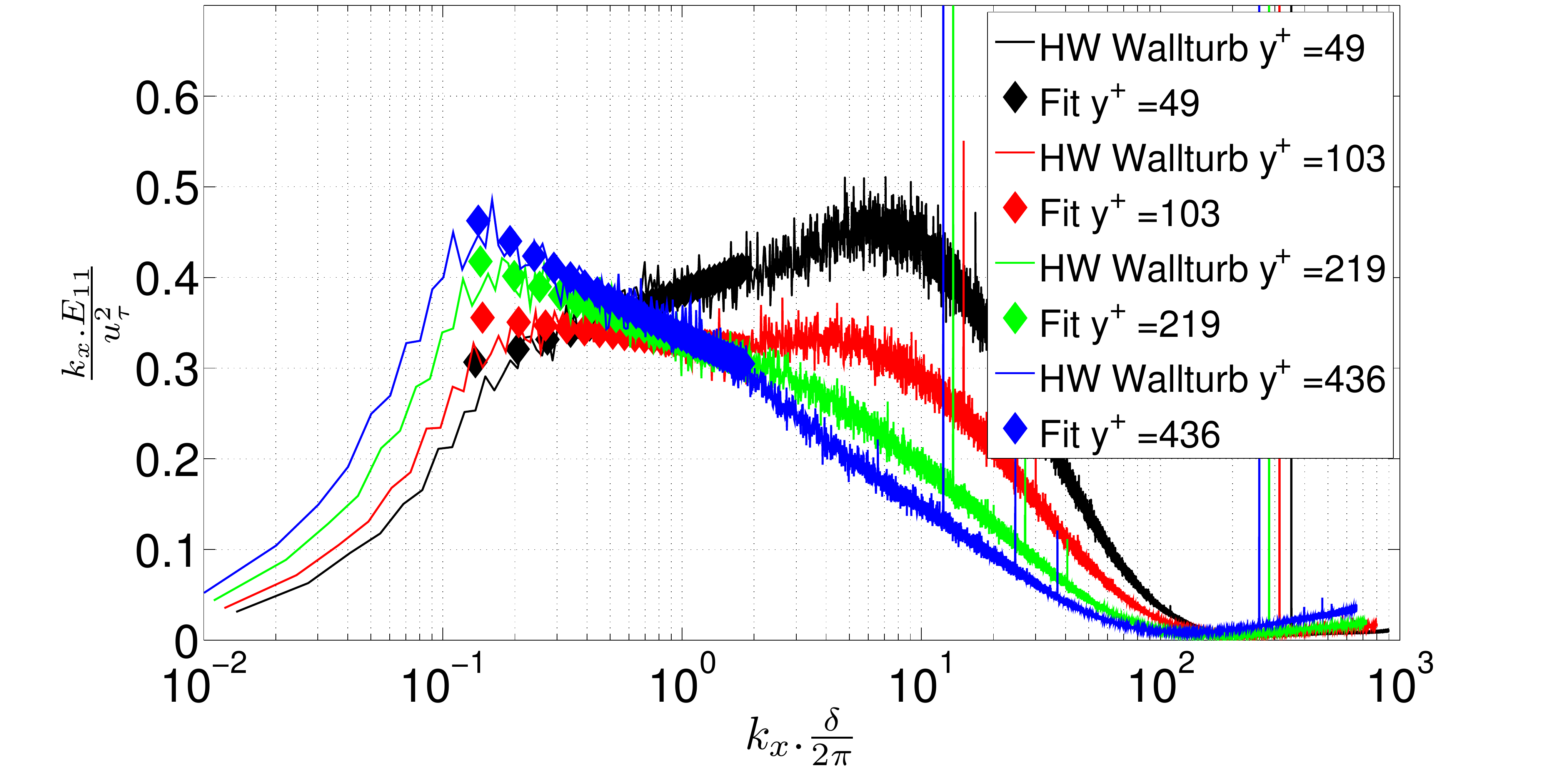}}
	   \centerline{\includegraphics[width=0.8\textwidth]{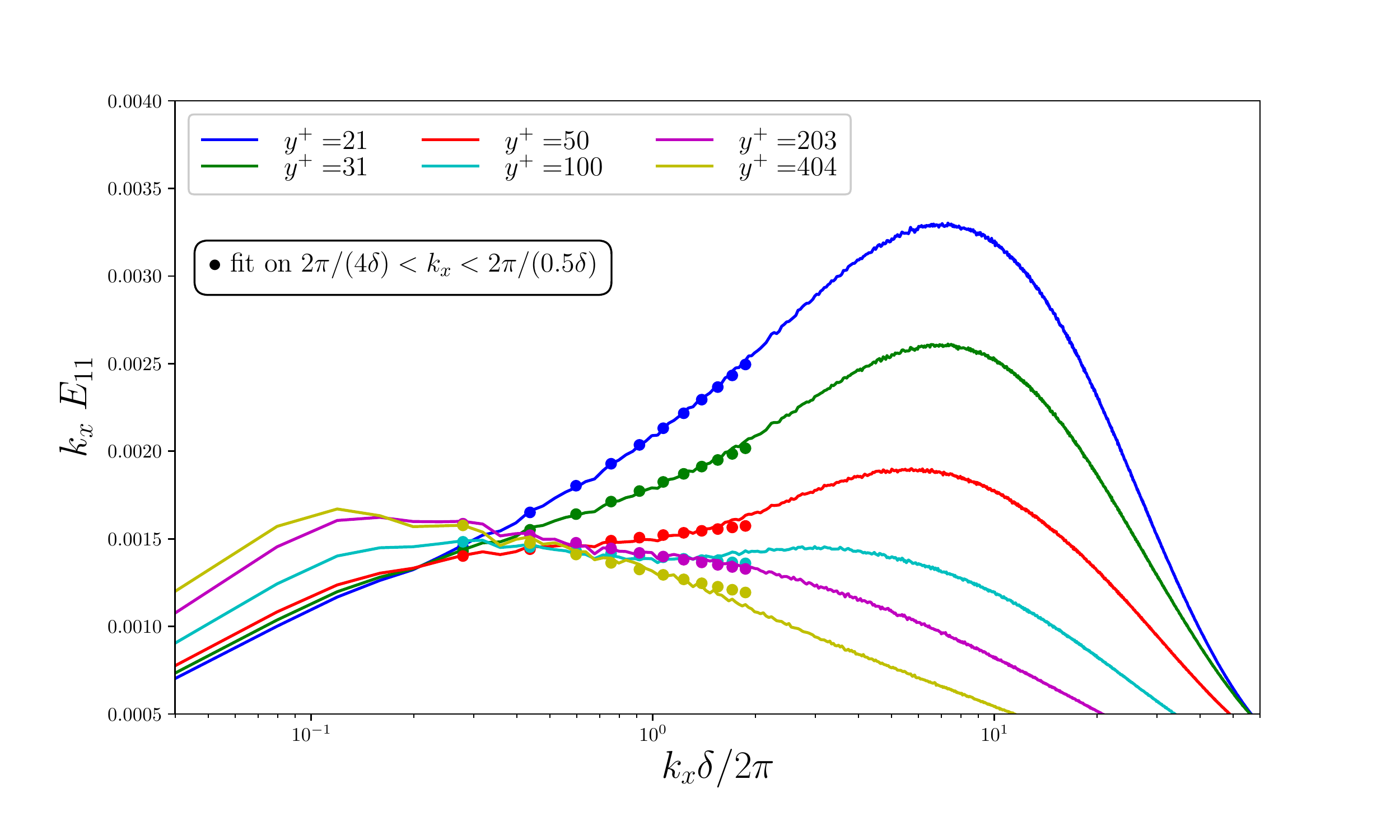}}
			\caption{From top to bottom: (a) Lin-log plots of pre-multiplied
                          streamwise energy spectra at selected wall
                          distances obtained with HWA in the
                          $Re_{\theta} = 19100$ turbulent boundary
                          layer of Tutkun \textit{et al.} [\onlinecite{tutkun09}]. Least-square fits
                          in a range bounded from above by
                          $k_{x}\delta /(2\pi) =2$ but extended to
                          wavenumbers as small as $k_{x}\delta/(2\pi)$
                          close to $10^{-1}$. (b) Lin-log plots of
                          pre-multiplied streamwise energy spectra at
                          selected wall distances from the
                          $Re_{\tau}=5200$ turbulent channel flow DNS
                          of Lee \& Moser [\onlinecite{leemoser15}]. Least-square fits in
                          the range $0.25\le k_{x}/(2\pi) \le 2$.}
			\label{fig:figureadd}
\end{figure}

Returning now to figures \ref{fig:figure9} to \ref{fig:figure12}, we have included best fits of power
law curves in the plots of $\overline{a^{2}}$ versus $\lambda/\delta$
and of $E_{11}$ versus $k_x$. These best fits are indicated in the
inserts of each plot and provide an estimation of the exponents $p$
and $q$ in $\overline{a^{2}} \sim (\lambda/\delta)^{p}$ and $E_{11}
(k_{x}) \sim k_{x}^{q}$. Figure \ref{fig:figure13} summarizes the information with
plots of $p$, $q$ and $p+q$ as functions of $y^+$. It is perhaps
remarkable that $p+q$ is very close to $-1$ (see figure \ref{fig:figure13}) as
predicted by (\ref{eq10})-(\ref{eq10bis}) for all examined values of $y^+$ and for both
Reynolds numbers $Re_{\theta}$.  Whereas this subsection's initial
interpretation in terms of the Townsend attached eddy model is limited
to $y^+$ larger or equal to $88$ (based on the log-log plots of figure
\ref{fig:figure7}), the lin-lin plots of figures \ref{fig:figure9} to \ref{fig:figure12} present a different and
consistent picture which covers both Reynolds numbers and all our
$y^+$ positions, including $y^+$ smaller than $88$. This picture is
confirmed by Hot Wire Anemometry (HWA) data from a turbulent boundary
layer in the same wind tunnel by Tutkun \textit{et al.} [\onlinecite{tutkun09}] and from the recent
Direct Numerical Simulations (DNS) of a turbulent channel flow at
$Re_{\tau}=5200$ by Lee \& Moser [\onlinecite{leemoser15}]. Indeed, these HWA and DNS data
show the same variation of the spectral exponent $q$ with $y^{+}$ that
we found from using (\ref{eq10})-(\ref{eq10bis}) on our PIV data; see figure \ref{fig:figureadd}, and
also figure \ref{fig:figure13}(b) where we collected the values of $q$ from different
data. The HWA data, in particular, provide a confirmation of our PIV
results because they extend to a wider range at the lower end of
wavenumbers (see figures \ref{fig:figureadd}(a) and also figure \ref{fig:figure13}(b) where it is shown
that the HWA's extended wavenumber range returns effectively same
values of $q$).

The much higher Reynolds number measurements of Vallikivi \textit{et al.} [\onlinecite{vallikivi2015spectral}] did not find support for the Townsend-Perry $k_x^{-1}$ spectrum either. However, these authors did find some agreement with the $k_{x}^{-1}\log(8\pi/k_{x}y)$ spectrum model of del \'Alamo \textit{et al.} [\onlinecite{alamo04}]. This approximate agreement was found in a range of wall-normal distances where we find positive values of $p$, i.e. in a region where the spectrum scales as $k_{x}^{-1} k_{x}^{-p}$ with values of $p$ above but close to $0$. It is quite difficult to distinguish between such a weak power law and $\log(8\pi/k_{x}y)$, so the two models qualitatively agree in this range of wall-normal distances. However, the model of del \'Alamo \textit{et al.} [\onlinecite{alamo04}] cannot account for
the scaling of the energy spectrum at closer distances to the wall where we find $p\le 0$, whereas our model fits the data in this region too.

\subsection{Discussion} \label{sec:discussion}

It is important to stress that the support for (\ref{eq10})-(\ref{eq10bis}) in figures
\ref{fig:figure9} to \ref{fig:figure12} cannot be obtained without the crucial last step of our
structure detection algorithm in subsection \ref{sec:Structure detection} which discards
structures that are not attached to the wall. The structures which do
not touch the wall are in fact less elongated and less intense
(i.e. smaller $\overline{a^{2}}$) on average. We have checked that if
we only consider them, we do not find anything close to $p+q=-1$,
i.e. (\ref{eq10bis}).

The attached eddy concept introduced by Townsend [\onlinecite{townsend1976structure}]
is therefore important for explaining $E_{11}(k_{x})$ but
the results of our analysis suggest that the Townsend-Perry model does
not hold \textcolor{black}{without some significant corrections}
because the turbulent kinetic energy content in these wall-attached
flow structures does not just scale with $U_{\tau}$. (If it did,
$\overline{a^{2}}$ would scale with $U_{\tau}$ and $p$ would be
uniformly 0.) At different $y$ inside such a structure, the level of
turbulent kinetic energy depends both on $U_{\tau}$ and on the
streamwise length of the structure at that height. Furthermore, this
dependence varies with height: $\overline{a^{2}}$ decreases with
increasing $\lambda/\delta$ very close to the wall, in the buffer
layer, and increases with increasing $\lambda$ further up. As
$\overline{a^{2}}$ transits smoothly from one dependence to the other,
a particular height exists where $\overline{a^{2}}$ is independent of
$\lambda$ and therefore depends only on $U_{\tau}$. At that very
particular height, $E_{11}(k_{x}) \sim k_{x}^{-1}$. However,
\textcolor{black}{strictly speaking,} this is not a Townsend-Perry
spectrum, it is just the spectrum at that particular distance from the
wall where the turbulent kinetic energy inside the streaky structures
transits from a decreasing to an increasing dependence on the length
of these structures. Our conclusion agrees with
Nickels \textit{et al.} [\onlinecite{nickels2007some}] in their statement that it is necessary to
take measurements close to the wall to observe a $k_x^{-1}$ behaviour,
in fact at $y^+$ between $100$ and $200$ as they also found. However,
these authors were not in possession of (\ref{eq10})-(\ref{eq10bis}) and therefore did
not measure $\overline{a^{2}}$ at various heights and for various
values of $\lambda/\delta$ which now allows us to see that the
$k_x^{-1}$ behaviour at the edge of the buffer layer is not the
Townsend-Perry spectrum but just a transitional instance of a more
involved spectral structure. In fact, the spectral picture which
emerges from our analysis is a unified picture which brings together
the buffer and inertial layers in a seemless way.

\begin{figure}
\centering

    \begin{subfigure}{\textwidth}
      \centerline{\includegraphics[width=0.75\textwidth]
{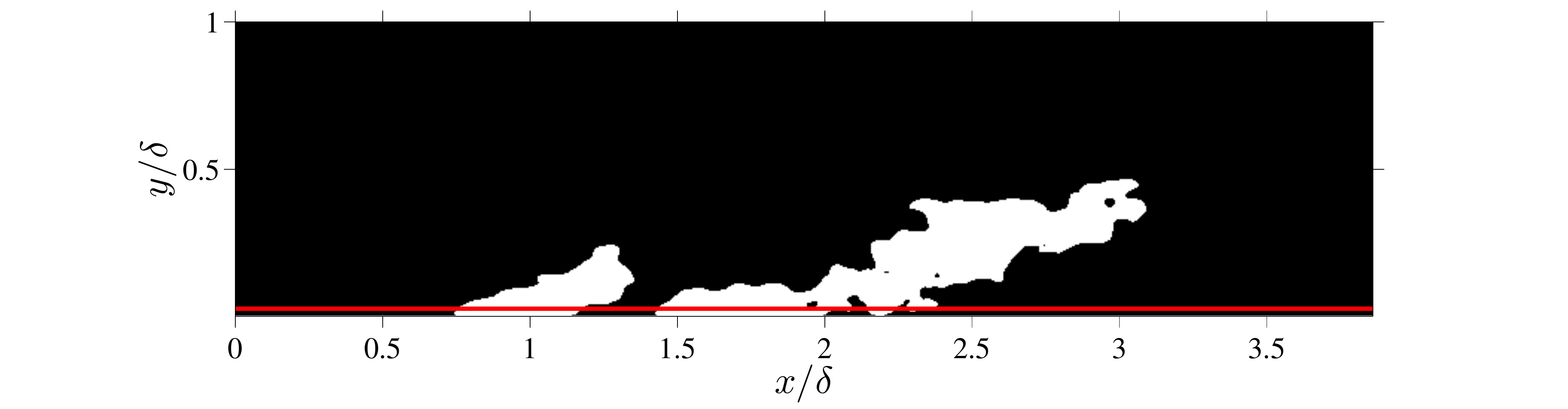}}

    \end{subfigure}  
    \quad
    \begin{subfigure}{\textwidth}
       \centerline{\includegraphics[width=0.75\textwidth]
{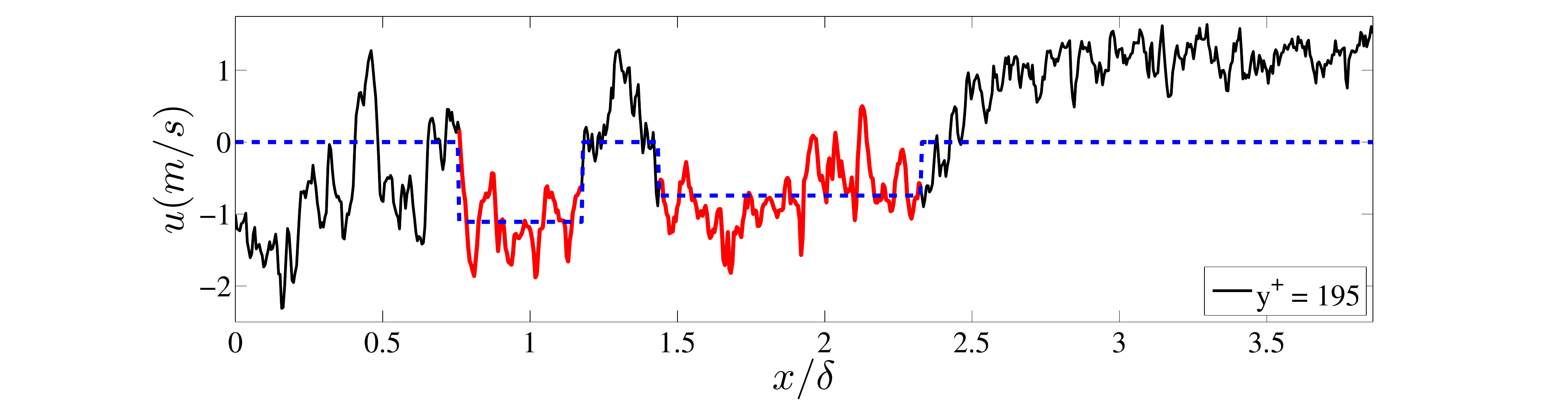}}

        \end{subfigure}
       
    \begin{subfigure}{\textwidth}
       \centerline{\includegraphics[width=0.75\textwidth]
{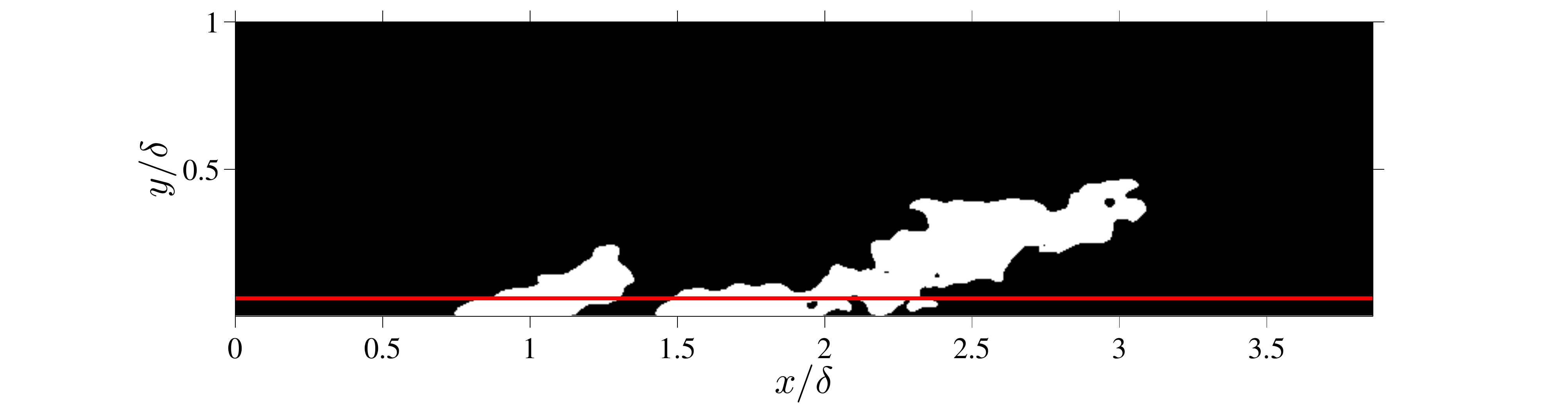}}

        \end{subfigure}
       
    \begin{subfigure}{\textwidth}
       \centerline{\includegraphics[width=0.75\textwidth]
{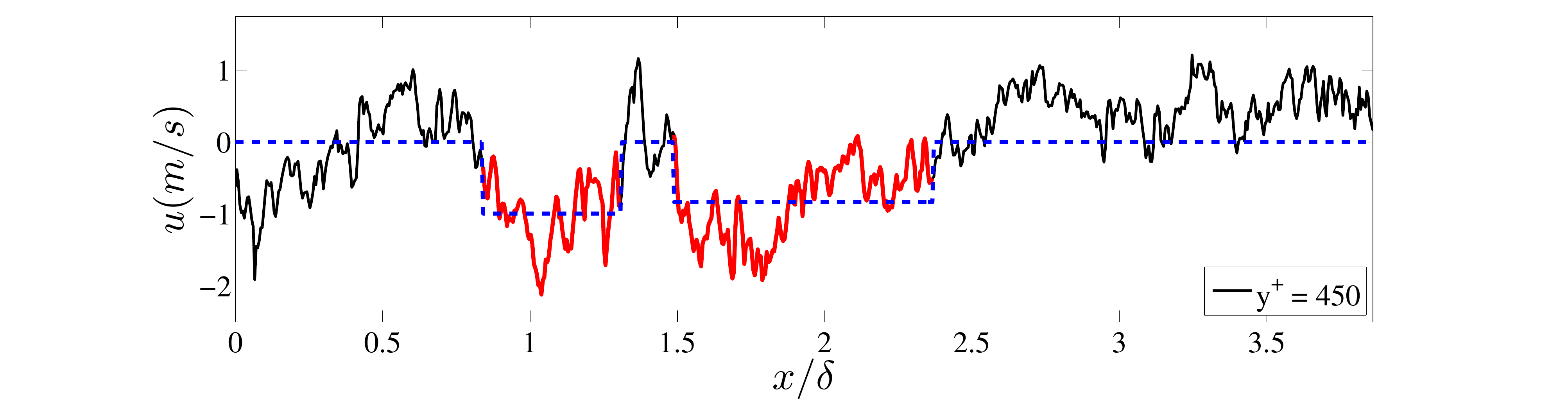}}

        \end{subfigure}
       
    \begin{subfigure}{\textwidth}
           \centerline{\includegraphics[width=0.75\textwidth]
{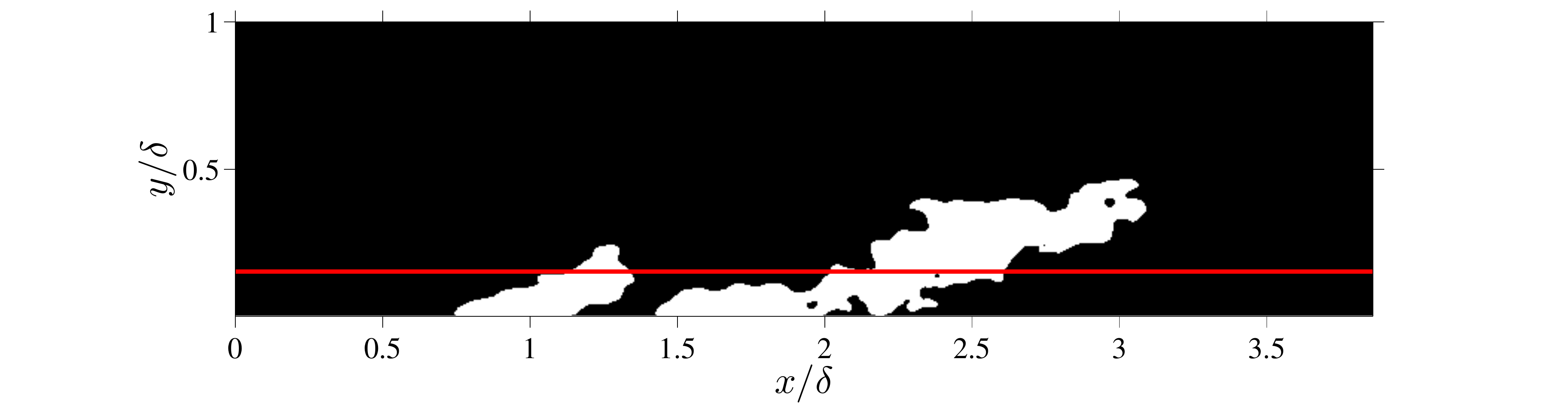}}

            \end{subfigure}

     \begin{subfigure}{\textwidth}
           \centerline{\includegraphics[width=0.75\textwidth]
{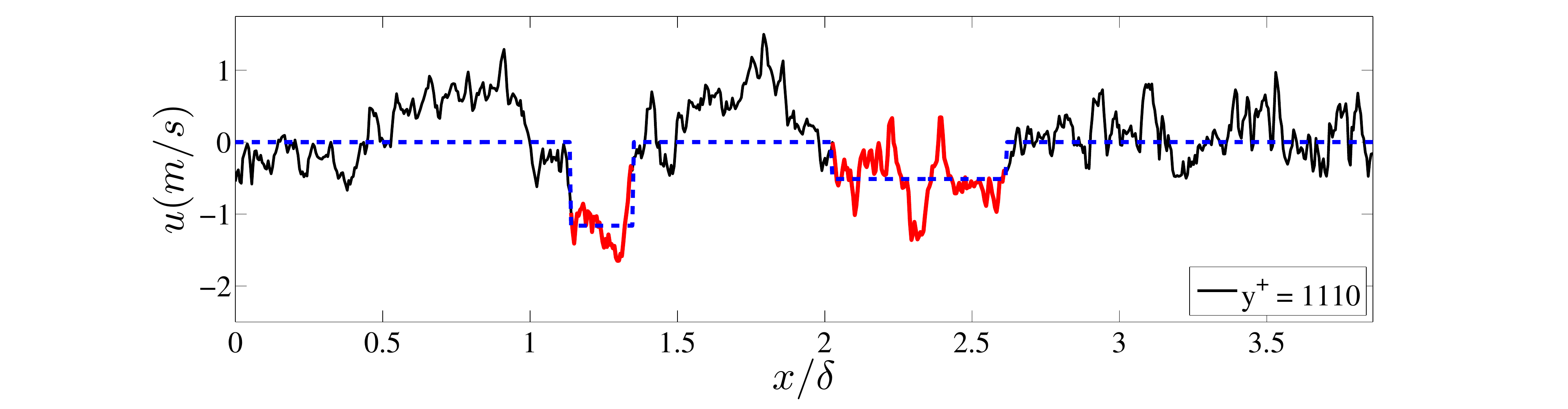}}

      \end{subfigure}
        \caption{An example of a detected wall-attached flow structure
          for $Re_{\theta}= 20600$ and the $u(x)$ signal through this
          structure at three different $y^+$ positions. The red line
          in the repeated binary image indicates the $y^+$ position
          where the signal $u(x)$ is recorded ($y^{+} = 195, 450,
          1110$). The black/red line in the $u(x)$ versus $x/\delta$
          plots is the raw (un-filtered) PIV fluctuating streamwise
          fluctuating velocity outside/inside the detected flow
          structures. The dashed blue line is our model signal, equal
          to 0 outside and to the average value of $u$ inside the
          detected structures.}
        \label{fig:figure15}
\end{figure}

In figure \ref{fig:figure15} we plot examples of measured streamwise velocity
fluctuations and the on-off signals with which we model them at
various heights from the wall. Our model on-off signals are clearly a
drastic simplification of the data but one gets the impression from
these plots that they capture the sharpest gradients in the signal and
therefore much of its spectral content at the length-scales considered
here. The lengths of the non-zero parts of the model signals
correspond to $\lambda$ and the actual values of the on-off signal in
these non-zero parts correspond to the average value $\alpha$ of the
streamwise fluctuating velocity component inside each
part. \textcolor{black}{We stress that it is enough that our on-off
  model agrees with the data in the way it does in figures \ref{fig:figure9} to \ref{fig:figure12} for a
  certain range of thresholds $u_{th}$ (see subsection \ref{sec:Structure detection}). Our model
  does not need to work for any arbitrary threshold; it only needs to
  work for those thresholds which effectively capture the spatial
  boundaries of the flow structure objects simulated by our on-off
  functions as mentioned at the end of the first paragraph of section
  \ref{sec: simplest possible model}.}

It is clear that a wider range of Reynolds numbers needs to be
examined to establish the scalings of the lower and upper bounds of
the range of wavenumbers where (\ref{eq10})-(\ref{eq10bis}) holds. One might expect the
upper bound to scale as $1/y$ because of the recent evidence
(see Hultmark \textit{et al.} [\onlinecite{hultmark2012turbulent}] and Laval \textit{et al.} [\onlinecite{laval2017comparison}]) that a Townsend-like
approximately logarithmic (or very weak power law) dependence on $y$
exists for the rms streamwise turbulent velocity in the outer part of
the inertial range of wall-distances. If one assumes the lower bound
to scale as $1/\delta$ and therefore an energy spectrum
$E_{11}(k_{x})$ of the form (i) $E_{11}(k_{x}) \sim
U_{\tau}^{2}\delta$ for $0\le k_{x}\le B_{1}/\delta$, (ii)
$E_{11}(k_{x}) \sim U_{\tau}^{2}\delta (k_{x}\delta)^{-1-p}$ for
$B_{1}/\delta \le k_{x} \le B_{2}/y$ (where $B_{1}$ and $B_{2}$ are
dimensionless constants and $p$ may be a function of $y$ as in figure
\ref{fig:figure13}(a) and (iii) comparatively negligible energy at wavenumbers $k_{x} >
B_{2}/y$, then we should have
\begin{equation}
(u'^{+})^{2} \sim 1 + {B_{1}^{p}\over p}(B_{1}^{-p} -
  (B_{2}\delta/y)^{-p}).
  \label{eq5.1}
\end{equation}
This expression for $(u'^{+})^{2}$ tends to
\begin{equation}
(u'^{+})^{2} \sim 1 + (\ln B_{1} - \ln (B_{2}\delta/y))
  \label{eq5.2}
\end{equation}
as $p\to0$ which is the Townsend logarithmic dependence on $y$
corresponding to $p\equiv 0$ (see Townsend [\onlinecite{townsend1976structure}], Perry \& Abell [\onlinecite{perry1977asymptotic}] and Perry \textit{et al} [\onlinecite{perry1986theoretical}])
. This
logarithmic dependence (\ref{eq5.2}) results from the assumption that
the upper bound of the range of wavenumbers where (\ref{eq10})-(\ref{eq10bis}) may hold
with $p\equiv 0$ scales as $1/y$. Slightly non-zero values of $p$ give
slight deviations from this logarithmic dependence, of the form
(\ref{eq5.1}).

Using the values of $p$ obtained in this work and plotted versus $y^+$
in figure \ref{fig:figure13} for our two values of $Re_{\theta}$, it is not possible
to fit (\ref{eq5.1}) to the data in the lower plot of figure \ref{fig:figure1} from
$y^{+}=41$ to $306$ in the $Re_{\theta} = 8100$ case and from
$y^{+}=90$ to $742$ in the $Re_{\theta} = 20600$ case. These are the
$y^+$ ranges where (\ref{eq10})-(\ref{eq10bis}) has been established for our data and
they should therefore also be the ranges where (\ref{eq5.1}) holds if
the spectral model of the previous paragraph is good enough. However,
in spite of the three adjustable dimensionless constants ($B_1$, $B_2$
and an overall constant of proportionality), (\ref{eq5.1}) cannot fit
the entire $y^+$ range for which this model has been designed, that is
a $y^+$ range which includes both the $p<0$ and the $p>0$ regions.

A most suspect part of the spectral model used to derive (\ref{eq5.1})
is its low wavenumber part.  Vassilicos \textit{et al.} [\onlinecite{vassilicos2015streamwise}] showed
that the second peak or plateau part of the $u'^{+}$ profile can be
reproduced by a spectral range between the very low wavenumber range
where $E_{11}(k_{x}) \sim U_{\tau}^{2}\delta$ and the wavenumber range
where $E_{11}(k_{x}) \sim U_{\tau}^{2}\delta (k_{x}\delta)^{-1-p}$. In
fact, Vassilicos \textit{et al.} [\onlinecite{vassilicos2015streamwise}] also showed that this extra
intermediate spectrum is necessary for a sufficiently fast growth of
the integral scale with distance from the wall. A complete model of
$E_{11}(k_{x})$ would therefore require the spectral range introduced
by Vassilicos \textit{et al.} [\onlinecite{vassilicos2015streamwise}] as well as the spectral range
studied here.

\section{Conclusion} \label{Conclusion}

We obtained well-resolved PIV data of a flat plate turbulent boundary
layer in a large field of view at two Reynolds numbers, $Re_{\theta} =
8100$ and $Re_{\theta} = 20600$. A direct inspection of log-log plots
of the streamwise energy spectrum would suggest $E_{11}(k_{x}) \sim
U_{\tau}^{2} k_{x}^{-1}$ in the range $2\pi/(4\delta) < k_{x} <
0.63/y$. However, a closer look assisted by relation (\ref{eq10})-(\ref{eq10bis})
reveals a significantly \textcolor{black}{subtler} behaviour. This
relation introduces a specific data analysis which involves the
extraction of wall-attached elongated streaky structures from PIV
data. The concurrent analysis of streamwise energy spectra and of the
relation between the turbulence levels inside streaky structures and
the length of these structures offers strong support for (\ref{eq10})-(\ref{eq10bis})
over a significant range of wavenumbers and length-scales. This range
covers LSMs and is comparable to the range where one might have
expected the Townsend-Perry attached eddy model spectra to be
present. Even though \textcolor{black}{$k_{x}^{-1}$ spectra are not,
  strictly speaking,} validated by our data, the streaky structures
which account for the scalings of $E_{11}(k_{x})$ do need to be
wall-attached for relation (\ref{eq10})-(\ref{eq10bis}) to hold.  Our conclusions agree
with the experiments of Vallikivi \textit{et al.} [\onlinecite{vallikivi2015spectral}] which actually
suggest that the Townsend-Perry $k_x^{-1}$ spectrum cannot be expected
even at very high Reynolds numbers. \textcolor{black}{The revised
  Townsend-Perry streamwise energy spectral form (\ref{eq10})-(\ref{eq10bis}) with
  $p=p(y^{+})$ given by figure \ref{fig:figure13}(a) appears to extend the validity of
  the attached eddy concept and its revised consequences to a wider
  range of Reynolds numbers and a wider range of wall distances.}

Finally, we stress that relation (\ref{eq10})-(\ref{eq10bis}) is predicated on these
wall-attached streaky structures being space-filling, i.e.  $D=1$ in
the notation of section \ref{sec: simplest possible model}. The pdf of the streamwise length of the
educed streaky structures does indeed follow a power law with exponent
$-1-D=-2$ over the range of scales which corresponds to the one where
(\ref{eq10})-(\ref{eq10bis}) holds.

Our work has shed some new light on the streamwise turbulence spectra
of wall turbulence by revealing that some of the inner structure of
wall-attached eddies is reflected in the scalings of these spectra via
$p(y^{+})$. An important implication of this structure is that the
friction velocity is not sufficient to scale the spectra. Future work
must now further probe the inner structure of wall-attached eddies,
attempt to explain it and extend our analysis to higher Reynolds
numbers so as to establish with certainty the ranges of the power laws
(exponents $p$ and $q$ in (\ref{eq10})-(\ref{eq10bis})) discussed in this paper. When
this will be done, a complete picture of streamwise energy spectra
will also need to integrate the spectral model of
Vassilicos \textit{et al.} [\onlinecite{vassilicos2015streamwise}].

\subsection*{Acknowledgements}
The work was carried out within the framework of the CNRS Research
Foundation on Ground Transport and Mobility, in articulation with the
ELSAT2020 project supported by the European Community, the French
Ministry of Higher Education and Research, the Hauts de France
Regional Council. The authors gratefully acknowledge the support of
these institutions. JCV also acknowledges the support of ERC Advanced
Grant 320560.

\appendix

\section{\textcolor{black}{Effects of threshold levels and sign}}\label{Appendix}

\begin{figure}
			\centerline{\includegraphics[width=0.8\textwidth]{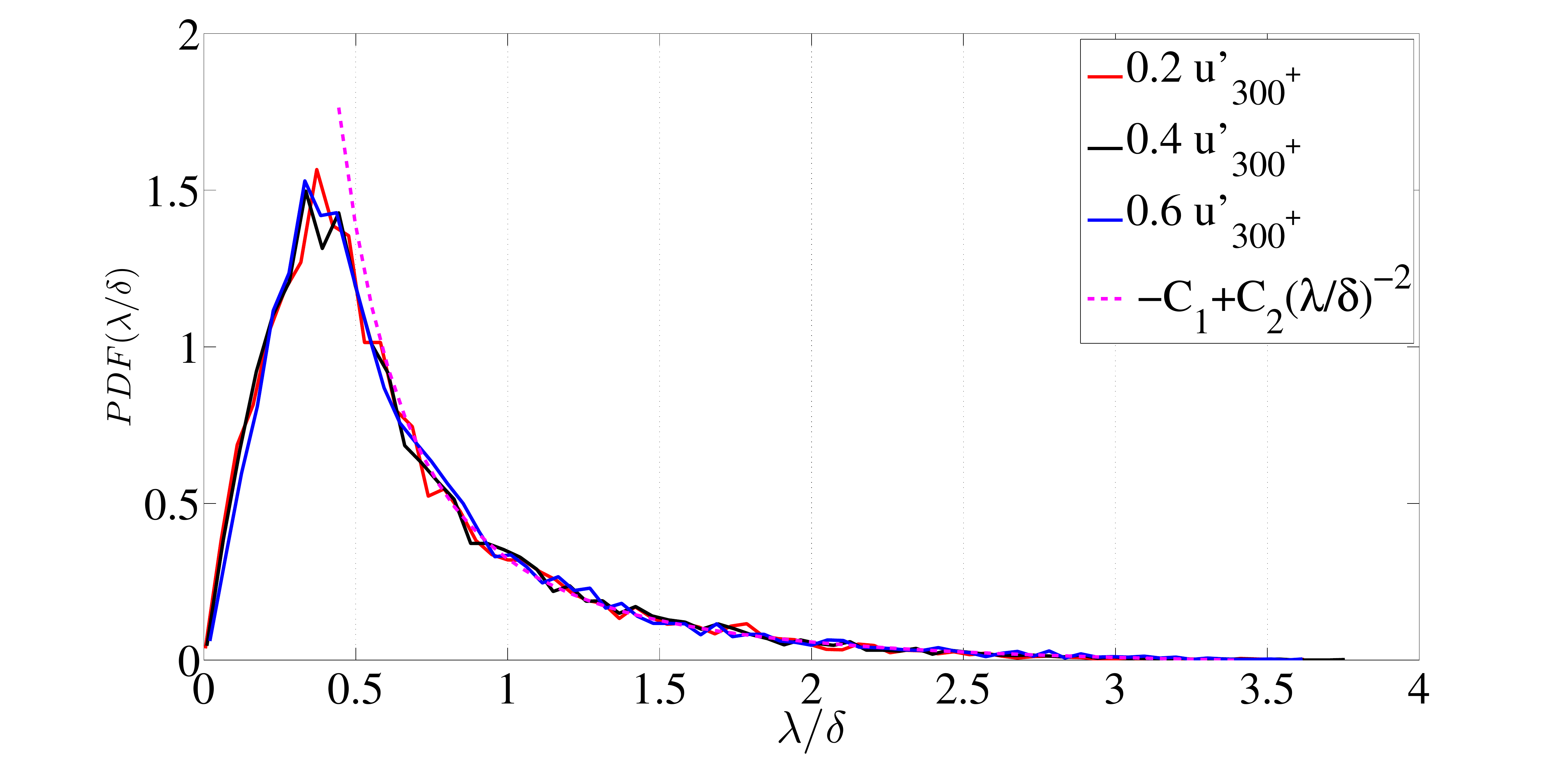}}

			\centerline{\includegraphics[width=0.8\textwidth]{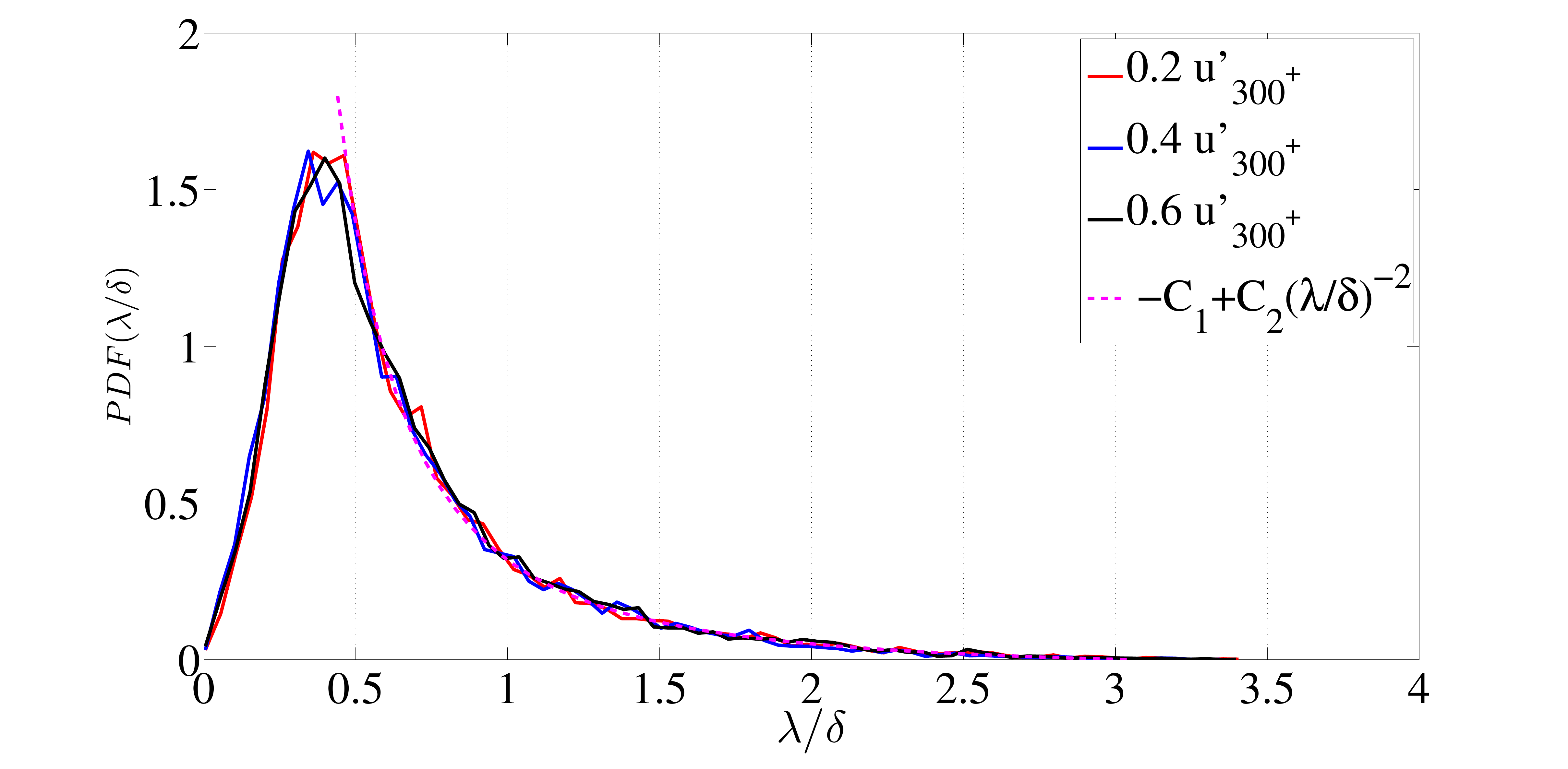}}
			\caption{PDFs of streamwise lengths $\lambda$
                          of wall-attached structures at $y^+ = 195$
                          for $Re_{\theta} = 20600$ (top) and $y^+ =
                          125$ for $Re_{\theta} = 8100$ (bottom) over
                          a set of thresholds}
			\label{fig:figure16}
\end{figure}

\textcolor{black}{Our results have no significant dependence on
  threshold $u_{th}$ in the range $-0.2 u'_{300^+}$ to $-0.6
  u'_{300^+}$. An example of this lack of threshold dependence can be
  seen in the PDFs of $\lambda/\delta$ which we plot in figure \ref{fig:figure16}. We
  also report in table \ref{tab:t2} the number of structures educed by the
  algorithm described in subsection \ref{sec:Structure detection} for the three negative
  threshold values $-0.2 u'_{300^+}$, $-0.4 u'_{300^+}$ and $-0.6
  u'_{300^+}$. Figures \ref{fig:figure9} to \ref{fig:figure13} have been obtained for $u_{th} = -0.4
  u'_{300^+}$ but we checked that they remain very similar without
  deviations from our conclusions if the threshold $u_{th}$ is chosen
  in the range $-0.2 u'_{300^+}$ to $-0.6 u'_{300^+}$.}

\begin{table}
\begin{center}
\def~{\hphantom{0}}
\begin{ruledtabular}
  \begin{tabular}{lcc}
	\multirow{1}{*}{$Re_\theta$} & \multicolumn{1}{c}{20600}
    & \multicolumn{1}{c}{8100} \\ \hline $-0.2 u'_{300^{+}}$ & 13517 &
    17338 \\
    $-0.4 u'_{300^{+}}$ & 14493 & 19576 \\
	$-0.6 u'_{300^{+}}$ & 13366 & 19290 \\
  \end{tabular}
  \caption{Number of structures detected over a set of three negative
    thresholds for $Re_{\theta} = 20600$ and $Re_{\theta} = 8100$ }
    \label{tab:t2}
    \end{ruledtabular}
  \end{center}
\end{table}

\begin{figure}
			\centerline{\includegraphics[width=0.8\textwidth]{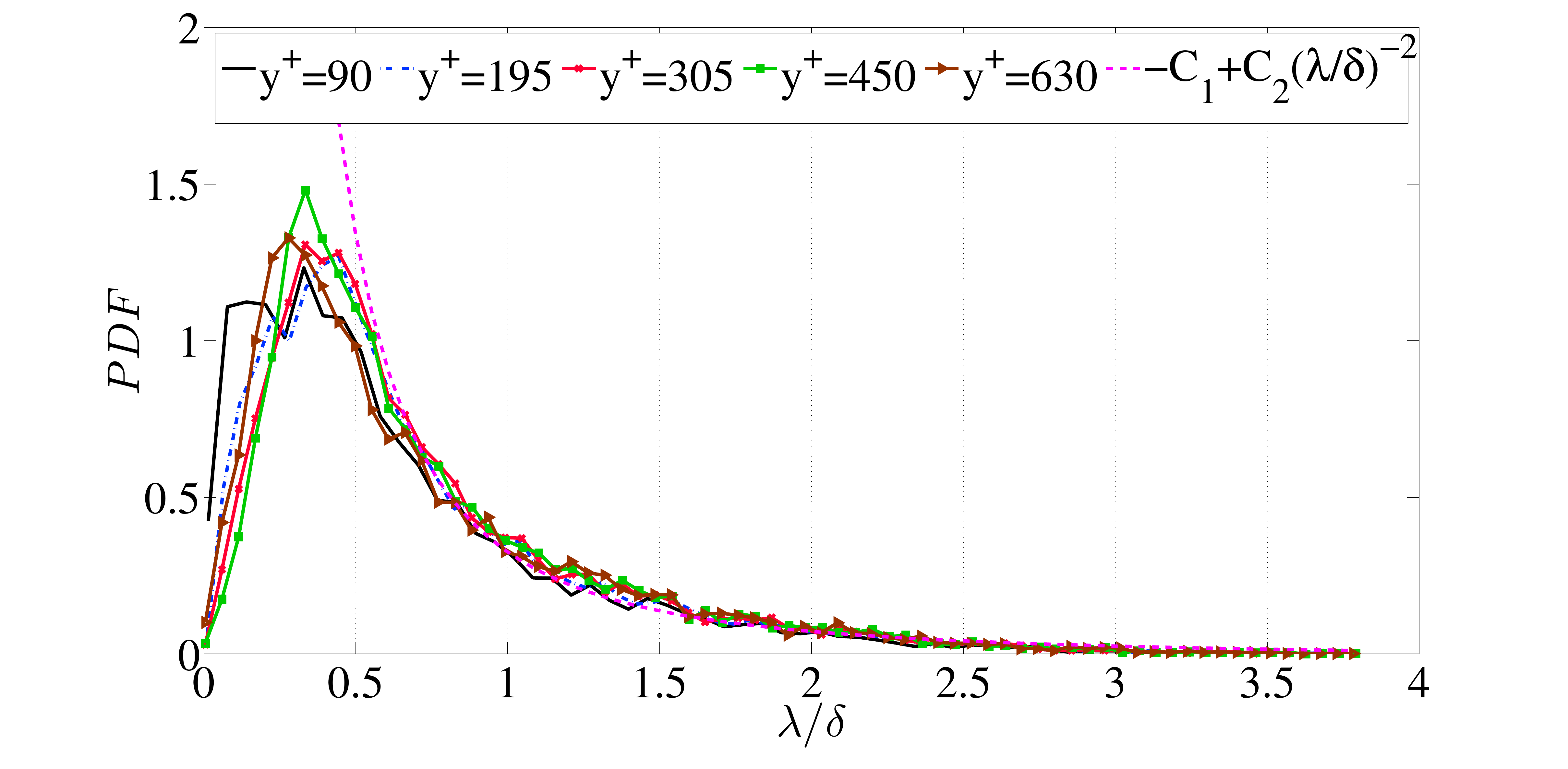}}

			\centerline{\includegraphics[width=0.8\textwidth]{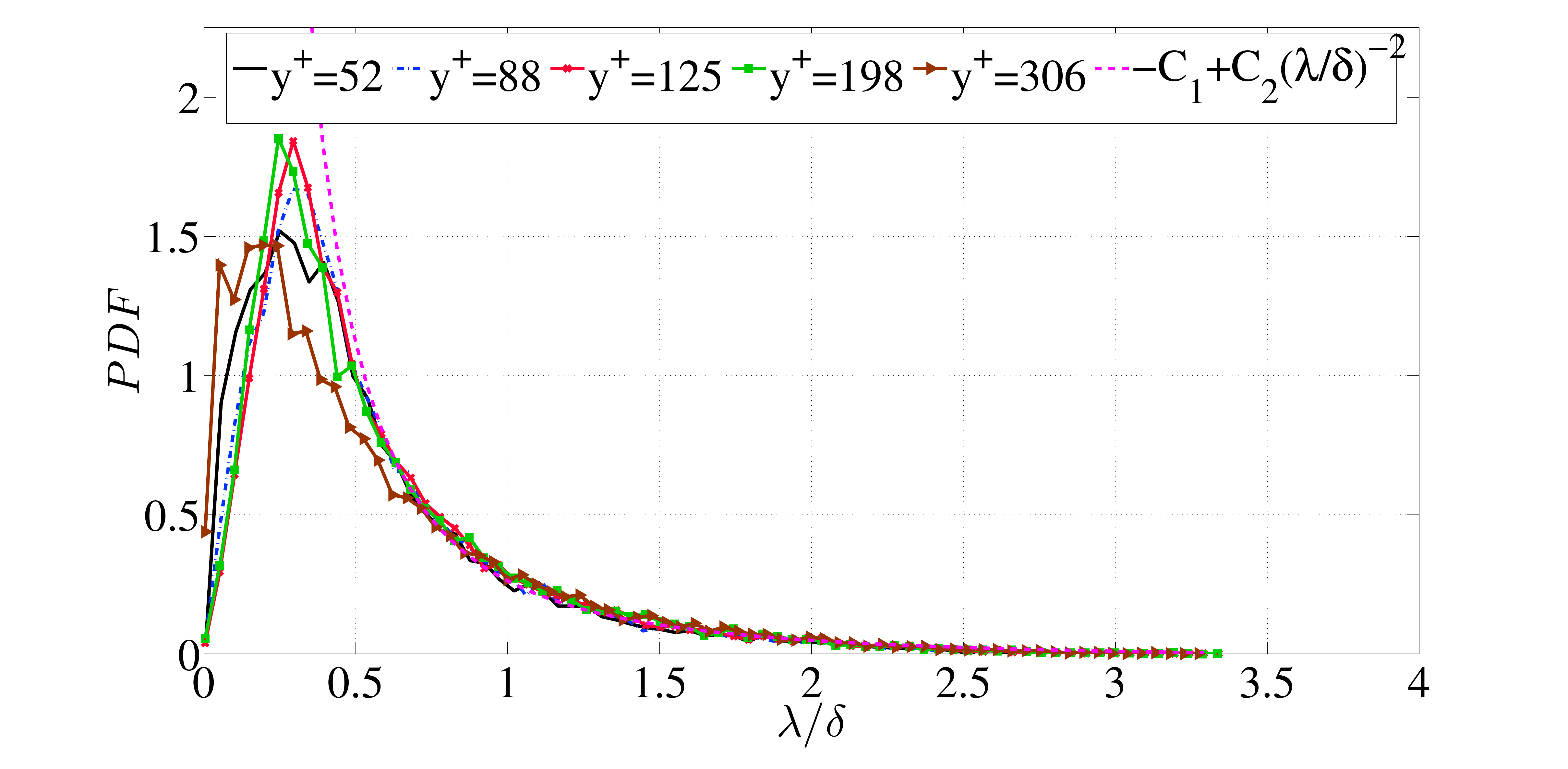}}
			\caption{PDFs of streamwise lengths $\lambda$
                          of wall-attached structures of positive
                          streamwise fluctuating velocity with $u_{th}
                          = u'_{300^{+}}$ at selected wall distances
                          for $Re_{\theta} = 20600$ (top) and
                          $Re_{\theta} = 8100$ (bottom). The fits
                          shown here are for $y^+ = 195$ at
                          $Re_{\theta} = 20600$ and $y^+ = 198$ at
                          $Re_{\theta} = 8100$.}
			\label{fig:figure17}
\end{figure}

\begin{figure}

       \centerline{\includegraphics[width=0.8\textwidth]{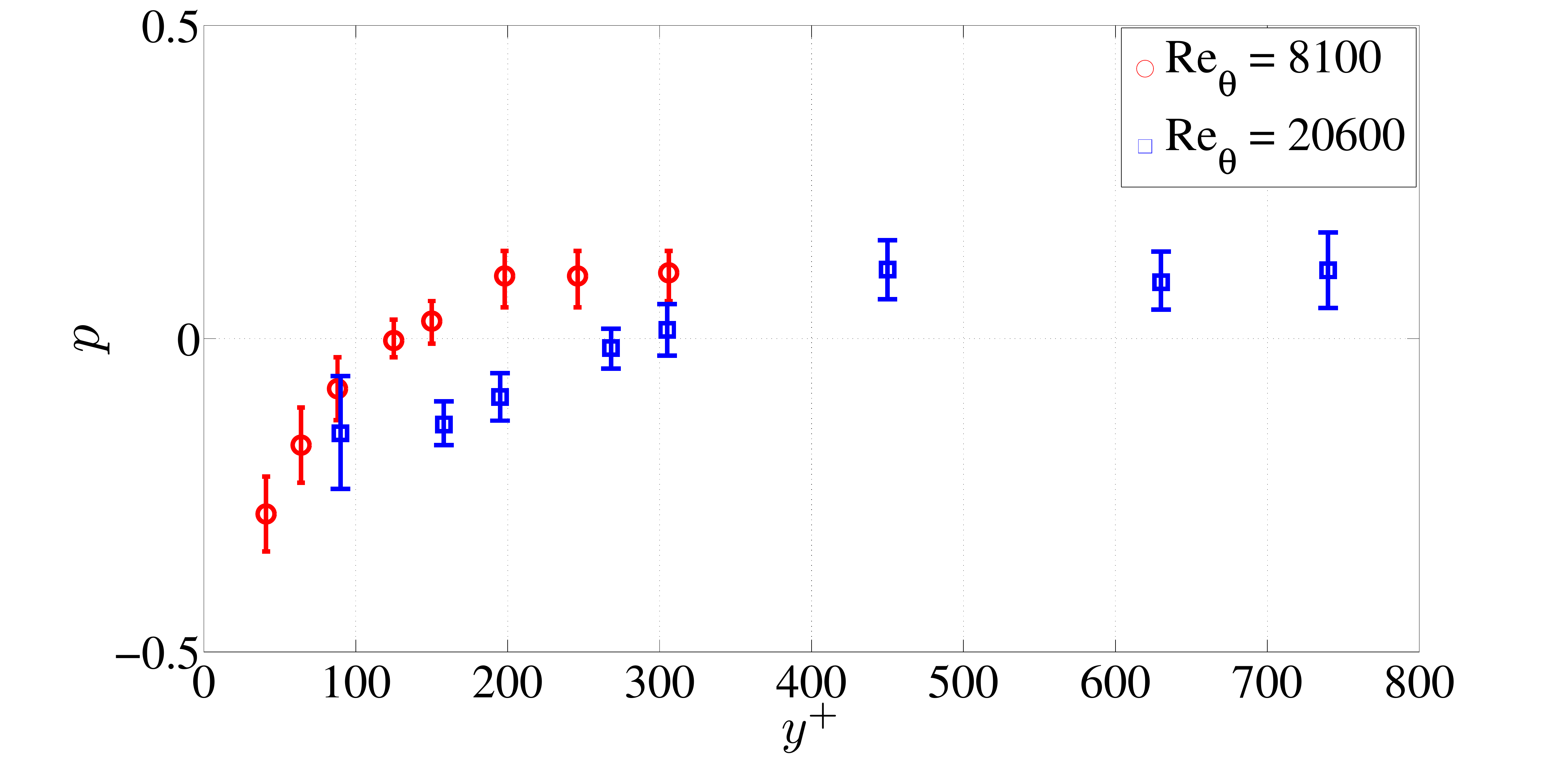}}
            \centerline{\includegraphics[width=0.8\textwidth]{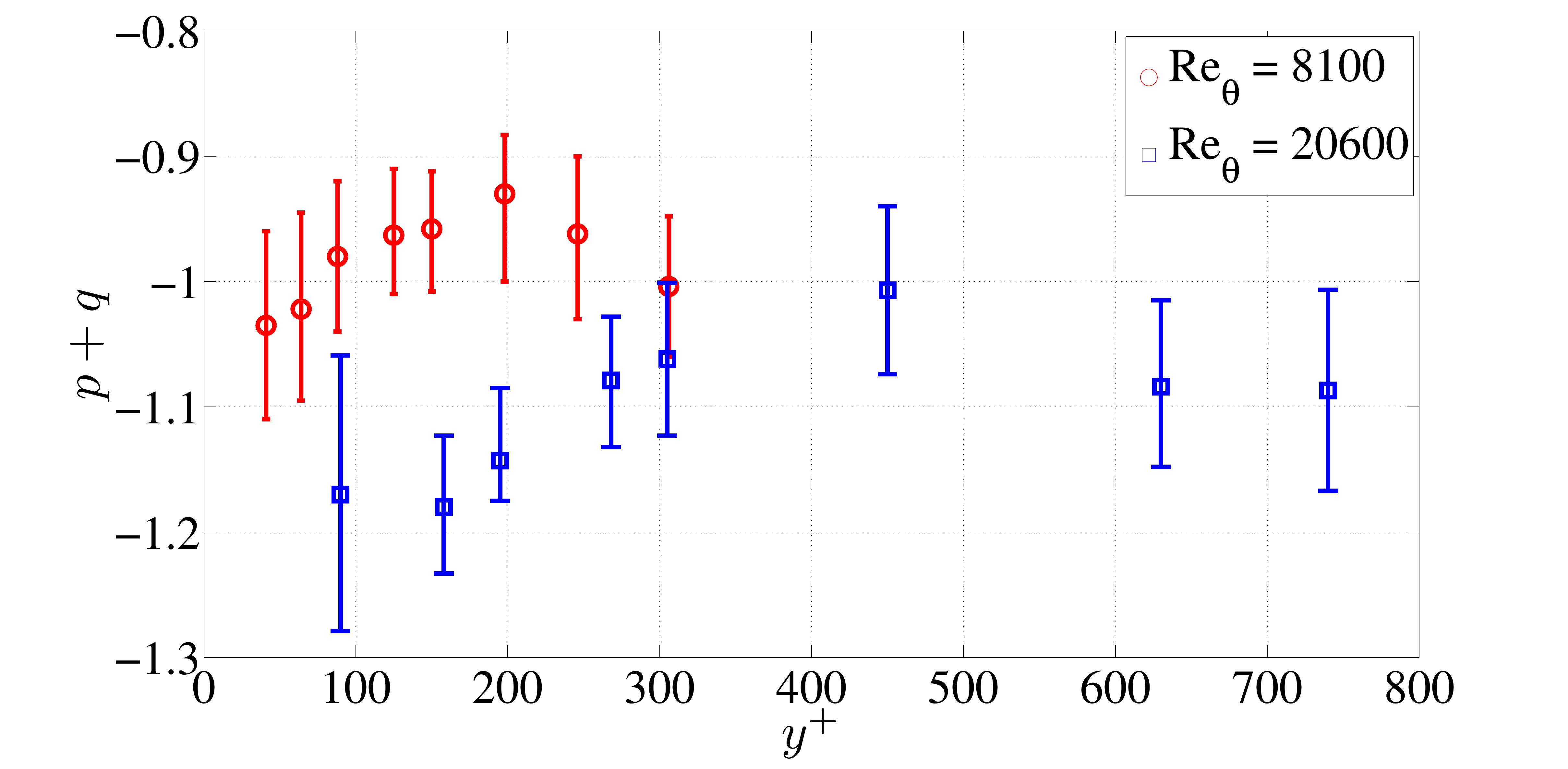}}
   
            \caption{From top to bottom: (a) Exponents $p$ obtained
              from the best power-law fit of $\overline{a^{2}} \sim
              (\lambda/\delta)^{p}$. (b) $p+q$ versus $y^+$. These
              fits are obtained over the range of scales investigated
              for the high-speed regions and the resulting exponents
              are plotted with the $95$\% confidence intervals for
              these fits. The $y^+$ positions and the two Reynolds
              numbers $Re_{\theta}$ are those in figures \ref{fig:figure9} to \ref{fig:figure12}.}
            \label{fig:figure18}
\end{figure}

\textcolor{black}{As mentioned in subsection \ref{sec:Structure detection}, this paper's
  analysis can be repeated equally well on structures of positive
  streamwise fluctuating velocity. We provide examples of results
  obtained with $u_{th} = 0.4 u'_{300^+}$ in figure \ref{fig:figure17} and table \ref{tab:t3}. There are indeed no significant differences in the results for
  the low and high speed attached flow regions, except for a lower but
  consistent value of $C_1$ and for a consistently lower value of
  $C_2$ in the lower $Re_{\theta}$ case. Figures \ref{fig:figure9} to \ref{fig:figure13} can be reproduced for this positive threshold $u_{th} = 0.4 u'_{300^+}$ and   show the exact same trend with $p$ increasing while $q$ is   decreasing with increasing $y^+$. However, whereas $p$ takes values   similar to those for $u_{th} = -0.4 u'_{300^+}$ in the lower   $Re_{\theta}$ case, it does not do so in the higher $Re_{\theta}$   case. As a result $p+q$ is quite close to $-1$ in the lower   $Re_{\theta}$ case but less so, and in fact closer to $-1.1$ on  average, for the higher $Re_{\theta}$ (see figure \ref{fig:figure18}).}

\begin{table}
\begin{center}
\def~{\hphantom{0}}
\begin{ruledtabular}
  \begin{tabular}{l|ccccc|ccccc}
    \multirow{1}{*}{$Re_\theta$} &
      \multicolumn{5}{c|}{20600} &
      \multicolumn{5}{c}{8100} \\   
    \hline
    $y^+$ & 90 & 195 & 305 & 450 & 630 & 52 & 88 & 125 & 198 & 306 \\
    $C_1$ & 0.02 & 0.02 & 0.01 & 0.02 & 0.02 & 0.02 & 0.02 & 0.02 & 0.03 & 0.02 \\
    $C_2$ & 0.35 & 0.36 & 0.34 & 0.38 & 0.37 & 0.28 & 0.29 & 0.28 & 0.31 & 0.29 \\
  \end{tabular}
  \caption{Values of the constants $C_1$ and $C_2$ in the form $-C_{1}
    + C_{2} (\lambda/\delta)^{-2}$ of the PDF of
    $\lambda/\delta$.}
    \label{tab:t3}
    \end{ruledtabular}
  \end{center}
\end{table}

\newpage \clearpage

\bibliography{biblio_old}

\end{document}